\documentclass[prd,%preprint, 
showpacs,superscriptaddress,preprintnumbers,amsmath,amssymb]{revtex4}

\usepackage{multirow} 
\usepackage{epsfig}
\usepackage{amsmath}
\usepackage{amsfonts}

\usepackage[normalem]{ulem}  % \sout{old text} for strikeout
\usepackage[dvips]{color} % For blue in-text comments and additions

\renewcommand\sout{\bgroup \color{red} \ULdepth=-.5ex \ULset}

\newcommand{\Ex}[2]{\ifmmode{#1\times10^{#2}}\else{$#1\times10^{#2}$}\fi}
\makeatletter
\renewcommand{\theequation}{\arabic{section}.\arabic{equation}}
\makeatother

\begin{document}
%\vspace*{-8mm}
\begin{flushright}
KEK-TH-1537 \\
RIKEN-QHP-17
\end{flushright}

%\vspace{-5mm}

\title{Non-Abelian statistics of vortices with non-Abelian Dirac fermions}

\vspace{4mm}

\author{Shigehiro Yasui}
\email{yasuis@post.kek.jp}
\affiliation{KEK Theory Center, Institute of Particle 
and Nuclear Studies,
High Energy Accelerator Research Organization (KEK),
1-1 Oho, Tsukuba, Ibaraki 305-0801, Japan}

\author{Yuji~Hirono}
\email{hirono@nt.phys.s.u-tokyo.ac.jp}
\affiliation{
Department of Physics, University of Tokyo, 
Hongo~7-3-1, Bunkyo-ku, Tokyo 113-0033, Japan}

\author{Kazunori Itakura} 
\email{kazunori.itakura@kek.jp}
\affiliation{KEK Theory Center, Institute of Particle 
and Nuclear Studies,
High Energy Accelerator Research Organization (KEK),
1-1 Oho, Tsukuba, Ibaraki 305-0801, Japan}
\affiliation{Department of Particle and Nuclear Studies, 
Graduate University for Advanced Studies (SOKENDAI), 
1-1 Oho, Tsukuba, Ibaraki 305-0801, Japan}

\author{Muneto Nitta}
\email{nitta@phys-h.keio.ac.jp}
\affiliation{Department of Physics, and Research and Education Center 
for Natural Sciences, Keio University, 4-1-1 Hiyoshi, Yokohama, 
Kanagawa 223-8521, Japan}

\date{\today}

\begin{abstract}
We extend our previous analysis on the exchange statistics of vortices 
having a single Dirac fermion trapped in each core, to the case where 
vortices trap two Dirac fermions with U(2) symmetry. Such a system 
of vortices with non-Abelian Dirac fermions appears in color 
superconductors at extremely high densities, and in supersymmetric QCD. 
We show that the exchange of two vortices having doublet Dirac fermions 
in each core is expressed by non-Abelian representations of a braid group, 
which is explicitly verified in the matrix representation 
of the exchange operators when the number of vortices is up to four.
We find that the result contains the matrices previously obtained for 
the vortices with a single Dirac fermion in each core as a special case.
The whole braid group does not immediately imply
%represent 
non-Abelian statistics of identical particles 
because it also contains exchanges between vortices 
with different numbers of Dirac fermions. 
However, we find that it does contain, as its subgroup, 
genuine non-Abelian statistics for the exchange of
%to exchange 
the identical particles, that is, vortices with the same number of 
Dirac fermions. 
This result is surprising compared with conventional understanding because   
all Dirac fermions are defined locally at each vortex, unlike 
the case of Majorana fermions for which Dirac fermions are defined 
non-locally by Majorana fermions %on
located at two spatially separated vortices.

\end{abstract}

\pacs{05.30.Pr, 74.25.Uv, 67.85.-d, 21.65.Qr, 03.67.-a}

\maketitle

\setcounter{page}{1}
\setcounter{footnote}{0}
\renewcommand{\thefootnote}{\arabic{footnote}}

\allowdisplaybreaks

\section{Introduction}

Topological insulators/superconductors have an attractive 
property that some of them possess quantum vortices which trap 
zero-energy, Majorana or Dirac, fermions in their cores 
\cite{SchnyderRFL:08,Kitaev:08}. 
The existence of such zero-energy fermions is topologically 
protected and is robust against small perturbations 
\cite{Volovik:1999,Roy:2010}. Thus, when we consider 
adiabatic manipulation of vortices such as interchanging the positions of 
two vortices, we can treat the vortices as objects that are always 
accompanied by zero-energy fermions. 
In particular, according to the recent discoveries, the exchange of such vortices can be represented by a non-trivial representation of a braid group, 
whose precise form is determined by the trapped zero-energy 
Majorana fermions \cite{Read:1999fn,Ivanov:2001,Wilczek:2009,Yasui:2010yh,
Yasui:2010yw,Hirono:2012ad} 
and Dirac fermions \cite{Yasui:2011gk}. 
The exchange of vortices with Majorana fermions 
gives non-Abelian statistics 
because they are all identical particles. 
The statistics is called non-Abelian because quantum states 
of two vortices transform non-diagonally under the exchange of two 
vortices (i.e., the exchange operation is described by non-diagonal 
matrices acting on the quantum states), and two adjacent exchange 
operations (such as those for vortex pairs (1,2) and (2,3)) do not 
commute with each other. 
This is highly contrasted with the ordinary 
statistics where only a phase factor ${\rm e}^{i\theta}$ 
appears under the exchange of two particles ($\theta=0$ for the 
Bose-Einstein, $\theta=\pi$ for the Fermi-Dirac, and others for 
the anyon statistics), and two adjacent operations are commutative. 
On the other hand, it is unclear whether 
the exchange of vortices with Dirac fermions gives a 
non-Abelian statistics or not because, in general, 
it exchanges different particles, that is, 
vortices with different numbers of Dirac fermions. 
For U(1) Dirac fermions, vortices can be distinguished 
by the occupancy of Dirac fermions. 

According to the classification of topological 
insulators/superconductors \cite{SchnyderRFL:08,Kitaev:08} and 
its extension to the case with topological defects \cite{Teo:2010,Roy:2010},
the vortices with the Majorana or Dirac fermions are categorized 
into different types: class D for the Majorana and classes C and DIII 
for the Dirac. However, the essential difference between the 
vortices with Majorana fermions and Dirac fermions is the parity 
of the number of zero-energy Majorana fermions trapped to a single 
vortex. Notice that a single Dirac fermion corresponds to two 
Majorana fermions. Thus, when a vortex traps an even number of the 
Majorana fermions, it should be regarded as the Dirac case, while  
an odd number of the Majorana fermions, just as the Majorana 
case \cite{Roy:2010}. 
So far, non-Abelian unitary transformations (non-Abelian representations 
of a braid group) are 
found both in the Majorana and Dirac cases. 
In the Majorana case, the non-Abelian statistics was first discovered 
when a vortex has a single Majorana fermion \cite{Ivanov:2001} 
and later when a vortex can have multiple 
Majorana fermions with non-Abelian symmetry \cite{Yasui:2010yh,
Yasui:2010yw,Hirono:2012ad}. 
In contrast, in the Dirac case, 
while non-Abelian representation of a braid group 
is analyzed when a vortex traps a single Dirac fermion \cite{Yasui:2011gk}, 
it is unclear if it gives genuine non-Abelian statistics.\footnote{
In the previous paper for the U(1) Dirac case \cite{Yasui:2011gk}, 
we used the word ``non-Abelian statistics" to imply non-Abelian 
representations of the braid group. However, in the present paper, we 
use the word ``non-Abelian statistics" only for exchanges of two 
identical states, which should be a subgroup of the whole representation.}
The present paper discusses 
the non-Abelian representation of a braid group for the 
exchange of vortices which have multiple Dirac fermions with 
non-Abelian symmetry, to complete the series of analyses. 
We then show that it indeed contains genuine non-Abelian 
statistics of the exchanges of identical particles, that is, vortices 
with the same numbers of Dirac fermions. This result is somewhat 
surprising compared with conventional understanding because 
non-Abelian statistics appears
in a system with
only locally-defined Dirac fermions. 

There is an important difference between the vortices with Majorana
fermions and Dirac fermions. Consider a system of vortices each of which 
traps only a single Majorana fermion. Recall that the Majorana fermion has 
a unique property that there is no distinction between a particle and a hole
(anti-particle) \cite{Majorana:1937vz}. Thus, in order to define the Fock 
vacuum, one has to introduce a Dirac fermion by using two Majorana fermions 
that belong to different vortices \cite{Read:1999fn,Ivanov:2001}. Therefore, 
the Dirac fermions thus constructed are non-local objects. When there 
are $2m$ vortices, the total Hilbert space of the zero-energy fermions 
has a dimension (degeneracy) of $2^m$, where each zero-energy 
Dirac fermion has the 
dimension 2 (empty or occupied). On the other hand, in the system of vortices 
each of which traps a single Dirac fermion, we can immediately construct the 
Hilbert space without introducing `non-local' Dirac fermions 
\cite{Yasui:2011gk}, and find 
a non-Abelian representation of a braid group 
(but not non-Abelian statistics) for the exchange of 
two vortices. When there are $2m$ vortices, the dimension of 
the Hilbert space is $2^{2m}=4^m$. 

In the Majorana case, increasing the number of Majorana 
fermions in a single vortex brings in an interesting nontrivial 
structure. When 
the multiple Majorana fermions are in the vector representations of 
SO(3) \cite{Yasui:2010yh} and, in general, SO($2N-1$) \cite{Hirono:2012ad},
the exchange matrices of two vortices are given as tensor products of 
the matrices that appear in the single Majorana case (called the Ivanov 
matrices) and generators of 
the Coxeter group of the $A_{2m-1}$ type (for $2m$ vortices). Besides, the 
dimension of the Hilbert space of zero-energy fermions becomes larger 
than that of the single Majorana case by the internal degrees of freedom.
For example, when three (triplet) Majorana fermions with SO(3) symmetry 
are trapped, the dimension is $2^{3m}$ which should be compared 
with $2^m$ for the case with a single Majorana fermion. It is not known 
if a similar structure exists and how the dimension of the Hilbert space 
is enlarged, when increasing the number of Dirac fermions at each vortex. 
The purpose of the present paper is to show the explicit forms of
the non-Abelian representation of a braid group for 
vortices with multiple Dirac fermions, 
and that it contains genuine non-Abelian statistics in the sectors 
of the exchange of vortices with the same numbers of Dirac fermions.
As the simplest but non-trivial example, we focus on the Dirac fermions 
with U(2) symmetry. Extension to general cases will become more 
complicated, but should be straightforward. 

It should be noticed that, since the vortices with zero-energy 
fermions are characterized by topology, they appear in many different 
quantum systems. For example, vortices with the Majorana fermions are 
realized in chiral $p$-wave superconductors \cite{Volovik:1999}, 
such as Sr$_2$RuO$_4$ \cite{Maeno:2003}, 
in chiral $p$-wave superfluids,  such as the A-phase of $^3$He 
in 2+1 dimensions, 
and also in other systems \cite{Sato:2003,Fu:2008}. 
Vortices with the Dirac fermions are identified with the integer (singular) 
vortices in the $^3$He A-phase in 2+1 dimensions \cite{Kawakami:2010} 
the normal ``$o$" vortices in the $^3$He B-phase 
in 3+1 dimensions \cite{Kopnin}
and also with dislocation lines in topological insulators \cite{Imura:2011}.
In these examples, the vortex has only a single 
zero-energy Majorana or Dirac fermion at 
its core. So far, there is no condensed-matter example of vortices with 
more than two Majorana fermions.
However, we know at least two examples in high-energy physics. In fact, 
the existence of zero-energy modes in the vortex-fermion system was 
discussed long time ago in the context of relativistic quantum field 
theory \cite{Jackiw:1981ee}. 

The primary example is the color superconductor 
in QCD which could exist in extremely high density matter such as in the 
cores of neutron stars \cite{Balachandran:2005ev}. 
In particular, vortices with non-Abelian symmetries appear in the 
color-flavor locked (CFL) phase where the original color SU(3) 
and flavor SU(3) symmetry in the vacuum breaks down to SU(3) 
with color and flavor degrees are locked \cite{Alford:2007xm}. 
The SU(3) CFL symmetry is further broken down to a U(2) symmetry 
in the core of non-Abelian vortices \cite{Nakano:2007dr}.
It has been recently shown \cite{Fujiwara:2011za} by using the index 
theorem that there appear two types of non-Abelian vortices: 
the vortex which traps a triplet of zero-energy 
Majorana fermions \cite{Yasui:2010yw} and the 
vortex which traps an U(2) doublet of zero-energy Dirac fermions. 
While the former gives an example of non-Abelian statistics of 
vortices with multiple Majorana fermions \cite{Yasui:2010yh}, 
the latter gives an example of non-Abelian 
representation of a braid group for the 
exchange of vortices with  
doublet Dirac fermions focused in the present paper. The secondary example
is the non-Abelian vortices in supersymmetric QCD with U($N$) gauge 
symmetry \cite{Hanany:2003hp} (see Ref.~\cite{review} for reviews).  
In this case, the color U($N$) symmetry and the flavor SU($N$) 
symmetry are spontaneously broken down to the SU($N$) color-flavor 
locked symmetry by the scalar quark condensates in the vacuum, 
and it is further broken down to a U($N-1$) symmetry in the core 
of non-Abelian vortices. These vortices contain one singlet and 
one (${N-1}$)-plet of zero-energy Dirac fermions in their core. 
Therefore, it gives an example of non-Abelian statistics of 
vortices with arbitrary number of Dirac fermions.

Before finishing Introduction, let us briefly comment on the 
potential application to quantum computers \cite{Nayak:2008zza,Kitaev:2006}. 
As discussed above, the system of vortices with zero-energy fermions 
is robust against small perturbations from environment, and has the 
Hilbert space with a large dimension. These are desirable properties as 
quantum computers. Comparing the Majorana and Dirac cases, it should be 
noticed that the Dirac case is simpler because we do not have to introduce 
non-local Dirac fermions. Since vortex systems with non-Abelian symmetry 
have larger dimensions, it would be worth to consider the case with 
non-Abelian Dirac fermions, even though it has not been realized in 
laboratory.

This paper is organized as follows.
In Section \ref{sec:single}, we briefly summarize the non-Abelian statistics 
for the vortices trapping a single Dirac fermion with U(1) symmetry 
at each core, as presented in Ref.~\cite{Yasui:2011gk}.
In Section \ref{sec:U(2)}, we discuss the non-Abelian 
representation of the braid group for the 
exchange of the 
vortices trapping doublet Dirac fermions with U(2) symmetry. 
In Section \ref{sec:relation}, we discuss the difference between 
non-Abelian representations of the braid group for the 
exchange of the Dirac fermion with U(1) symmetry 
and that with U(2) symmetry. 
In Section \ref{sec:genuine}, we show that the whole braid group contains 
genuine non-Abelian statistics as its subgroup, 
in the sectors of the vortices with the same number of Dirac fermions. 
Section \ref{sec:summary} is devoted to 
a summary. In Appendices, we present detailed supplementary information. 
In Appendix \ref{sec:U(1)tau}, we give the transformation matrices 
for U(1) (singlet) Dirac fermions.
In Appendix \ref{sec:n=4} we give the Hilbert space and the exchange matrices 
for $n=4$ U(2) Dirac vortices.
In Appendix \ref{sec:square}, we discuss restricted Hilbert subspaces, in which two successive exchanges of vortices is equal to the identity.

%%%%%%%%%%%%%%%%%%%%%%%%%%%%%%%%%%%%%%%%%%%%%%%%%%%%%%%%%%%%%%%
\section{Non-Abelian representation of braid group for exchange of 
U(1) Dirac vortices}\label{sec:single}
\setcounter{equation}{0}

Let us first explain how the non-Abelian 
representation of the braid group appears 
in the system of vortices having a single Dirac fermion in each 
core, which corresponds to considering the Dirac fermion with
the U(1) symmetry. This is a brief summary of the recent work done 
by three of us~\cite{Yasui:2011gk}. 
We highlight the similarities to and differences from the case with 
Majorana fermions. Below, we call vortices with the Dirac fermions
``the Dirac vortices'', while vortices with the Majorana fermions, 
``the Majorana vortices''.

Consider $n$ Dirac vortices which are labelled by $k=1,\cdots, n$. 
The number of vortices, $n$, can be arbitrary in contrast with the 
case of Majorana vortices where we define Dirac fermions by using 
two Majorana fermions, and thus the total fermion number is even. 
The operator $\hat\psi_k$ denotes the Dirac fermion of the $k$-th 
vortex. Together with its hermitian conjugate $\hat\psi^\dag_k$, 
they satisfy the following algebra:
\begin{eqnarray}
\{ \hat{\psi}_{k}, \hat{\psi}_{\ell}^{\dag} \}=\delta_{k\ell}, \quad
\{ \hat{\psi}_{k}, \hat{\psi}_{\ell} \}=0,  \quad
\{ \hat{\psi}_{k}^{\dag}, \hat{\psi}_{\ell}^{\dag} \}=0. 
\label{eq:anticommutation}
\end{eqnarray}
We regard $\hat\psi_k$ and $\hat\psi_k^\dag$ as the 
annihilation and creation operators, respectively.
Exchange of the $k$-th and $(k+1)$-th vortices, 
which is denoted by $T_{k}$, induces the following exchange of
$\hat{\psi}_{k}$ and $\hat{\psi}_{k+1}$,
\begin{eqnarray}
T_k : \left\{ 
\begin{array}{l}
 \hat{\psi}_{k} \quad \rightarrow\, \hat{\psi}_{k+1} \\
 \hat{\psi}_{k+1} \rightarrow -\hat{\psi}_{k}  
\end{array}
\right. ,
\label{eq:T_k_1}
\end{eqnarray}
with the rest $\hat{\psi}_{\ell}$ ($\ell \neq k$ and $k+1$) unchanged.
We note that $T_{k}$'s satisfy the braid relations, 
\begin{eqnarray}
\mbox{(i) } &&T_{k}T_{\ell}T_{k}=T_{\ell}T_{k}T_{\ell}
\quad \mbox{ for } \quad |k-\ell|=1\, , \\ 
\mbox{(ii) } && T_{k}T_{\ell}=T_{\ell}T_{k} 
\quad \mbox{ for }\quad |k-\ell|>1\, ,
\end{eqnarray} 
as a general rule of exchange operations. 
Notice that these relations and the transformation 
(\ref{eq:T_k_1}) are the same as in the case of the Majorana vortices,
while the fermion operators $\hat\psi_k$ and $\hat\psi^\dag_k$ 
now satisfy Eq.~(\ref{eq:anticommutation}), instead of the Clifford 
algebra $\{\hat{\gamma}_k,\hat{\gamma}_\ell\}=2\delta_{k\ell}$ 
satisfied by the Majorana fermion operators $\hat{\gamma}_k$'s. 
Therefore,
one has the same property of the operator $T_k$, namely, four-time
successive application of $T_k$ is equivalent to the identity, $(T_k)^4=1$.
This fact by itself suggests that the exchange of these vortices shows 
a representation of the braid group different from the ordinary Bose-Einstein, 
Fermi-Dirac or Abelian anyon one which shows Abelian representation of 
the braid group. 
In fact, the exchange statistics of the Majorana vortices 
showing the same property $(T_k)^4=1$ turned out to be non-Abelian
\cite{Ivanov:2001,Yasui:2010yh, Hirono:2012ad}, 
and we will discuss below if the same is true for the Dirac vortices.

The transformation~(\ref{eq:T_k_1}) can be represented by the
following unitary operator 
\begin{eqnarray}
\hat{\tau}_{k}^{\rm s} \equiv 1+ \hat{\psi}^{}_{k+1} \hat{\psi}_{k}^{\dag} 
                    + \hat{\psi}_{k+1}^{\dag} \hat{\psi}^{}_{k} 
                    - \hat{\psi}_{k+1}^{\dag} \hat{\psi}^{}_{k+1} 
                    - \hat{\psi}_{k}^{\dag} \hat{\psi}^{}_{k}  
                    + 2 \hat{\psi}_{k+1}^{\dag}\hat{\psi}_{k+1}^{} 
                        \hat{\psi}_{k}^{\dag} \hat{\psi}_{k}^{}\, ,
\label{eq:exchange_Dirac_rep1}
\end{eqnarray}
so that
$\hat{\tau}_{k}^{\rm s} \hat{\psi}_{\ell} (\hat{\tau}_{k}^{\rm s})^{-1}$ 
($\ell=1, \cdots, n$) reproduces the transformation law.
The superscript ``$\mathrm{s}$'' implies the singlet Dirac fermion.
One can also confirm by a straightforward calculation that 
$\hat{\tau}_{k}$'s satisfy the braid relations, 
\begin{eqnarray}
\mbox{(i')} && \hat{\tau}_{k}^{\rm s} \hat{\tau}_{\ell}^{\rm s}
 \hat{\tau}_{k}^{\rm s} = \hat{\tau}_{\ell}^{\rm s} \hat{\tau}_{k}^{\rm s}
 \hat{\tau}_{\ell}^{\rm s} \quad \mbox{ for } \quad |k-\ell|=1\, ,\\
\mbox{(ii')} && \hat{\tau}_{k}^{\rm s} \hat{\tau}_{\ell}^{\rm s} =
 \hat{\tau}_{\ell}^{\rm s} \hat{\tau}_{k}^{\rm s} 
\quad \mbox{ for } \quad |k-\ell|>1\, .
\end{eqnarray}
Having the explicit form of the exchange operator 
(\ref{eq:exchange_Dirac_rep1}), we are able to check 
the representation of the braid group for the 
exchange of the Dirac vortices. First of all, as we mentioned above,
four successive exchanges of the $k$-th and $(k+1)$-th vortices indeed 
yield the identity:
\begin{equation} 
(\hat{\tau}_{k}^{\rm s})^{4}=1\, , \label{eq:fourth}
\end{equation}
while two successive exchanges do not, 
\begin{equation}
 (\hat{\tau}_{k}^{\rm s})^{2} =
  (1-2\hat{\psi}_{k}^{\dag}\hat{\psi}^{}_{k})
  (1-2\hat{\psi}_{k+1}^{\dag}\hat{\psi}^{}_{k+1})
  \neq 1 \, .\label{eq:second}
\end{equation} 
Next, it should be noted that $\hat{\tau}_{k}^{\rm s}$ and
$\hat{\tau}_{k+1}^{\rm s}$ are not commutative; 
\begin{equation}
\left[\hat{\tau}_{k}^{\rm s}\, ,
\hat{\tau}_{k+1}^{\rm s}\right] \neq 0.
\end{equation}
Therefore, one observes at the operator level that the exchange 
operation of two Dirac vortices is in general 
non-Abelian. However, this observation 
does not necessarily mean that the exchange is always non-Abelian. 
In order to confirm whether the exchange operation is indeed non-Abelian, 
we have to check if the commutator 
$\left[\hat{\tau}_{k}^{\rm s}\, , \hat{\tau}_{k+1}^{\rm s}\right]$ does not 
vanish in the matrix representation.

Let us consider the matrix representation of the operator 
$\hat{\tau}_{k}^{\rm s}$.
As a basis of the Hilbert space, we choose the Fock states defined by 
the Dirac fermion operators  $\hat{\psi}_{k}$'s. 
One of the merits of the Dirac vortices is that we 
can construct the Hilbert space naturally by using {\it locally-defined} 
Dirac fermions, which is in clear contrast with the Majorana case.
We first define the Fock vacuum state $|0\rangle$ by 
\begin{eqnarray} 
 \hat{\psi}_{\ell}|0\rangle=0 \quad \mbox{ for all } \ \ell.
\end{eqnarray}
Then, by acting successively the creation operators 
$\hat{\psi}_{\ell}^{\dag}$'s on the vacuum,
we obtain the $f$-particle state ($0 \le f \le n$)
\allowdisplaybreaks[0]
\begin{eqnarray}
&&|0\cdots 0 1 \cdots 1 \cdots 1 0\cdots 0\rangle
 =\hat{\psi}^{\dag}_{\ell_{1}} \dots \hat{\psi}^{\dag}_{\ell_{i}} \dots
 \hat{\psi}^{\dag}_{\ell_{f}} |0\rangle\, , \\ 
&&\ \check{1}\qquad    \check{\ell}_{1}\quad \, \check{\ell}_{i}\quad \,
 \check{\ell}_{f}\quad \ \, \check{n}\nonumber 
\end{eqnarray}
\allowdisplaybreaks
in which the $\ell_{i}$-th ($i=1, \cdots, f$) vortex is occupied by a
Dirac fermion, while the other vortices are empty. 

When we have only one vortex, $n=1$ ($k$-th vortex), there are 
two Fock states, $|0\rangle$ and $|1\rangle \equiv 
\hat{\psi}_{k}^{\dag}|0\rangle$.

When we have two vortices, $n=2$ ($k$-th and $(k+1)$-th vortices), 
there are $2^2=4$ Fock states,
$|00\rangle \equiv |0\rangle$, 
$|10\rangle \equiv \hat{\psi}_{k}^{\dag}|0\rangle$,
$|01\rangle \equiv \hat{\psi}_{k+1}^{\dag}|0\rangle$ and
$|11\rangle \equiv \hat{\psi}_{k}^{\dag} \hat{\psi}_{k+1}^{\dag}|0\rangle$.

We can similarly obtain the Fock states for any number of vortices $n$.
The basis of the whole Hilbert space for the $n$-vortex system 
is given by a tensor product of the 4 Fock states constructed at each vortex. 
Because the fermion number operator $\hat{f}^{\mathrm{s}}
\equiv\sum_{i=1}^{n} \hat{\psi}_{i}^{\dag}\hat{\psi}^{}_{i}$ commutes 
with $\hat{\tau}_{\ell}^{\mathrm{s}}$ for $\ell=1, \cdots,n-1$, 
the fermion number $f$ (an eigenvalue of $\hat{f}^{\mathrm{s}}$) 
is a conserved quantity under the exchange of vortices. 
Thus, the whole Hilbert space $\mathbb{H}^{(n)}$ for $n$ vortices 
(with the dimension $2^n$) can be decomposed into subspaces 
$\mathbb{H}^{(n,f)}$ labelled further by
the total fermion number $f$: 
$\mathbb{H}^{(n)} = \oplus_{f=1}^{n} \mathbb{H}^{(n,f)}$.
Then, in each 
subspace, the operators $\hat{\tau}_{\ell}^{\rm s}$'s are represented by 
matrices, whose explicit expressions 
are shown up to $n=4$ in Appendix \ref{sec:U(1)tau}. 
One can confirm that the matrix representations of the exchange 
operators $\hat\tau_{\ell}$'s are indeed non-Abelian for $n\ge 3$.

%\com{COMMENT}

%%%%%%%%%%%%%%%%%%%%%%%%%%%%%%%%%%%%%%%%%%%%%%%%%%%%%%%%%%%%%%%%%%%%%%%
\section{Non-Abelian representation of  braid group for exchange of 
U(2) Dirac vortices}\label{sec:U(2)}
\setcounter{equation}{0}

Now let us turn to the case with vortices which trap two massless 
Dirac fermions having the ``pseudo-spin'' U(2) symmetry. In particular,
we consider Dirac fermions in the doublet of U(2), which are denoted by 
$\hat{\psi}_{k}^{a}$ $(a=1,2)$ for the $k$-th vortex.
We also use the vector notation for the doublet 
$(\hat{\psi}_{k}^{1},\hat{\psi}_{k}^{2})^{\rm t}$. The exchange of
 the $k$-th and $(k+1)$-th vortices induces transformation of 
the Dirac fermions. Here we consider the simplest transformation  
similarly as the single Dirac fermion shown in Eq.~(\ref{eq:T_k_1}):
\begin{eqnarray}
T_k : \left\{ 
\begin{array}{l}
 \hat{\psi}_{k}^{a} \quad \rightarrow\, \hat{\psi}_{k+1}^{a} \\
 \hat{\psi}_{k+1}^{a} \rightarrow -\hat{\psi}_{k}^{a}  
\end{array}
\right. ,
\label{eq:T_k}
\end{eqnarray}
for each $a=1,2$, with the rest $\hat{\psi}_{\ell}^{a}$ 
($\ell \neq k$ and $k+1$) unchanged. 
This transformation is expressed by the unitary operator
\begin{eqnarray}
\hat{\tau}_{k} &\equiv & \prod_{a=1,2} \left(1 + \hat{\psi}_{k+1}^{a}\hat{\psi}_{k}^{a\dag} + \hat{\psi}_{k+1}^{a\dag}\hat{\psi}_{k}^{a} - \hat{\psi}_{k+1}^{a\dag}\hat{\psi}_{k+1}^{a} - \hat{\psi}_{k}^{a\dag}\hat{\psi}_{k}^{a} + 2\hat{\psi}_{k+1}^{a\dag}\hat{\psi}_{k+1}^{a} \hat{\psi}_{k}^{a\dag}\hat{\psi}_{k}^{a}\right),
\label{eq:tau_full}
\end{eqnarray}
which is invariant under the U(2) transformation, 
$(\hat{\psi}_{k}^{1},\hat{\psi}_{k}^{2})^{\rm t} \rightarrow 
\exp(i\varphi) \exp(i\vec{\theta} \cdot \vec{\sigma }/2) 
(\hat{\psi}_{k}^{1},\hat{\psi}_{k}^{2})^{\rm t}$ with $\varphi$ and 
$\vec \theta$ being parameters and $\vec \sigma$ the Pauli matrices. 
We confirm that $\hat{\tau}^{}_{k} \hat{\psi}_{\ell}^{a} \hat{\tau}_{k}^{-1}$
($a=1$, $2$) reproduces the transformation (\ref{eq:T_k}), and that
$\hat{\tau}_{k}$'s satisfy the braid relations, (i')
$\hat{\tau}_{k}\hat{\tau}_{\ell}\hat{\tau}_{k}=\hat{\tau}_{\ell}\hat{\tau}_{k}\hat{\tau}_{\ell}$
for  $|k-\ell|=1$ and (ii')
$\hat{\tau}_{k}\hat{\tau}_{\ell}=\hat{\tau}_{\ell}\hat{\tau}_{k}$ for
$|k-\ell|>1$.
Similarly as in the U(1) case (see Eqs.~(\ref{eq:fourth}) and 
(\ref{eq:second})), the exchange operator $\hat \tau_k$ satisfies
\begin{eqnarray}
(\hat{\tau}_{k})^{4} = 1\, ,
\label{eq:tau^4}
\end{eqnarray}
while two successive exchanges do not go back to the identity 
at the operator level:
\begin{eqnarray}
(\hat{\tau}_{k})^{2} = (1-2\hat{\psi}_{k}^{1\,\dag}\hat{\psi}_{k}^{1}) (1-2\hat{\psi}_{k}^{2\,\dag}\hat{\psi}_{k}^{2}) (1-2\hat{\psi}_{k+1}^{1\,\dag}\hat{\psi}_{k+1}^{1}) (1-2\hat{\psi}_{k+1}^{2\,\dag}\hat{\psi}_{k+1}^{2}) \neq 1\, .
\label{eq:tau^2}
\end{eqnarray}
We can also check that $\hat \tau_k$ and $\hat \tau_{k+1}$ are 
non-commutative, which suggests that the representation of the braid group 
for the exchange of U(2) Dirac vortices is non-Abelian.

We define the number operator of the Dirac fermions 
\begin{eqnarray}
\hat{f} = \sum_{\ell=1}^{n} \sum_{a=1,2} \hat{\psi}_{\ell}^{a\,\dag} \hat{\psi}_{\ell}^{a},
\end{eqnarray}
whose eigenvalues $f$ give the total number of zero-energy Dirac
fermions  in the vortex system ($0 \le f \le 2n$).
We note that $\sum_{a=1,2}\psi_{k}^{a\,\dag} \psi_{k}^{a}$ is U(2)
invariant, hence $\hat{f}$ is also U(2) invariant.
We also note that $\hat{f}$ commutes with $\hat\tau_{\ell}$ for $\ell=1,\cdots,n-1$, hence $f$ is
a conserved quantity under the exchange of two Dirac vortices.

We can construct the Hilbert space by successively applying
$\hat{\psi}_{\ell}^{a\dag}$ to the Fock vacuum $|0\rangle$ defined by
\begin{eqnarray}
\hat{\psi}_{\ell}^{a}|0\rangle=0\ \mbox{ for all \ $\ell$ \ and \ $a=1$, $2$}.
\end{eqnarray}
Below, we explicitly construct the Hilbert spaces when the numbers of 
vortices are $n=1,2,3$, and see the non-Abelian properties of the exchange 
operation in the matrix representations for $n=2$ and 3. The results for 
$n=4$ vortices are quite involved and are shown in Appendix~\ref{sec:n=4}.

%%%%%%%%%%%%%%%%%%%%%%%%%%%%%%%%%%
\subsection{The case of $n=1$}\label{subsec:n=1}

When we have only one vortex, we are not able to discuss the exchange 
of vortices. However, 
let us consider this case to demonstrate how to construct the 
Hilbert space for the U(2) Dirac vortex. We have two massless 
Dirac fermion operators
$\hat \psi_k^1$ and $\hat \psi^2_k$ (to avoid notational confusion, we use the 
label $k$ to specify this single vortex). 
By applying the creation operators $\hat{\psi}^{1\dag}_{k}$ and 
$\hat{\psi}^{2\dag}_{k}$ to the Fock vacuum $| 0 \rangle$, we obtain 
$2\times 2=4$ energetically degenerate states (2 for empty/occupied, 
and another 2 for $a=1,2$). 
Let us introduce the notation $\vert {\cal R}_f \rangle $ where
${\cal R}$ is the representation of U(2) group and $f$ is the total 
number of the Dirac fermions. Namely, 
\begin{eqnarray}
| {\bf 1}_{0} \rangle &\equiv& | 0 \rangle, \nonumber \\
| {\bf 2}_{1} \rangle &\equiv&
\left(
\begin{array}{c}
 \hat{\psi}_{k}^{1\dag} \\
 \hat{\psi}_{k}^{2\dag} 
\end{array}
\right) | 0 \rangle, \label{eq:states_n=1}\\
| {\bf 1}_{2} \rangle &\equiv& \hat{\psi}_{k}^{1\dag} \hat{\psi}_{k}^{2\dag} | 0 \rangle.
\nonumber 
\end{eqnarray}
The bold numbers {\bf 1, 2} imply the singlet and doublet 
representations of the U(2) group.
Notice that the fully occupied state 
$\hat{\psi}_{k}^{1\dag} \hat{\psi}_{k}^{2\dag} | 0 \rangle $ is invariant
under the U(2) transformation, thus it belongs to the singlet 
representation. Hence, there are two singlet states 
$| {\bf 1}_{0} \rangle$ (empty) and $| {\bf 1}_{2} \rangle$ 
(fully-occupied by two fermions), and one doublet state 
$| {\bf 2}_{1} \rangle$ (occupied by one fermion).
We use these states as the basis to span the whole Hilbert space 
of the zero-energy states.
Therefore, we have decomposed the representations of the U(2) pseudo-spin 
as 
\begin{equation}
{\bf 1}_{0} + {\bf 2}_{1} + {\bf 1}_{2}\, , \label{eq:single}
\end{equation}
and correspondingly the whole Hilbert space of a single vortex 
$\mathbb{H}^{\{n=1\}}$ into 
 a direct sum
\begin{eqnarray}
\mathbb{H}^{\{n=1\}} = \mathbb{H}^{{\bf 1}_{[0]}} \oplus 
\mathbb{H}^{{\bf 2}_{[1]}} \oplus \mathbb{H}^{{\bf 1}_{[2]}},
\end{eqnarray}
where $\mathbb{H}^{{\bf 1}_{[0]}} \equiv \{ | {\bf 1}_{0} \rangle \}$,
$\mathbb{H}^{{\bf 2}_{[1]}} \equiv \{ | {\bf 2}_{1} \rangle \}$ and
$\mathbb{H}^{{\bf 1}_{[2]}} \equiv \{ | {\bf 1}_{2} \rangle \}$.

\subsection{The case of $n=2$}\label{subsec:n=2}
Consider the case when we have two Dirac vortices which are 
respectively labelled by $k$ and $k+1$. First of all, we can use the 
decomposition  for a single Dirac vortex,  Eq.~(\ref{eq:single}),
to decompose the whole Hilbert space of two 
Dirac vortices into representations. 
Since each vortex contains the representations 
$({\bf 1}_{0} + {\bf 2}_{1} + {\bf 1}_{2})_{{\rm vortex}\,\ell}$ 
($\ell=k$, $k+1$),
the total representation is obtained as a tensor product of them, 
which can be decomposed as
\begin{eqnarray}
&&\hspace{-2cm}({\bf 1}_{0} + {\bf 2}_{1} + {\bf 1}_{2})_{{\rm vortex}\, k} \otimes ({\bf 1}_{0} + {\bf 2}_{1} + {\bf 1}_{2})_{{\rm vortex}\, k+1} \nonumber \\
&=& {\bf 1}_{00} + {\bf 1}_{11} + {\bf 1}_{20} + {\bf 1}_{02} + {\bf 1}_{22} \nonumber \\
&+& {\bf 2}_{10} + {\bf 2}_{01} + {\bf 2}_{21} + {\bf 2}_{12} \nonumber \\
&+& {\bf 3}_{11}, \label{eq:two}
\end{eqnarray}
where the bold numbers denote representations, and
the subscript $n_{k}n_{k+1}$ ($n_{k}$, $n_{k+1}=0$, $1$, $2$) denotes 
the number of the Dirac fermions, $n_{k}$ and $n_{k+1}$, at the $k$-th 
and $(k+1)$-th vortices, respectively. 
Next, one obtains 
the basis of the whole Hilbert space by applying 
$\hat{\psi}_{\ell}^{a\dag}$ ($\ell=k$, $k+1$ and $a=1$, $2$) 
successively to the Fock vacuum $| 0 \rangle$ defined by  
$\hat{\psi}_{\ell}^{a} | 0 \rangle = 0$ for all $\ell$ and $a=1$, 2. 
Then, one reorganizes the Fock states according to the 
decomposition into representations.

Below, we explicitly show the basis of the Hilbert space according 
to the decomposition in Eq.~(\ref{eq:two}). We introduce the notations 
$\vert {\cal R}_{n_{k} n_{k+1}}\rangle $ for the basis states and 
$\mathbb{H}^{{\cal R}_{[N_{1} N_{2}]}}$ for the subspaces of the Hilbert space.
Here ${\cal R}$ denotes the representation of pseudo-spin, and the subscript $n_{k}n_{k+1}$ ($n_{k}, n_{k+1}=0,1,2$) denotes the number of the Dirac fermions at the $k$-th and $(k+1)$-th vortices, respectively. 
In the subspaces, we do not distinguish $n_{k}$ and $n_{k+1}$ for the reasons 
mentioned later, and thus $N_1$ and $N_2$ are defined as 
$N_1=\max \{n_{k},n_{k+1}\}$ and $N_2=\min \{n_{k},n_{k+1}\}$. 
For example, $(N_1,N_2)=(1,0)$ contains two cases $(n_{k},n_{k+1})=(1,0), 
(0,1)$. 
Then, the whole Hilbert space of two Dirac vortices is decomposed into seven 
subspaces:
\begin{eqnarray}
\mathbb{H}^{\{n=2\}} 
 &=& \mathbb{H}^{{\bf 1}_{[00]}} \oplus \mathbb{H}^{{\bf 1}_{[11]}} \oplus \mathbb{H}^{{\bf 1}_{[20]}} \oplus \mathbb{H}^{{\bf 1}_{[22]}} \nonumber \\
& \oplus & \mathbb{H}^{{\bf 2}_{[10]}} \oplus \mathbb{H}^{{\bf 2}_{[21]}}  
\nonumber \\
& \oplus & \mathbb{H}^{{\bf 3}_{[11]}}\, . \label{decomp:two}
\end{eqnarray}
This should be compared with the decomposition (\ref{eq:two}).

The first line in Eq.~(\ref{decomp:two}) 
corresponds to the singlet subspaces. They are defined by the basis states as
$\mathbb{H}^{{\bf 1}_{[00]}} \equiv \{ | {\bf 1}_{00} \rangle \}$, 
$\mathbb{H}^{{\bf 1}_{[11]}} \equiv \{ | {\bf 1}_{11} \rangle \}$, 
$\mathbb{H}^{{\bf 1}_{[20]}} \equiv \{ | {\bf 1}_{20} \rangle, | {\bf 1}_{02} \rangle \}$,
and $\mathbb{H}^{{\bf 1}_{[22]}} \equiv \{ | {\bf 1}_{22} \rangle \}$. 
Explicit forms of the basis states are given as follows:
\begin{eqnarray}
| {\bf 1}_{00} \rangle &\equiv& | 0 \rangle,\\
| {\bf 1}_{11} \rangle &\equiv& \frac{1}{\sqrt{2}} ( \hat{\psi}_{k}^{1\dag} \hat{\psi}_{k+1}^{2\dag} - \hat{\psi}_{k}^{2\dag} \hat{\psi}_{k+1}^{1\dag} ) | 0 \rangle,\\
| {\bf 1}_{20} \rangle &\equiv& \hat{\psi}_{k}^{1\dag} \hat{\psi}_{k}^{2\dag} | 0 \rangle,  \\
| {\bf 1}_{02} \rangle &\equiv& \hat{\psi}_{k+1}^{1\dag} \hat{\psi}_{k+1}^{2\dag} | 0 \rangle,\\
| {\bf 1}_{22} \rangle &\equiv& \hat{\psi}_{k}^{1\dag} \hat{\psi}_{k}^{2\dag} \hat{\psi}_{k+1}^{1\dag} \hat{\psi}_{k+1}^{2\dag} | 0 \rangle.
\end{eqnarray}

The second line in Eq.~(\ref{decomp:two}) 
corresponds to the doublet subspaces. They are defined by the basis states as
$\mathbb{H}^{{\bf 2}_{[10]}} \equiv \{ | {\bf 2}_{10} \rangle, | {\bf 2}_{01} \rangle \}$ and $\mathbb{H}^{{\bf 2}_{[21]}} \equiv \{ | {\bf 2}_{21} \rangle, | {\bf 2}_{12} \rangle \}$. Explicit forms of the basis states are given as follows:
\begin{eqnarray}
| {\bf 2}_{10} \rangle &\equiv& 
\left(
\begin{array}{c}
 \hat{\psi}_{k}^{1\dag} \\
 \hat{\psi}_{k}^{2\dag}
\end{array}
\right) | 0 \rangle, \nonumber \\
| {\bf 2}_{01} \rangle &\equiv&
\left(
\begin{array}{c}
 \hat{\psi}_{k+1}^{1\dag} \\
 \hat{\psi}_{k+1}^{2\dag}
\end{array}
\right) | 0 \rangle,\\
| {\bf 2}_{21} \rangle &\equiv& \hat{\psi}_{k}^{1\dag} \hat{\psi}_{k}^{2\dag}
\left(
\begin{array}{c}
 \hat{\psi}_{k+1}^{1\dag} \\
 \hat{\psi}_{k+1}^{2\dag}
\end{array}
\right) | 0 \rangle, \nonumber \\
| {\bf 2}_{12} \rangle &\equiv&
\left(
\begin{array}{c}
 \hat{\psi}_{k}^{1\dag} \\
 \hat{\psi}_{k}^{2\dag}
\end{array}
\right) \hat{\psi}_{k+1}^{1\dag} \hat{\psi}_{k+1}^{2\dag} | 0 \rangle.
\end{eqnarray}
Here, in the last two expressions, we have 
factored out the singlet part 
so that the vector structure becomes manifest.

Lastly, the third line in Eq.~(\ref{decomp:two}) 
corresponds to the triplet subspace. It is defined by the basis states 
as $\mathbb{H}^{{\bf
3}_{[11]}} \equiv \{ | {\bf 3}_{11} \rangle \}$ with 
\begin{eqnarray}
| {\bf 3}_{11} \rangle \equiv 
\left(
\begin{array}{c}
 \hat{\psi}_{k}^{1\dag} \hat{\psi}_{k+1}^{1\dag} \\
 \frac{1}{\sqrt{2}} ( \hat{\psi}_{k}^{1\dag} \hat{\psi}_{k+1}^{2\dag} +
  \hat{\psi}_{k}^{2\dag} \hat{\psi}_{k+1}^{1\dag} ) \\
 \hat{\psi}_{k}^{2\dag} \hat{\psi}_{k+1}^{2\dag}
\end{array}
\right) | 0 \rangle. 
\end{eqnarray}

Now we have prepared to discuss the exchange of vortices.
We recall that the exchange of $\hat{\psi}_{k}^{a}$ and $\hat{\psi}_{k+1}^{a}$ is expressed by the operator $\hat{\tau}_{k}$ defined 
in Eq.~(\ref{eq:tau_full}) as a unitary transformation
$\hat{\tau}_{k} \hat{\psi}_{\ell}^{a} \hat{\tau}_{k}^{-1}$ 
($\ell=k$, $k+1$ and $a=1$, $2$).
With the explicit forms of the basis states in Hilbert subspaces, 
we are able to express the operator $\hat{\tau}_{k}$ as matrices. 
In the singlet subspaces $\mathbb{H}^{{\bf 1}_{[00]}}$, 
$\mathbb{H}^{{\bf 1}_{[11]}}$, 
$\mathbb{H}^{{\bf 1}_{[20]}}$ and $\mathbb{H}^{{\bf 1}_{[22]}}$, 
the corresponding matrices are 
\begin{eqnarray}
\tau_{k}^{{\bf 1}_{[00]}} &=& 1, \nonumber \\
\tau_{k}^{{\bf 1}_{[11]}} &=& -1, \nonumber \\
\tau_{k}^{{\bf 1}_{[20]}} &=&
\left(
\begin{array}{cc}
 0 & 1 \\
 1 & 0
\end{array}
\right), \nonumber \\
\tau_{k}^{{\bf 1}_{[22]}} &=& 1\, . \nonumber
\end{eqnarray}
In the doublet subspaces $\mathbb{H}^{{\bf 2}_{[10]}}$ and 
$\mathbb{H}^{{\bf 2}_{[21]}}$, the corresponding matrices are
\begin{eqnarray}
\tau_{k}^{{\bf 2}_{[10]}} &=&
\left(
\begin{array}{cc}
 0 & -1 \\
 1 & 0
\end{array}
\right), \nonumber \\
\tau_{k}^{{\bf 2}_{[21]}} &=&
\left(
\begin{array}{cc}
 0 & 1 \\
 -1 & 0
\end{array}
\right)\, . \nonumber
\end{eqnarray}
Lastly, in the triplet subspace $\mathbb{H}^{{\bf 3}_{[11]}}$,
the matrix is
\begin{eqnarray}
\tau_{k}^{{\bf 3}_{[11]}} &=& 1\, . \nonumber
\end{eqnarray}
Since $\mathbb{H}^{{\bf 1}_{[00]}}$, 
$\mathbb{H}^{{\bf 1}_{[11]}}$, $\mathbb{H}^{{\bf 1}_{[22]}}$
in the singlet subspace 
and $\mathbb{H}^{{\bf 3}_{[11]}}$ in the triplet subspace
consist of only one basis state, we have one-dimensional 
representations. In contrast, the other subspaces 
$\mathbb{H}^{{\bf 1}_{[20]}}$, $\mathbb{H}^{{\bf 2}_{[10]}}$,
and $\mathbb{H}^{{\bf 2}_{[21]}}$ have two basis states, thus 
yielding two-dimensional representations. Notice that 
the exchange matrices in these subspaces have off-diagonal 
elements. However, this does not mean that 
the representation of the braid group is non-Abelian.  
Rather, this simply implies that two basis states are mixed with each other 
by the exchange operation (this is the reason why we did 
not specify the order of $n_{k}$ and $n_{k+1}$ in defining the subspaces). 
In fact, one can choose appropriate basis states so that the exchange 
matrices are expressed as diagonal.
For example, $\tau_{k}^{{\bf 1}_{[20]}}$ is diagonalized with 
eigenvalues $\pm 1$ showing the 
Abelian representation of the braid group, while 
$\tau_{k}^{{\bf 2}_{[10]}}$ and $\tau_{k}^{{\bf 2}_{[21]}}$ are 
diagonalized with $\pm i$, showing anyon-like 
Abelian representation of the braid group.

\subsection{The case of $n=3$: emergence of non-Abelian representation}

Let us finally consider the case of three Dirac 
vortices which are respectively labelled by $k$, $k+1$ and $k+2$. 
Basically we will follow the same procedures presented in the case 
of two Dirac vortices, but as easily expected, the analysis becomes
quite involved. Still, we present here all the information since 
this is the simplest case where the non-Abelian 
representation of the braid group appears. In fact,
when we have only two vortices ($k$-th and $(k+1)$-th vortices), there is  
only one exchange operator $\hat\tau_k$, thus we are not able to 
discuss the non-commutativity of two exchange operations. It makes 
sense only when we have three or more vortices. When we have three vortices 
($k$-th, $(k+1)$-th and $(k+2)$-th vortices), we can
define two exchange operators $\hat\tau_k$ and $\hat\tau_{k+1}$, and 
thus we can check if these two are commutative or not.

First of all, by using Eq.~(\ref{eq:single}) for a single Dirac 
vortex, we can decompose the pseudo-spin structure 
made of three Dirac vortices into representations of U(2). We again 
introduce the notation ${\cal R}_{n_{k}n_{k+1}n_{k+2}}$ similar to 
the $n=2$ case. Here the subscript $n_{k}n_{k+1}n_{k+2}$ 
($n_{k}$, $n_{k+1}$, $n_{k+2}=0$, $1$, $2$) 
denotes the number of Dirac fermions at the $k$-th, $(k+1)$-th and 
$(k+2)$-th vortices, respectively.
Then, one finds
\begin{eqnarray}
&&\hspace{-1cm}
({\bf 1}_{0} + {\bf 2}_{1} + {\bf 1}_{2})_{{\rm vortex}\, k} \otimes
 ({\bf 1}_{0} + {\bf 2}_{1} + {\bf 1}_{2})_{{\rm vortex}\, k+1} \otimes
 ({\bf 1}_{0} + {\bf 2}_{1} + {\bf 1}_{2})_{{\rm vortex}\, k+2}
 \nonumber \\
&=& {\bf 1}_{000} + ({\bf 1}_{020} + {\bf 1}_{200} + {\bf 1}_{002}) 
+ ({\bf 1}_{110} + {\bf 1}_{011} + {\bf 1}_{101})
+ ({\bf 1}_{220} + {\bf 1}_{022} + {\bf 1}_{202}) 
+ ({\bf 1}_{112} + {\bf 1}_{211} + {\bf 1}_{121}) + {\bf 1}_{222} \nonumber \\
&+& ({\bf 2}_{100} + {\bf 2}_{010} + {\bf 2}_{001}) 
+ ({\bf 2}_{210} + {\bf 2}_{120} + {\bf 2}_{012} + {\bf 2}_{102}
+ {\bf 2}_{021} + {\bf 2}_{201})
+ ({\bf 2}_{212} + {\bf 2}_{122} + {\bf
 2}_{221}) + ({\bf 2}_{{\rm A}\,\underline{11}1} + {\bf 2}_{{\rm S}\,\underline{11}1})
 \nonumber \\
&+& ({\bf 3}_{110} + {\bf 3}_{011} + {\bf 3}_{101}) + 
({\bf 3}_{112} + {\bf 3}_{211} + {\bf 3}_{121}) \nonumber \\
&+& {\bf 4}_{{\rm S}\,\underline{11}1}\, .
\end{eqnarray}
Here we have also introduced new notations 
${\bf 2}_{{\rm A}\,\underline{11}1}$, 
${\bf 2}_{{\rm S}\,\underline{11}1}$ and ${\bf 4}_{{\rm S}\,\underline{11}1}$, 
meaning that the first two subscripts with underlines are made 
{\bf A}symmetric ({\bf S}ymmetric) with respect to the indices. We will note
again when we present the explicit forms of the basis states. Each term 
corresponds to the basis state in the Hilbert subspace. Notice that we 
have already grouped the representations so as not to distinguish the 
ordering of $n_{k}$, $n_{k+1}$ and $n_{k+2}$. Thus, one can easily see that 
the whole Hilbert space $\mathbb{H}^{\{n=3\}}$ of three U(2) 
Dirac vortices can be decomposed into a direct sum
\begin{eqnarray}
\mathbb{H}^{\{n=3\}} 
 &=& \mathbb{H}^{{\bf 1}_{[000]}} \oplus \mathbb{H}^{{\bf 1}_{[200]}} \oplus \mathbb{H}^{{\bf 1}_{[110]}} \oplus \mathbb{H}^{{\bf 1}_{[220]}} \oplus \mathbb{H}^{{\bf 1}_{[211]}} \oplus \mathbb{H}^{{\bf 1}_{[222]}} \nonumber \\
& \oplus & \mathbb{H}^{{\bf 2}_{[100]}} \oplus \mathbb{H}^{{\bf 2}_{[210]}} \oplus \mathbb{H}^{{\bf 2}_{[221]}} \oplus \mathbb{H}^{{\bf 2}_{[111]}} \nonumber \\
& \oplus & \mathbb{H}^{{\bf 3}_{[110]}} \oplus  \mathbb{H}^{{\bf 3}_{[211]}} \nonumber \\
& \oplus & \mathbb{H}^{{\bf 4}_{[111]}}.\label{decompose:three}
\end{eqnarray}
Next, one can obtain the basis states of the Hilbert space 
by applying $\hat{\psi}_{\ell}^{a\dag}$ ($\ell=k$, $k+1$, $k+2$ and 
$a=1$, $2$) successively to the vacuum $| 0 \rangle$ defined by 
 $\hat{\psi}_{\ell}^{a} | 0 \rangle = 0$ for all $\ell$ and $a=1$, $2$.

The first line in the decomposition (\ref{decompose:three}) 
corresponds to the singlet subspaces. They are defined by the basis states as
$\mathbb{H}^{{\bf 1}_{[000]}} \equiv \{ | {\bf 1}_{000} \rangle \}$, 
$\mathbb{H}^{{\bf 1}_{[200]}} \equiv \{ | {\bf 1}_{020} \rangle, | {\bf 1}_{200} \rangle, | {\bf 1}_{002} \rangle \}$,
$\mathbb{H}^{{\bf 1}_{[110]}} \equiv \{ | {\bf 1}_{110} \rangle, | {\bf 1}_{011} \rangle, | {\bf 1}_{101} \rangle \}$,
$\mathbb{H}^{{\bf 1}_{[220]}} \equiv \{ | {\bf 1}_{220} \rangle, | {\bf 1}_{022} \rangle, | {\bf 1}_{202} \rangle \}$,
$\mathbb{H}^{{\bf 1}_{[211]}} \equiv \{ | {\bf 1}_{112} \rangle, | {\bf 1}_{211} \rangle, | {\bf 1}_{121} \rangle \}$ and 
$\mathbb{H}^{{\bf 1}_{[222]}} \equiv \{ | {\bf 1}_{222} \rangle \}$. 
Notice that all these states have even numbers of Dirac fermions. 
Explicit forms of the basis states are given as follows:
\begin{eqnarray}
\mathbb{H}^{{\bf 1}_{[000]}}  :\qquad && \quad \,
| {\bf 1}_{000} \rangle \equiv | 0 \rangle,
 \\
 \mathbb{H}^{{\bf 1}_{[200]}}: \qquad &&
\left\{ 
\begin{array}{l}
| {\bf 1}_{020} \rangle \equiv \hat{\psi}_{k+1}^{1\dag} \hat{\psi}_{k+1}^{2\dag} | 0 \rangle,  \\
| {\bf 1}_{200} \rangle \equiv \hat{\psi}_{k}^{1\dag} \hat{\psi}_{k}^{2\dag} | 0 \rangle,  \\
| {\bf 1}_{002} \rangle \equiv \hat{\psi}_{k+2}^{1\dag} \hat{\psi}_{k+2}^{2\dag} | 0 \rangle,
\end{array}
\right. \\
\mathbb{H}^{{\bf 1}_{[110]}}:\qquad &&
\left\{
\begin{array}{l}
| {\bf 1}_{110} \rangle \equiv \frac{1}{\sqrt{2}} ( \hat{\psi}_{k}^{1\dag} \hat{\psi}_{k+1}^{2\dag} - \hat{\psi}_{k}^{2\dag} \hat{\psi}_{k+1}^{1\dag} ) | 0 \rangle,  \\
| {\bf 1}_{011} \rangle \equiv \frac{1}{\sqrt{2}} ( \hat{\psi}_{k+1}^{1\dag} \hat{\psi}_{k+2}^{2\dag} - \hat{\psi}_{k+1}^{2\dag} \hat{\psi}_{k+2}^{1\dag} ) | 0 \rangle,  \\
| {\bf 1}_{101} \rangle \equiv \frac{1}{\sqrt{2}} ( \hat{\psi}_{k}^{1\dag} \hat{\psi}_{k+2}^{2\dag} - \hat{\psi}_{k}^{2\dag} \hat{\psi}_{k+2}^{1\dag} ) | 0 \rangle,
\end{array}
\right. \\
\mathbb{H}^{{\bf 1}_{[220]}}:\qquad &&
\left\{
\begin{array}{l}
| {\bf 1}_{220} \rangle \equiv \hat{\psi}_{k}^{1\dag} \hat{\psi}_{k}^{2\dag} \hat{\psi}_{k+1}^{1\dag} \hat{\psi}_{k+1}^{2\dag} | 0 \rangle,  \\
| {\bf 1}_{022} \rangle \equiv \hat{\psi}_{k+1}^{1\dag} \hat{\psi}_{k+1}^{2\dag} \hat{\psi}_{k+2}^{1\dag} \hat{\psi}_{k+2}^{2\dag} | 0 \rangle,  \\
| {\bf 1}_{202} \rangle \equiv \hat{\psi}_{k}^{1\dag} \hat{\psi}_{k}^{2\dag} \hat{\psi}_{k+2}^{1\dag} \hat{\psi}_{k+2}^{2\dag} | 0 \rangle,
\end{array}
\right.\\
\mathbb{H}^{{\bf 1}_{[211]}}: \qquad &&
\left\{
\begin{array}{l}
| {\bf 1}_{112} \rangle \equiv  \frac{1}{\sqrt{2}} ( \hat{\psi}_{k}^{1\dag} \hat{\psi}_{k+1}^{2\dag} - \hat{\psi}_{k}^{2\dag} \hat{\psi}_{k+1}^{1\dag} ) \hat{\psi}_{k+2}^{1\dag} \hat{\psi}_{k+2}^{2\dag} | 0 \rangle,  \\
| {\bf 1}_{211} \rangle \equiv \frac{1}{\sqrt{2}} \hat{\psi}_{k}^{1\dag} \hat{\psi}_{k}^{2\dag} ( \hat{\psi}_{k+1}^{1\dag} \hat{\psi}_{k+2}^{2\dag} - \hat{\psi}_{k+1}^{2\dag} \hat{\psi}_{k+2}^{1\dag} )  | 0 \rangle,  \\
| {\bf 1}_{121} \rangle \equiv \frac{1}{\sqrt{2}} ( \hat{\psi}_{k}^{1\dag} \hat{\psi}_{k+2}^{2\dag} - \hat{\psi}_{k}^{2\dag} \hat{\psi}_{k+2}^{1\dag} ) \hat{\psi}_{k+1}^{1\dag} \hat{\psi}_{k+1}^{2\dag} | 0 \rangle,
\end{array}
\right.\\
\mathbb{H}^{{\bf 1}_{[222]}} : \qquad && \quad \,
| {\bf 1}_{222} \rangle \equiv \hat{\psi}_{k}^{1\dag} \hat{\psi}_{k}^{2\dag} \hat{\psi}_{k+1}^{1\dag} \hat{\psi}_{k+1}^{2\dag} \hat{\psi}_{k+2}^{1\dag} \hat{\psi}_{k+2}^{2\dag} | 0 \rangle. 
\end{eqnarray}
It is interesting to notice that the structures in 
$\mathbb{H}^{{\bf 1}_{[110]}}$ and $\mathbb{H}^{{\bf 1}_{[211]}}$,
or in $\mathbb{H}^{{\bf 1}_{[200]}}$ and $\mathbb{H}^{{\bf 1}_{[220]}}$
are similar. This is the symmetry between ``occupied" and ``empty" states
(or ``particles" and ``holes"),
reflecting the ambiguity in defining the creation operator (i.e., we could
define $\hat\psi_k^a$ as a creation operator, instead of the 
annihilation operator).

The second line of Eq.~(\ref{decompose:three}) 
corresponds to the doublet subspaces. 
These four subspaces are respectively defined by the basis states as 
$\mathbb{H}^{{\bf 2}_{[100]}} \equiv \{ | {\bf 2}_{100}
\rangle, | {\bf 2}_{010} \rangle, | {\bf 2}_{001} \rangle \}$,  
$\mathbb{H}^{{\bf 2}_{[210]}} \equiv \{ | {\bf 2}_{210} \rangle, | {\bf 2}_{120} \rangle, | {\bf 2}_{012} \rangle, | {\bf 2}_{102} \rangle, | {\bf 2}_{021} \rangle, | {\bf 2}_{201} \rangle \}$, 
$\mathbb{H}^{{\bf 2}_{[221]}} \equiv \{ | {\bf 2}_{212} \rangle, | {\bf 2}_{122} \rangle, | {\bf 2}_{221} \rangle \}$, 
and $\mathbb{H}^{{\bf 2}_{[111]}} \equiv \{ | {\bf 2}_{{\rm A}\,\underline{11}1} \rangle, | {\bf 2}_{{\rm S}\,\underline{11}1} \rangle \}$. 
This time, the total numbers of Dirac fermions are odd in these states.
Explicit forms of the basis states 
are given as follows:
\begin{eqnarray}
\mathbb{H}^{{\bf 2}_{[100]}} : \qquad &&
\left\{
\begin{array}{l}
| {\bf 2}_{100} \rangle \equiv
\left(
\begin{array}{c}
 \hat{\psi}_{k}^{1\dag} \\
 \hat{\psi}_{k}^{2\dag}
\end{array}
\right) | 0 \rangle,  \\
| {\bf 2}_{010} \rangle \equiv
\left(
\begin{array}{c}
 \hat{\psi}_{k+1}^{1\dag} \\
 \hat{\psi}_{k+1}^{2\dag}
\end{array}
\right) | 0 \rangle,  \\
| {\bf 2}_{001} \rangle \equiv
\left(
\begin{array}{c}
 \hat{\psi}_{k+2}^{1\dag} \\
 \hat{\psi}_{k+2}^{2\dag}
\end{array}
\right) | 0 \rangle,
\end{array}
\right. \\
\mathbb{H}^{{\bf 2}_{[210]}}:\qquad &&
\left\{ 
\begin{array}{l}
| {\bf 2}_{210} \rangle \equiv \hat{\psi}_{k}^{1\dag} \hat{\psi}_{k}^{2\dag}
\left(
\begin{array}{c}
 \hat{\psi}_{k+1}^{1\dag} \\
 \hat{\psi}_{k+1}^{2\dag}
\end{array}
\right) | 0 \rangle,  \\
| {\bf 2}_{120} \rangle \equiv
\left(
\begin{array}{c}
 \hat{\psi}_{k}^{1\dag} \\
 \hat{\psi}_{k}^{2\dag}
\end{array}
\right) \hat{\psi}_{k+1}^{1\dag} \hat{\psi}_{k+1}^{2\dag} | 0 \rangle, \\
| {\bf 2}_{012} \rangle \equiv 
\left(
\begin{array}{c}
 \hat{\psi}_{k+1}^{1\dag} \\
 \hat{\psi}_{k+1}^{2\dag}
\end{array}
\right) \hat{\psi}_{k+2}^{1\dag} \hat{\psi}_{k+2}^{2\dag} | 0 \rangle,  \\
| {\bf 2}_{102} \rangle \equiv
\left(
\begin{array}{c}
 \hat{\psi}_{k}^{1\dag} \\
 \hat{\psi}_{k}^{2\dag}
\end{array}
\right) \hat{\psi}_{k+2}^{1\dag} \hat{\psi}_{k+2}^{2\dag} | 0 \rangle, \\
| {\bf 2}_{021} \rangle \equiv \hat{\psi}_{k+1}^{1\dag} \hat{\psi}_{k+1}^{2\dag}
\left(
\begin{array}{c}
 \hat{\psi}_{k+2}^{1\dag} \\
 \hat{\psi}_{k+2}^{2\dag}
\end{array}
\right) | 0 \rangle,  \\
| {\bf 2}_{201} \rangle \equiv \hat{\psi}_{k}^{1\dag} \hat{\psi}_{k}^{2\dag}
\left(
\begin{array}{c}
 \hat{\psi}_{k+2}^{1\dag} \\
 \hat{\psi}_{k+2}^{2\dag}
\end{array}
\right) | 0 \rangle,
\end{array}
\right.\\
\mathbb{H}^{{\bf 2}_{[221]}}:\qquad &&
\left\{
\begin{array}{l}
| {\bf 2}_{212} \rangle \equiv \hat{\psi}_{k}^{1\dag} \hat{\psi}_{k}^{2\dag}
\left(
\begin{array}{c}
 \hat{\psi}_{k+1}^{1\dag} \\
 \hat{\psi}_{k+1}^{2\dag}
\end{array}
\right) \hat{\psi}_{k+2}^{1\dag} \hat{\psi}_{k+2}^{2\dag} | 0 \rangle, \\
| {\bf 2}_{122} \rangle \equiv 
\left(
\begin{array}{c}
 \hat{\psi}_{k}^{1\dag} \\
 \hat{\psi}_{k}^{2\dag}
\end{array}
\right) \hat{\psi}_{k+1}^{1\dag} \hat{\psi}_{k+1}^{2\dag} \hat{\psi}_{k+2}^{1\dag} \hat{\psi}_{k+2}^{2\dag} | 0 \rangle, \\
| {\bf 2}_{221} \rangle \equiv \hat{\psi}_{k}^{1\dag} \hat{\psi}_{k}^{2\dag} \hat{\psi}_{k+1}^{1\dag} \hat{\psi}_{k+1}^{2\dag}
\left(
\begin{array}{c}
 \hat{\psi}_{k+2}^{1\dag} \\
 \hat{\psi}_{k+2}^{2\dag}
\end{array}
\right) | 0 \rangle,
\end{array}
\right.\\
\mathbb{H}^{{\bf 2}_{[111]}}:\qquad &&
\left\{
\begin{array}{l}
| {\bf 2}_{{\rm A}\,\underline{11}1} \rangle \equiv \frac{1}{\sqrt{2}} ( \hat{\psi}_{k}^{1\dag} \hat{\psi}_{k+1}^{2\dag} - \hat{\psi}_{k}^{2\dag} \hat{\psi}_{k+1}^{1\dag} )
\left(
\begin{array}{c}
  \hat{\psi}_{k+2}^{1\dag} \\
  \hat{\psi}_{k+2}^{2\dag}
\end{array}
\right) | 0 \rangle,  \\
| {\bf 2}_{{\rm S}\,\underline{11}1} \rangle \equiv 
\frac{1}{\sqrt6} \left(
\begin{array}{c}
 2\hat{\psi}_{k}^{1\dag} \hat{\psi}_{k+1}^{1\dag} \hat{\psi}_{k+2}^{2\dag} 
- ( \hat{\psi}_{k}^{1\dag} \hat{\psi}_{k+1}^{2\dag} + \hat{\psi}_{k}^{2\dag} \hat{\psi}_{k+1}^{1\dag} ) \hat{\psi}_{k+2}^{1\dag} \\
 ( \hat{\psi}_{k}^{1\dag} \hat{\psi}_{k+1}^{2\dag} + \hat{\psi}_{k}^{2\dag} \hat{\psi}_{k+1}^{1\dag} ) \hat{\psi}_{k+2}^{2\dag} - 2 \hat{\psi}_{k}^{2\dag} \hat{\psi}_{k+1}^{2\dag} \hat{\psi}_{k+2}^{1\dag}
\end{array}
\right) | 0 \rangle\, .
\end{array}
\right.  \label{eq:H2(111)}
\end{eqnarray}
Now the meaning of the notations A and S is evident. For example, 
in $| {\bf 2}_{{\rm A}\,\underline{11}1} \rangle$ the first ($k$-th) 
and second (($k+1$)-th)
vortices form anti-symmetric combination with respect to the indices.

The third line of the decomposition (\ref{decompose:three}) 
corresponds to the triplet subspaces. 
They are defined by the basis states as 
$\mathbb{H}^{{\bf 3}_{[110]}} \equiv \{ 
| {\bf 3}_{110} \rangle, 
| {\bf 3}_{011} \rangle, 
| {\bf 3}_{101} \rangle \}$ and 
$\mathbb{H}^{{\bf 3}_{[211]}} \equiv \{ 
| {\bf 3}_{112} \rangle, 
| {\bf 3}_{211} \rangle, 
| {\bf 3}_{121} \rangle \}$. 
Total fermion numbers are now even. 
Explicit forms of the basis states are given as follows:
\begin{eqnarray}
\mathbb{H}^{{\bf 3}_{[110]}}:\qquad &&
\left\{ 
\begin{array}{l}
| {\bf 3}_{110} \rangle \equiv 
\left(
\begin{array}{c}
 \hat{\psi}_{k}^{1\dag} \hat{\psi}_{k+1}^{1\dag} \\
 \frac{1}{\sqrt{2}} ( \hat{\psi}_{k}^{1\dag} \hat{\psi}_{k+1}^{2\dag} + \hat{\psi}_{k}^{2\dag} \hat{\psi}_{k+1}^{1\dag} ) \\
 \hat{\psi}_{k}^{2\dag} \hat{\psi}_{k+1}^{2\dag}
\end{array}
\right) | 0 \rangle,  \\
| {\bf 3}_{011} \rangle \equiv
\left(
\begin{array}{c}
 \hat{\psi}_{k+1}^{1\dag} \hat{\psi}_{k+2}^{1\dag} \\
 \frac{1}{\sqrt{2}} ( \hat{\psi}_{k+1}^{1\dag} \hat{\psi}_{k+2}^{2\dag} + \hat{\psi}_{k+1}^{2\dag} \hat{\psi}_{k+2}^{1\dag} ) \\
 \hat{\psi}_{k+1}^{2\dag} \hat{\psi}_{k+2}^{2\dag}
\end{array}
\right) | 0 \rangle,  \\
| {\bf 3}_{101} \rangle \equiv
\left(
\begin{array}{c}
 \hat{\psi}_{k}^{1\dag} \hat{\psi}_{k+2}^{1\dag} \\
 \frac{1}{\sqrt{2}} ( \hat{\psi}_{k}^{1\dag} \hat{\psi}_{k+2}^{2\dag} + \hat{\psi}_{k}^{2\dag} \hat{\psi}_{k+2}^{1\dag} ) \\
 \hat{\psi}_{k}^{2\dag} \hat{\psi}_{k+2}^{2\dag}
\end{array}
\right) | 0 \rangle,
\end{array}
\right. \\
\mathbb{H}^{{\bf 3}_{[211]}}:\qquad &&
\left\{ 
\begin{array}{l}
| {\bf 3}_{112} \rangle \equiv
\left(
\begin{array}{c}
 \hat{\psi}_{k}^{1\dag} \hat{\psi}_{k+1}^{1\dag} \\
 \frac{1}{\sqrt{2}} ( \hat{\psi}_{k}^{1\dag} \hat{\psi}_{k+1}^{2\dag} + \hat{\psi}_{k}^{2\dag} \hat{\psi}_{k+1}^{1\dag} ) \\
 \hat{\psi}_{k}^{2\dag} \hat{\psi}_{k+1}^{2\dag}
\end{array}
\right) \hat{\psi}_{k+2}^{1\dag} \hat{\psi}_{k+2}^{2\dag} | 0 \rangle, \\
| {\bf 3}_{211} \rangle \equiv \hat{\psi}_{k}^{1\dag} \hat{\psi}_{k}^{2\dag}
\left(
\begin{array}{c}
 \hat{\psi}_{k+1}^{1\dag} \hat{\psi}_{k+2}^{1\dag} \\
 \frac{1}{\sqrt{2}} ( \hat{\psi}_{k+1}^{1\dag} \hat{\psi}_{k+2}^{2\dag} + \hat{\psi}_{k+1}^{2\dag} \hat{\psi}_{k+2}^{1\dag} ) \\
 \hat{\psi}_{k+1}^{2\dag} \hat{\psi}_{k+2}^{2\dag}
\end{array}
\right) | 0 \rangle,  \\
| {\bf 3}_{121} \rangle \equiv
\left(
\begin{array}{c}
 \hat{\psi}_{k}^{1\dag} \hat{\psi}_{k+2}^{1\dag} \\
 \frac{1}{\sqrt{2}} ( \hat{\psi}_{k}^{1\dag} \hat{\psi}_{k+2}^{2\dag} + \hat{\psi}_{k}^{2\dag} \hat{\psi}_{k+2}^{1\dag} ) \\
 \hat{\psi}_{k}^{2\dag} \hat{\psi}_{k+2}^{2\dag}
\end{array}
\right) \hat{\psi}_{k+1}^{1\dag} \hat{\psi}_{k+1}^{2\dag} | 0 \rangle. 
\end{array}
\right.
\end{eqnarray}

The last line of the decomposition (\ref{decompose:three}) 
corresponds to the quartet subspace. It is defined by the basis states as 
$\mathbb{H}^{{\bf 4}_{[111]}} \equiv 
\{ | {\bf 4}_{{\rm S}\,\underline{11}1} \rangle \}$. Explicitly,
\begin{eqnarray}
\mathbb{H}^{{\bf 4}_{[111]}}: \qquad
| {\bf 4}_{{\rm S}\,\underline{11}1} \rangle \equiv
\left(
\begin{array}{c}
 \hat{\psi}_{k}^{1\dag} \hat{\psi}_{k+1}^{1\dag} \hat{\psi}_{k+2}^{1\dag} \\
 \frac{1}{\sqrt{3}} \hat{\psi}_{k}^{1\dag} \hat{\psi}_{k+1}^{1\dag}
  \hat{\psi}_{k+2}^{2\dag} + \frac{1}{\sqrt{3}} ( \hat{\psi}_{k}^{1\dag}
  \hat{\psi}_{k+1}^{2\dag} + \hat{\psi}_{k}^{2\dag}
  \hat{\psi}_{k+1}^{1\dag} ) \hat{\psi}_{k+2}^{1\dag} \\
 \frac{1}{\sqrt{3}} ( \hat{\psi}_{k}^{1\dag} \hat{\psi}_{k+1}^{2\dag} +
  \hat{\psi}_{k}^{2\dag} \hat{\psi}_{k+1}^{1\dag} ) \hat{\psi}_{k+2}^{2\dag}
  + \frac{1}{\sqrt{3}} \hat{\psi}_{k}^{2\dag} \hat{\psi}_{k+1}^{2\dag}
  \hat{\psi}_{k+2}^{1\dag} \\
 \hat{\psi}_{k}^{2\dag} \hat{\psi}_{k+1}^{2\dag}
  \hat{\psi}_{k+2}^{2\dag}
\end{array}
\right) | 0 \rangle. \nonumber
\end{eqnarray}

Now we can finally discuss the exchange of U(2) Dirac vortices 
in the Hilbert space constructed for three Dirac vortices.
As in the case of $n=2$, we express the operators $\hat{\tau}_{k}$ 
and $\hat{\tau}_{k+1}$ as matrices with the basis states presented above.

In the singlet subspaces $\mathbb{H}^{{\bf 1}_{[000]}}$, $\mathbb{H}^{{\bf 1}_{[200]}}$, $\mathbb{H}^{{\bf 1}_{[110]}}$, $\mathbb{H}^{{\bf 1}_{[220]}}$, $\mathbb{H}^{{\bf 1}_{[211]}}$ and $\mathbb{H}^{{\bf 1}_{[222]}}$, the 
exchange matrices are
\begin{eqnarray}
&& \tau_{k}^{{\bf 1}_{[000]}} = \tau_{k+1}^{{\bf 1}_{[000]}} = 1, \nonumber \\
&& \tau_{k}^{{\bf 1}_{[200]}} =
\left(
\begin{array}{ccc}
 0 & 1 & 0 \\
 1 & 0 & 0 \\
 0 & 0 & 1  
\end{array}
\right), \quad
\tau_{k+1}^{{\bf 1}_{[200]}} =
\left(
\begin{array}{ccc}
 0 & 0 & 1 \\
 0 & 1 & 0 \\
 1 & 0 & 0  
\end{array}
\right), \nonumber \\
&& \tau_{k}^{{\bf 1}_{[110]}} =
\left(
\begin{array}{ccc}
 -1 & 0 & 0 \\
 0 & 0 & 1 \\
 0 & -1 & 0  
\end{array}
\right), \quad 
\tau_{k+1}^{{\bf 1}_{[110]}} =
\left(
\begin{array}{ccc}
 0 & 0 & -1 \\
 0 & -1 & 0 \\
 1 & 0 & 0  
\end{array}
\right), \nonumber \\
&& \tau_{k}^{{\bf 1}_{[220]}} =
\left(
\begin{array}{ccc}
 1 & 0 & 0 \\
 0 & 0 & 1 \\
 0 & 1 & 0  
\end{array}
\right), \quad 
\tau_{k+1}^{{\bf 1}_{[220]}} =
\left(
\begin{array}{ccc}
 0 & 0 & 1 \\
 0 & 1 & 0 \\
 1 & 0 & 0  
\end{array}
\right), \nonumber \\
&& \tau_{k}^{{\bf 1}_{[211]}} =
\left(
\begin{array}{ccc}
 -1 & 0 & 0 \\
 0& 0 & 1 \\
 0 & -1 & 0  
\end{array}
\right), \quad 
\tau_{k+1}^{{\bf 1}_{[211]}} =
\left(
\begin{array}{ccc}
 0 & 0 & -1 \\
 0 & -1 & 0 \\
 1 & 0 & 0  
\end{array}
\right), \nonumber \\
&& \tau_{k}^{{\bf 1}_{[222]}} = \tau_{k+1}^{{\bf 1}_{[222]}} = 1\, .
\end{eqnarray}

In the doublet subspaces, $\mathbb{H}^{{\bf 2}_{[100]}}$, $\mathbb{H}^{{\bf 2}_{[210]}}$, $\mathbb{H}^{{\bf 2}_{[221]}}$ and $\mathbb{H}^{{\bf 2}_{[111]}}$, the exchange matrices are
\begin{eqnarray}
&& \tau_{k}^{{\bf 2}_{[100]}} =
\left(
\begin{array}{ccc}
 0 & -1 & 0 \\
 1 & 0 & 0 \\
 0 & 0 & 1  
\end{array}
\right), \quad
\tau_{k+1}^{{\bf 2}_{[100]}} =
\left(
\begin{array}{ccc}
 1 & 0 & 0 \\
 0 & 0 & -1 \\
 0 & 1 & 0  
\end{array}
\right), \nonumber \\
&& \tau_{k}^{{\bf 2}_{[210]}} =
\left(
\begin{array}{cccccc}
 0 & 1 & 0 & 0 & 0 & 0 \\
 -1 & 0 & 0 & 0 & 0 & 0 \\
 0 & 0 & 0 & 1 & 0 & 0 \\
 0 & 0 & -1 & 0 & 0 & 0 \\
 0 & 0 & 0 & 0 & 0 & 1 \\
 0 & 0 & 0 & 0 & 1 & 0
\end{array}
\right), \quad
 \tau_{k+1}^{{\bf 2}_{[210]}} =
\left(
\begin{array}{cccccc}
 0 & 0 & 0 & 0 & 0 & -1 \\
 0 & 0 & 0 & 1 & 0 & 0 \\
 0 & 0 & 0 & 0 & -1 & 0 \\
 0 & 1 & 0 & 0 & 0 & 0 \\
 0 & 0 & 1 & 0 & 0 & 0 \\
 1 & 0 & 0 & 0 & 0 & 0
\end{array}
\right), \nonumber \\
&& \tau_{k}^{{\bf 2}_{[221]}} =
\left(
\begin{array}{ccc}
 0 & 1 & 0 \\
 -1 & 0 & 0 \\
 0 & 0 & 1  
\end{array}
\right), \quad
\tau_{k+1}^{{\bf 2}_{[221]}} =
\left(
\begin{array}{ccc}
 0 & 0 & -1 \\
 0 & 1 & 0 \\
 1 & 0 & 0  
\end{array}
\right), \nonumber \\
&& \tau_{k}^{{\bf 2}_{[111]}} =
\left(
\begin{array}{cc}
 -1 & 0 \\
 0 & 1
\end{array}
\right), \quad
\tau_{k+1}^{{\bf 2}_{[111]}} =
\left(
\begin{array}{cc}
 \frac{1}{2} & \frac{\sqrt{3}}{2} \\
 \frac{\sqrt{3}}{2} & -\frac{1}{2}
\end{array}
\right)\, . \label{eq:basisH(111)}
\end{eqnarray}

In the triplet subspaces, $\mathbb{H}^{{\bf 3}_{[110]}}$ 
and $\mathbb{H}^{{\bf 3}_{[211]}}$,   the exchange matrices are
\begin{eqnarray}
&& \tau_{k}^{{\bf 3}_{[110]}} =
\left(
\begin{array}{ccc}
 1 & 0 & 0 \\
 0 & 0 & 1 \\
 0 & -1 & 0  
\end{array}
\right), \quad
\tau_{k+1}^{{\bf 3}_{[110]}} =
\left(
\begin{array}{ccc}
 0 & 0 & -1 \\
 0 & 1 & 0 \\
 1 & 0 & 0  
\end{array}
\right), \nonumber \\
&& \tau_{k}^{{\bf 3}_{[211]}} =
\left(
\begin{array}{ccc}
 1 & 0 & 0 \\
 0 & 0 & 1 \\
 0 & -1 & 0  
\end{array}
\right), \quad
\tau_{k+1}^{{\bf 3}_{[211]}} =
\left(
\begin{array}{ccc}
 0 & 0 & -1 \\
 0 & 1 & 0 \\
 1 & 0 & 0  
\end{array}
\right)\, .
\end{eqnarray}

Lastly, in the quartet subspace $\mathbb{H}^{{\bf 4}_{[111]}}$, 
the exchange matrices are
\begin{eqnarray}
&& \tau_{k}^{{\bf 4}_{[111]}} = \tau_{k+1}^{{\bf 4}_{[111]}} =1\, .
\end{eqnarray}

We now find that, except for the one-dimensional representations 
(in $\mathbb{H}^{{\bf 1}_{[000]}}$, $\mathbb{H}^{{\bf 1}_{[222]}}$ and 
$\mathbb{H}^{{\bf 4}_{[111]}}$), all the exchange matrices are non-commutative;
$[\tau_{k}, \tau_{k+1}] \neq 0$. Therefore, the exchange 
operations in these subspaces exhibit non-Abelian representation of the 
braid group.

Recall that we have shown that the exchange operators 
$\hat \tau_{\ell}$'s satisfy $(\hat \tau_{\ell})^4=1$ at the operator level
(see Eq.~(\ref{eq:tau^4})). 
Of course this is satisfied by all the exchange matrices obtained above. 
However, interestingly, a stronger relation $(\hat \tau_{\ell})^2=1$ is 
satisfied in some subspaces. 
In addition to the trivial one-dimensional representations in 
$\mathbb{H}^{{\bf 1}_{[000]}}$, 
$\mathbb{H}^{{\bf 1}_{[222]}}$ and $\mathbb{H}^{{\bf 4}_{[111]}}$,
we observe that the exchange matrices in $\mathbb{H}^{{\bf 1}_{[200]}}$, 
$\mathbb{H}^{{\bf 1}_{[220]}}$ and $\mathbb{H}^{{\bf 2}_{[111]}}$
satisfy this stronger relation. More details are discussed in 
Appendix~\ref{sec:square}.

We can continue the construction of the Hilbert space when the number of 
vortices is more than three. The dimension of the total Hilbert space
of $n$-vortex system is $(2^2)^n$.
We present the results for $n=4$ vortices in Appendix~\ref{sec:n=4}, which
also show non-Abelian representation of the braid group.

%%%%%%%%%%%%%%%%%%%%%%%%%%%%%%%%%%%%%%%%%%%%%%%%%%%%%%%%%%%%%%
\section{U(1) Dirac structure embedded in U(2) Dirac vortices}
\label{sec:relation}
\setcounter{equation}{0}

Recall that we already found in the previous paper 
\cite{Yasui:2011gk} that the system of U(1) Dirac vortices shows
non-Abelian representation of the braid group. Then, it is natural 
to ask how the previous results are related to the present results.
In this section, we discuss that we can indeed identify the same 
structure as the U(1) vortices by a simple reduction of the U(2) 
system.

So far, we have not specified any detail of the interaction, 
but suppose that one can turn on an interaction which breaks the 
global U(2) symmetry so that only the U(1) symmetry is preserved. 
Then, in the presence of such an interaction, 
only one Dirac fermion would remain massless. We can realize such a case 
by simply ignoring the lower component $\hat{\psi}_{\ell}^{2}$
($\ell=1, \cdots, n$) of the U(2) Dirac fermions. 
Under this simple procedure,
the Hilbert space $\mathbb{H}^{\{n\}}$ of the U(2)
Dirac vortices reduces to
the Hilbert space $\mathbb{H}^{(n)}$ of the U(1) Dirac vortices 
discussed in the previous paper \cite{Yasui:2011gk} (and also in 
Sec.~\ref{sec:single} and Appendix~\ref{sec:U(1)tau}). 
Let us discuss the cases of $n=2$ and $3$, separately.

Consider first the case of $n=2$ (see Sec.~\ref{subsec:n=2}).  
One finds that only the following states survive after the reduction: 
\begin{equation}
|{\bf 1}_{00}\rangle =|0\rangle,\quad 
|{\bf 2}_{10}\rangle = \left(
\begin{matrix}
\hat\psi^{1\dag}_k\\ 0
\end{matrix}
\right)|0\rangle,\quad 
|{\bf 2}_{01}\rangle = \left(
\begin{matrix}
\hat\psi^{1\dag}_{k+1}\\ 0
\end{matrix}
\right)|0\rangle,\quad 
|{\bf 3}_{11}\rangle = \left(
\begin{matrix}
\hat\psi^{1\dag}_k\hat\psi^{1\dag}_{k+1}\\ 0\\ 0
\end{matrix}
\right)|0\rangle .
\end{equation}
This means that the whole Hilbert space $\mathbb{H}^{\{2\}}$ shrinks into its
subspaces $\mathbb{H}^{{\bf 1}_{[00]}}$, $\mathbb{H}^{{\bf 2}_{[10]}}$ and $\mathbb{H}^{{\bf 3}_{[11]}}$. Notice that these four states are equivalent to the 
basis states for
$\mathbb{H}^{(2,0)}$, $\mathbb{H}^{(2,1)}$ and $\mathbb{H}^{(2,2)}$ in
the U(1) Dirac vortices. 
Thus, it is not surprising that the exchange matrices in 
the reduced U(2) Dirac vortices are equivalent to those of the U(1) vortices:
\begin{eqnarray}
 \tau_{k}^{{\bf 1}_{[00]}} &=& \tau_{k}^{(2,0)}, \\
 \tau_{k}^{{\bf 2}_{[10]}} &=& \tau_{k}^{(2,1)}, \\
 \tau_{k}^{{\bf 3}_{[11]}} &=& \tau_{k}^{(2,2)}.
\end{eqnarray}

Similarly, in the case of $n=3$, one finds that only 
the subspaces $\mathbb{H}^{{\bf 1}_{[000]}}$, $\mathbb{H}^{{\bf 2}_{[100]}}$, 
$\mathbb{H}^{{\bf 3}_{[110]}}$ and $\mathbb{H}^{{\bf 4}_{[111]}}$ survive 
after the reduction of the lower component. These Hilbert subspaces 
correspond, respectively, 
to $\mathbb{H}^{(3,0)}$, $\mathbb{H}^{(3,1)}$, $\mathbb{H}^{(3,2)}$ 
and $\mathbb{H}^{(3,3)}$ in the U(1) Dirac vortices.
The matrices of these two cases are equivalent to each other:
\begin{eqnarray}
 \tau_{\ell}^{{\bf 1}_{[000]}} &=& \tau_{\ell}^{(3,0)}, \\
 \tau_{\ell}^{{\bf 2}_{[100]}} &=& \tau_{\ell}^{(3,1)}, \\
 \tau_{\ell}^{{\bf 3}_{[110]}} &=& \tau_{\ell}^{(3,2)}, \\
 \tau_{\ell}^{{\bf 4}_{[111]}} &=& \tau_{\ell}^{(3,3)}, 
\end{eqnarray}
with $\ell=k$, $k+1$.

By repeating the same arguments for general $n$, one can easily 
find that there will be a correspondence between the Hilbert subspaces 
$\mathbb{H}^{{\cal M}_{[1\cdots10\cdots0]}}$ and $\mathbb{H}^{(n,f)}$.
The matrices in these two subspaces will be equivalent;
\begin{eqnarray}
 \tau_{\ell}^{{\cal M}_{[1\cdots10\cdots0]}} = \tau_{\ell}^{(n,f)},
\end{eqnarray}
with $\ell = 1, \cdots, n-1$.
Here, the U(2) representation ${\cal
M}_{[\underbrace{1\cdots1}_{f}\underbrace{0\cdots0}_{n-f}]}$ contains
$f$ Dirac fermions, and the dimension of ${\cal M}_{[1\cdots10\cdots0]}$
is equal to $f+1$.
Thus, the Hilbert space $\mathbb{H}^{\{n\}}$ of the U(2) Dirac vortices
covers $\mathbb{H}^{(n)}$ of the U(1) Dirac vortices. 
This is confirmed also for $n=4$ vortices in Appendix \ref{sec:n=4}.

%%%%%%%%%%%%%%%%%%%%%%%%%%%%%%%%%%%%%
\section{Genuine non-Abelian statistics}\label{sec:genuine}

\subsection{Sectors of exchanges of the identical vortices}
We have constructed the braid group made of exchanges of U(2) Dirac 
vortices. One may be able to regard it as a kind of exchange 
statistics since all the Dirac vortices to be exchanged are energetically 
the same. However, this is true only if we are allowed to 
neglect the difference between the occupation and absence of 
Dirac fermions in each vortex. 
Genuine non-Abelian statistics in a strict sense 
should appear in
exchanges of identical particles, 
which implies that we should distinguish 
the number of Dirac fermions in each vortex.  
In the case of vortices with U(1) Dirac fermions, the whole braid 
group representation is non-Abelian, but does not have genuine non-Abelian 
statistics when restricted to a subspace of exchanges of identical states 
(i.e., between two occupied states or two un-occupied states). 
What is truly remarkable in the U(2) case is
that the braid group which we have found
indeed contains genuine non-Abelian statistics as 
its subgroup where we consider only
the exchanges between the identical vortices, the vortices with 
the same numbers of Dirac fermions.
Below, we explicitly show them in subspaces with $n=3$ 
and $4$ vortices.

\bigskip
\begin{enumerate}
\item
The doublet Hilbert subspace $\mathbb{H}^{{\bf 2}_{[111]}}$ for
three U(2) Dirac vortices ($n=3$; $k$-th, $(k+1)$-th and $(k+2)$-th vortices) 
with single Dirac fermion occupations.
The minimum genuine non-Abelian statistics appears in the doublet Hilbert subspace 
$\mathbb{H}^{{\bf 2}_{[111]}}$ in $n=3$ vortices. 
The two-dimensional Hilbert subspace $\mathbb{H}^{{\bf 2}_{[111]}}$ 
is spanned by the basis $| {\bf 2}_{{\rm A}\,\underline{11}1} \rangle$ and $| {\bf 2}_{{\rm S}\,\underline{11}1} \rangle$ defined in Eq.~(\ref{eq:H2(111)}). 
The Dirac fermion numbers in each U(2) Dirac vortex are one so that 
they are all identical vortices.
Two exchange matrices, i.e., $\tau_{k}^{{\bf 2}_{[111]}}$
for the exchange of the $k$-th and $(k+1)$-th vortices, and 
$\tau_{k+1}^{{\bf 2}_{[111]}}$ for the exchange of the $(k+1)$-th and 
$(k+2)$-th vortices, can be found in the last line in 
Eq.~(\ref{eq:basisH(111)}):  
\begin{eqnarray}
&& \tau_{k}^{{\bf 2}_{[111]}} =
\left(
\begin{array}{cc}
 -1 & 0 \\
 0 & 1
\end{array}
\right), \quad
\tau_{k+1}^{{\bf 2}_{[111]}} =
\left(
\begin{array}{cc}
 \frac{1}{2} & \frac{\sqrt{3}}{2} \\
 \frac{\sqrt{3}}{2} & -\frac{1}{2}
\end{array}
\right)\, . \label{eq:genuine1}
\end{eqnarray}
They are non-commutative:   
$\tau_{k}^{{\bf 2}_{[111]}}\tau_{k+1}^{{\bf 2}_{[111]}} \neq \tau_{k+1}^{{\bf 2}_{[111]}}\tau_{k}^{{\bf 2}_{[111]}}$, showing genuine 
non-Abelian statistics.

\item
The non-Abelian transformation equivalent to
Eq.~(\ref{eq:genuine1}) are embedded in a sector of four vortices. 
For instance, let us consider four vortices three of which contain one 
Dirac fermion at each vortex but the rest of which 
contains no Dirac fermion, 
then 
the exchange of the identical vortices with the one Dirac fermion 
occupation leads to
the same result as in Eq.~(\ref{eq:genuine1}).
To see this more explicitly,
consider the doublet subspace with the basis 
$|{\bf 2}_{\mathrm{A}\,01\underline{11}}\rangle$ and 
 $|{\bf 2}_{\mathrm{S}\,01\underline{11}}\rangle$
defined in Eq.~(\ref{eq:H2(1110)}).
We can see that the submatrices made of the fifth and sixth columns and rows 
in $\tau_{k+1}^{{\bf 2}_{[1110]}}$ and $\tau_{k+2}^{{\bf 2}_{[1110]}}$ 
in Eq.~(\ref{eq:tau(1110)}) 
are nothing but the matrices in Eq.~(\ref{eq:genuine1}).
The rest corresponding submatrix 
in $\tau_{k}^{{\bf 2}_{[1110]}}$ in Eq.~(\ref{eq:tau(1110)}) is 
consistently zero, 
because we are considering the exchange of only three identical vortices 
with one Dirac fermion occupations.

\item
The singlet Hilbert space $\mathbb{H}^{{\bf 1}_{[1111]}}$ for
four U(2) Dirac vortices ($n=4$; $k$-th, $(k+1)$-th, $(k+2)$-th and $(k+3)$-th vortices) 
with single Dirac fermion occupations.
We have now
four identical vortices each of which contains one Dirac fermion. 
The singlet Hilbert space $\mathbb{H}^{{\bf 1}_{[1111]}}$ is a two-dimensional subspace with the basis $| {\bf 1}_{\mathrm{AA}\,\underline{11}\,\underline{11}} \rangle$ and $| {\bf 1}_{\mathrm{SS}\,\underline{11}\,\underline{11}} \rangle$ defined in Eq.~(\ref{eq:wf_1_1111}). 
The non-Abelian exchange matrices can be found in
 Eq.~(\ref{eq:matrix_1_1111}) as
\begin{eqnarray}
\tau_{k}^{{\bf 1}_{[1111]}} =
\left(
\begin{array}{cc}
 -1 & 0 \\
 0 & 1
\end{array}
\right), \,
\tau_{k+1}^{{\bf 1}_{[1111]}} =
\left(
\begin{array}{cc}
 \frac{1}{2} & \frac{\sqrt{3}}{2} \\
 \frac{\sqrt{3}}{2} & -\frac{1}{2}
\end{array}
\right), \,
\tau_{k+2}^{{\bf 1}_{[1111]}} =
\left(
\begin{array}{cc}
 -1 & 0 \\
 0 & 1
\end{array}
\right), 
\end{eqnarray}
The exchange of the Dirac vortices are non-Abelian; 
$\tau_{\ell}^{{\bf 1}_{[1111]}} \tau_{\ell+1}^{{\bf 1}_{[1111]}} \neq \tau_{\ell+1}^{{\bf 1}_{[1111]}} \tau_{\ell}^{{\bf 1}_{[1111]}}$ for $\ell=k$, $k+1$.

\item
The triplet  Hilbert space $\mathbb{H}^{{\bf 3}_{[1111]}}$ for 
four U(2) Dirac vortices ($n=4$; $k$-th, $(k+1)$-th, $(k+2)$-th and $(k+3)$-th vortices) 
with single Dirac fermion occupations.
The four identical Dirac vortices with single Dirac fermion occupations allow 
for another sector having genuine non-Abelian statistics. 
The triplet Hilbert space $\mathbb{H}^{{\bf 3}_{[1111]}}$ is three dimensional with the basis $|{\bf 3}_{\mathrm{AS}\,\underline{11}\,\underline{11}}\rangle$, $|{\bf 3}_{\mathrm{SA}\,\underline{11}\,\underline{11}}\rangle$ and $|{\bf 3}_{\mathrm{SS}\,\underline{11}\,\underline{11}}\rangle$ in Eq.~(\ref{eq:wf_3_1111}). 
The exchange matrices can be found in Eq.~(\ref{eq:matrix_3_1111}) as
\begin{eqnarray}
\tau_{k}^{{\bf 3}_{[1111]}} = 
\left(
\begin{array}{ccc}
 -1 & 0 & 0 \\
 0 & 1 & 0 \\
 0 & 0 & 1
\end{array}
\right), \,
\tau_{k+1}^{{\bf 3}_{[1111]}} = 
\left(
\begin{array}{ccc}
 \frac{1}{2} & -\frac{1}{2} & \frac{1}{\sqrt{2}} \\
 -\frac{1}{2} & \frac{1}{2} & \frac{1}{\sqrt{2}} \\
 \frac{1}{\sqrt{2}} & \frac{1}{\sqrt{2}} & 0
\end{array}
\right), \,
\tau_{k+2}^{{\bf 3}_{[1111]}} = 
\left(
\begin{array}{ccc}
 1 & 0 & 0 \\
 0 & -1 & 0 \\
 0 & 0 & 1
\end{array}
\right), 
\end{eqnarray}
The exchange of these Dirac vortices are non-Abelian; 
$\tau_{\ell}^{{\bf 3}_{[1111]}} \tau_{\ell+1}^{{\bf 3}_{[1111]}} \neq \tau_{\ell+1}^{{\bf 3}_{[1111]}} \tau_{\ell}^{{\bf 3}_{[1111]}}$ for $\ell=k$, $k+1$.

\end{enumerate}

From the above examples,  we find that 
the U(2) Dirac vortices with one Dirac fermion occupation 
give non-Abelian quantum statistics of identical particles.

\subsection{Entanglement in the Hilbert space}

If one reminds of the case of Majorana vortices showing the 
genuine non-Abelian statistics, one may critically wonder why 
we have found the genuine non-Abelian statistics even though 
Dirac fermions are defined {\it locally} at each vortex. 
As is well known, for vortices with Majorana fermions, Dirac 
fermions have to be defined {\it non-locally} by using two 
Majorana fermions localized at spatially separated vortices, 
and, thus far, it has been commonly thought that emergence 
of non-Abelian statistics is attributed to such a non-locality 
of Dirac fermions. In contrast, in our case, Dirac fermions 
are introduced from the beginning, and are of course defined 
locally at each vortex. Then, it is natural to ask 
why we get non-Abelian statistics from local Dirac fermions, 
or to raise question if the non-locality is really essential 
for the emergence of non-Abelian statistics. 

Our answer is that the non-locality is indeed essential but it 
does not come from the definition of Dirac fermions but rather 
from the definition of the basis of the Hilbert space. Namely,
the use of entangled states, which was absent in the previous 
U(1) Dirac case, is essential in the U(2) case. For instance, 
in the first example of the previous subsection, the basis 
$| {\bf 2}_{{\rm A}\,\underline{11}1} \rangle$ and 
$| {\bf 2}_{{\rm S}\,\underline{11}1} \rangle$ in Eq.~(\ref{eq:H2(111)}) 
are entangled
because of (anti-)symmetric combination of indices.
Such entangled states
are necessary because the basis of the Hilbert space 
have to belong to the irreducible representations of U(2) symmetry. 
Therefore, genuine non-Abelian statistics does not appear in U(1) 
Dirac vortices but it first appears in U(2) Dirac vortices because 
of non-Abelian U(2) symmetry group acting on the doublet Dirac 
fermions on the vortices. 

 From these considerations, we conclude that some kind of non-locality is 
necessary to obtain the genuine non-Abelian statistics: It is the non-local
definition of the Dirac fermions in the case of Majorana vortices, 
while it is the entanglement in the Hilbert space in the case of Dirac 
vortices.

%%%%%%%%%%%%%%%

\section{Summary}\label{sec:summary}
\setcounter{equation}{0}

We have considered the simplest system of non-Abelian Dirac vortices, 
namely the system of vortices in which U(2) doublet zero-energy Dirac 
fermions are trapped. 
We have constructed the non-Abelian representation of the braid group 
for the exchange of vortices with U(2) Dirac fermions. 
This was confirmed both at the operator 
level and in matrix representations of the exchange operations. 
In particular, the whole Hilbert spaces for $n=2$, 3, 4 vortices are 
decomposed into subspaces according to representations of U(2), and the 
matrix forms of the exchange operations were obtained in these subspaces. 
We have found  they have off-diagonal elements 
in some subspaces. This analysis is an extension of the previous one for 
the U(1) (one component) Dirac vortices \cite{Yasui:2011gk}. 
By using a simple reduction of the U(2) results, we have indeed 
identified the same matrices as in the U(1) Dirac vortices.
This way of identifying the U(1) structure is rather 
different from the one in the non-Abelian Majorana vortices where 
one can extract the U(1) part as a tensor product. It is not trivial
at this point if a similar identification is possible in the Dirac 
vortices.
Finally, we have found that the whole braid group contains a subgroup 
of genuine non-Abelian statistics for the exchange of
the vortices with one Dirac fermion. 
The reason why non-Abelian statistics appears only from locally defined 
Dirac fermions is that the basis of the Hilbert space are entangled
because of the representation of U(2) acting on the Dirac fermions. 
In this sense, the `spatial' non-locality of Dirac fermions 
is not needed to have non-Abelian statistics but rather
the non-locality, namely entanglement,  
in representations is needed.

\begin{center}
\begin{table}[ht]
\begin{tabular}{|c|c|c|c|c|}
\hline
 & \multicolumn{2}{|c|}{{\large Majorana vortices}} 
 & \multicolumn{2}{|c|}{{\large Dirac vortices}}  \\ 
 & \multicolumn{2}{|c|}{{\large $\hat\gamma^{}_k=\hat\gamma^\dag_k$}} 
 & \multicolumn{2}{|c|}{{\large $\hat\psi^{}_k\neq \hat\psi^\dag_k$}} 
 \\ \hline \hline
\# of trapped fermions & 1 & 3,\qquad\qquad 2$N$-1  & 1 & 2 \\ 
and symmetry & non &  SO(3),\quad SO($2N-1$) & U(1) & U(2) \\ \hline
examples & $p$-wave SC, etc & CFL phase of color SC & A-phase of $^3$He, etc 
& CFL phase of color SC, etc 
\\ \hline
dimension of & $2^{n/2}$& $(2^3)^{n/2}$,\quad $(2^{2N-1})^{n/2}$ & $2^n$& 
$(2^2)^n$ \\  
$n$-vortex system & for even $n$ & for even $n$& & \\ \hline
representation & non-Abelian & non-Abelian & non-Abelian & non-Abelian \\ 
of braid group& (Ivanov matrices) & (Ivanov $\times$ Coxeter) & & \\
\hline
exchange & non-Abelian & non-Abelian & Abelian & non-Abelian \\ 
statistics& (Ivanov matrices) & (Ivanov $\times$ Coxeter) & & \\
\hline
references & \cite{Ivanov:2001} & 
SO(3) \cite{Yasui:2010yw}, 
SO($2N-1$) \cite{Hirono:2012ad}& 
\cite{Yasui:2011gk}& this paper \\ \hline
\end{tabular} 
\caption{Summary of the results on non-Abelian representations of the braid 
group and non-Abelian statistics of 
Majorana/Dirac vortices with Abelian/non-Abelian symmetry. 
Note that the braid group acts on all particles while the non-Abelian statistics concerns only on identical particles. 
Multiple zero-energy fermions are in the vector representations of 
the groups shown in the table. In the Majorana vortices, the exchange 
matrices for a single fermion are called Ivanov matrices, and those 
for multiple fermions are decomposed into the Ivanov matrices and 
generators of the Coxeter group.}
\end{table}
\end{center}

In Table~I, we have summarized the results of 
Majorana/Dirac vortices with Abelian/non-Abelian symmetry. 
All of them exhibit non-Abelian representations of the braid group. 
The right most column corresponds to the present result. 
The exchange of Majorana fermions automatically gives non-Abelian statistics, 
while the exchange of Dirac fermions contains non-Abelian statistics 
when one
exchanges identical particles, 
that is possible at least for U(2).

So far, we know only high-energy-physics examples of 
vortices with multiple Dirac fermions , {\it i.e.}, non-Abelian
vortices in dense QCD and supersymmetric QCD.
As future studies, examples in condensed matter physics and
experimental observation of the non-Abelian
statistics in laboratory will be important.
Applications to quantum computing should also be explored.

\begin{appendix}

\renewcommand{\theequation}{A.\arabic{equation}}
%%%%%%%%%%%%%%%%%%%%%%%%%%%%%%%%%%%%%
\section{Exchange matrices for U(1) Dirac vortices} \label{sec:U(1)tau}
\setcounter{equation}{0}

In this Appendix, we present the matrix representations 
of the exchange operators $\tau_{\ell}$'s ($\ell=1,\cdots,n-1$) 
for the U(1) Dirac vortices which was first obtained in 
Ref.~\cite{Yasui:2011gk} (below we omit the superscript ``s" 
for notational simplicity).
The Hilbert space of $n$ vortices, $\mathbb{H}^{(n)}$, is 
decomposed into subspaces $\mathbb{H}^{(n,f)}$ which are specified 
by the total fermion number $f$.

In the case of $n=2$ ($k$-th and $(k+1)$-th vortices), we have only one 
exchange operation $T_{k}$. The Hilbert space $\mathbb{H}^{(2)}$ is 
decomposed into a direct sum of three subspaces: 
$\mathbb{H}^{(2,0)}\equiv\{ |00\rangle \}$, 
$\mathbb{H}^{(2,1)}\equiv\{ |10\rangle, |01\rangle \}$ and 
$\mathbb{H}^{(2,2)}\equiv\{ |11\rangle \}$.
Thus, $\mathbb{H}^{(2)} = \mathbb{H}^{(2,0)} \oplus \mathbb{H}^{(2,1)} \oplus \mathbb{H}^{(2,2)}$.
Then, we obtain the matrix representations of $\hat \tau_{k}$ as
\begin{eqnarray}
\tau_{k}^{(2,0)} &=& 1, \\
\tau_{k}^{(2,1)} &=&
\left(
\begin{array}{cc}
 0 & -1 \\
 1 & 0
\end{array}
\right), \\
\tau_{k}^{(2,2)} &=& 1.
\end{eqnarray}

In the case of $n=3$ ($k$-th, $(k+1)$-th and $(k+2)$-th vortices), we have two operations $T_{k}$ and $T_{k+1}$.
The Hilbert space $\mathbb{H}^{(3)}$ is decomposed into 
a direct sum of $\mathbb{H}^{(3,0)}\equiv\{ |000\rangle \}$, $\mathbb{H}^{(3,1)}\equiv\{ |100\rangle, |010\rangle, |001\rangle \}$, $\mathbb{H}^{(3,2)}\equiv\{ |110\rangle, |011\rangle, |101\rangle \}$ and $\mathbb{H}^{(3,3)}\equiv\{ |111\rangle \}$.
Namely, $\mathbb{H}^{(3)} = \mathbb{H}^{(3,0)} \oplus \mathbb{H}^{(3,1)} \oplus \mathbb{H}^{(3,2)} \oplus \mathbb{H}^{(3,3)}$.
We thus have the matrix representations of $\hat \tau_{k}$ and $\hat \tau_{k+1}$ as
\begin{eqnarray}
\tau_{k}^{(3,0)} &=& \tau_{k+1}^{(3,0)} = 1, \label{eq:single_3_0} \\ 
\tau_{k}^{(3,1)} &=&
\left(
\begin{array}{ccc}
 0 & -1 & 0 \\
 1 & 0 & 0 \\
 0 & 0 & 1
\end{array}
\right), \,
\tau_{k+1}^{(3,1)} =
\left(
\begin{array}{ccc}
 1 & 0 & 0 \\
 0 & 0 & -1 \\
 0 & 1 & 0
\end{array}
\right), \label{eq:single_3_1} \\
\tau_{k}^{(3,2)} &=&
\left(
\begin{array}{ccc}
 1 & 0 & 0 \\
 0 & 0 & 1 \\
 0 & -1 & 0
\end{array}
\right), \,
\tau_{k+1}^{(3,2)} =
\left(
\begin{array}{ccc}
 0 & 0 & -1 \\
 0 & 1 & 0 \\
 1 & 0 & 0
\end{array}
\right), \label{eq:single_3_2} \\
\tau_{k}^{(3,3)} &=& \tau_{k+1}^{(3,3)} = 1. \label{eq:single_3_3}
\end{eqnarray}

In the case of $n=4$ ($k$-th, $(k+1)$-th, $(k+2)$-th and $(k+3)$-th vortices), we have three operations $T_{k}$, $T_{k+1}$ and $T_{k+2}$.
The Hilbert space $\mathbb{H}^{(4)}$ is decomposed into a direct 
sum of five sectors: $\mathbb{H}^{(4,0)}\equiv\{ |0000\rangle \},$
$\mathbb{H}^{(4,1)}$
$\equiv\{ |1000\rangle, 
|0100\rangle, |0010\rangle, |0001\rangle \}$, 
$\mathbb{H}^{(4,2)}\equiv\{ |1100\rangle$, 
$|1010\rangle, |1001\rangle, |0110\rangle, |0101\rangle, |0011\rangle \}$, 
$\mathbb{H}^{(4,3)}\equiv\{ |1110\rangle$,  
$|1101\rangle, |1011\rangle, |0111\rangle \}$ and 
$\mathbb{H}^{(4,4)}\equiv\{ |1111\rangle \}$.
Namely, $\mathbb{H}^{(4)} = \mathbb{H}^{(4,0)} \oplus \mathbb{H}^{(4,1)} \oplus \mathbb{H}^{(4,2)} \oplus \mathbb{H}^{(4,3)} \oplus \mathbb{H}^{(4,4)}$.
The matrix representations of $\hat \tau_{k}$, $\hat \tau_{k+1}$ and $\hat \tau_{k+2}$ are
\begin{eqnarray}
\tau_{k}^{(4,0)} &=& \tau_{k+1}^{(4,0)} = \tau_{k+2}^{(4,0)} = 1, \label{eq:single_4_0} \\
\tau_{k}^{(4,1)} &=&
\left(
\begin{array}{cccc}
 0 & -1 & 0 & 0 \\
 1 & 0 & 0 & 0 \\
 0 & 0 & 1 & 0 \\
 0 & 0 & 0 & 1
\end{array}
\right), \,
\tau_{k+1}^{(4,1)} =
\left(
\begin{array}{cccc}
 1 & 0 & 0 & 0 \\
 0 & 0 & -1 & 0 \\
 0 & 1 & 0 & 0 \\
 0 & 0 & 0 & 1
\end{array}
\right), \,
\tau_{k+2}^{(4,1)} =
\left(
\begin{array}{cccc}
 1 & 0 & 0 & 0 \\
 0 & 1 & 0 & 0 \\
 0 & 0 & 0 & -1 \\
 0 & 0 & 1 & 0
\end{array}
\right),  \label{eq:single_4_1} \\
\tau_{k}^{(4,2)} &=&
\left(
\begin{array}{cccccc}
 1 & 0 & 0 & 0 & 0 & 0 \\
 0 & 0 & 0 & -1 & 0 & 0 \\
 0 & 0 & 0 & 0 & -1 & 0 \\
 0 & 1 & 0 & 0 & 0 & 0 \\
 0 & 0 & 1 & 0 & 0 & 0 \\
 0 & 0 & 0 & 0 & 0 & 1
\end{array}
\right), \,
\tau_{k+1}^{(4,2)} =  
\left(
\begin{array}{cccccc}
 0 & -1 & 0 & 0 & 0 & 0 \\
 1 & 0 & 0 & 0 & 0 & 0 \\
 0 & 0 & 1 & 0 & 0 & 0 \\
 0 & 0 & 0 & 1 & 0 & 0 \\
 0 & 0 & 0 & 0 & 0 & -1 \\
 0 & 0 & 0 & 0 & 1 & 0
\end{array}
\right), \,
\tau_{k+2}^{(4,2)} =
\left(
\begin{array}{cccccc}
 1 & 0 & 0 & 0 & 0 & 0 \\
 0 & 0 & -1 & 0 & 0 & 0 \\
 0 & 1 & 0 & 0 & 0 & 0 \\
 0 & 0 & 0 & 0 & -1 & 0 \\
 0 & 0 & 0 & 1 & 0 & 0 \\
 0 & 0 & 0 & 0 & 0 & 1
\end{array}
\right), \label{eq:single_4_2} \\
\tau_{k}^{(4,3)} &=&
\left(
\begin{array}{cccc}
 1 & 0 & 0 & 0 \\
 0 & 1 & 0 & 0 \\
 0 & 0 & 0 & -1 \\
 0 & 0 & 1 & 0
\end{array}
\right), \,
\tau_{k+1}^{(4,3)} =
\left(
\begin{array}{cccc}
 1 & 0 & 0 & 0 \\
 0 & 0 & -1 & 0 \\
 0 & 1 & 0 & 0 \\
 0 & 0 & 0 & 1
\end{array}
\right), \,
\tau_{k+2}^{(4,3)} =
\left(
\begin{array}{cccc}
 0 & -1 & 0 & 0 \\
 1 & 0 & 0 & 0 \\
 0 & 0 & 1 & 0 \\
 0 & 0 & 0 & 1
\end{array}
\right), \label{eq:single_4_3} \\ 
\tau_{k}^{(4,4)} &=& \tau_{k+1}^{(4,4)} 
= \tau_{k+2}^{(4,4)} = 1. \label{eq:single_4_4}
\end{eqnarray}

%%%%%%%%%%%%%%%%%%%%%%%%%%%%%%%%%%%%%
\section{The case of $n=4$ for the U(2) Dirac vortices}\label{sec:n=4}

\renewcommand{\theequation}{B.\arabic{equation}}

\setcounter{equation}{0}

We consider the system of $n=4$ vortices, which is an ensemble of $k$-th,
$(k+1)$-th, $(k+2)$-th, and $(k+3)$-th vortices.

The Hilbert space is constructed straightforwardly from the $n=2$ and $n=3$ cases.
The representations of the U(2) symmetry are decomposed as
\begin{eqnarray}
&&({\bf 1}_{0} + {\bf 2}_{1} + {\bf 1}_{2})_{{\rm vortex}\, k} \otimes ({\bf 1}_{0} + {\bf 2}_{1} + {\bf 1}_{2})_{{\rm vortex}\, k+1} \otimes ({\bf 1}_{0} + {\bf 2}_{1} + {\bf 1}_{2})_{{\rm vortex}\, k+2} \otimes ({\bf 1}_{0} + {\bf 2}_{1} + {\bf 1}_{2})_{{\rm vortex}\, k+3} \nonumber \\
&=& {\bf 1}_{0000} + {\bf 1}_{0020} + {\bf 1}_{0002} + {\bf 1}_{2000} + {\bf 1}_{0200} \nonumber \\
&+& {\bf 1}_{0011} + {\bf 1}_{1100} + {\bf 1}_{1010} + {\bf 1}_{1001} + {\bf 1}_{0110} + {\bf 1}_{0101}  \nonumber \\
&+& {\bf 1}_{0022} + {\bf 1}_{2020} + {\bf 1}_{2002} + {\bf 1}_{0220} + {\bf 1}_{0202} + {\bf 1}_{2200} \nonumber \\
&+& {\bf 1}_{1120} + {\bf 1}_{1102} + {\bf 1}_{2011} + {\bf 1}_{0211} + {\bf 1}_{1021} + {\bf 1}_{1012} + {\bf 1}_{0121} + {\bf 1}_{0112} + {\bf 1}_{2110} + {\bf 1}_{2101} + {\bf 1}_{1210} + {\bf 1}_{1201} \nonumber \\
&+& {\bf 1}_{2022} + {\bf 1}_{0222} + {\bf 1}_{2220} + {\bf 1}_{2202} \nonumber \\  
&+& {\bf 1}_{\mathrm{AA}\;\underline{11}\,\underline{11}} + {\bf 1}_{\mathrm{SS}\;\underline{11}\,\underline{11}} \nonumber \\
&+& {\bf 1}_{1122} + {\bf 1}_{2211} + {\bf 1}_{2121} + {\bf 1}_{2112} + {\bf 1}_{1221} + {\bf 1}_{1212} \nonumber \\
&+& {\bf 1}_{2222} \nonumber \\
&+& {\bf 2}_{1000} + {\bf 2}_{0100} + {\bf 2}_{0010} + {\bf 2}_{0001} \nonumber \\
&+& {\bf 2}_{0021} + {\bf 2}_{0012} + {\bf 2}_{2010} + {\bf 2}_{2001}  + {\bf 2}_{0210} + {\bf 2}_{0201} + {\bf 2}_{1020} + {\bf 2}_{1002}  + {\bf 2}_{0120} + {\bf 2}_{0102}  + {\bf 2}_{2100}  + {\bf 2}_{1200} \nonumber \\
&+& {\bf 2}_{\mathrm{A}\,\underline{11}10} + {\bf 2}_{\mathrm{A}\,\underline{11}01} + {\bf 2}_{\mathrm{A}\,10\underline{11}} + {\bf 2}_{\mathrm{S}\,10\underline{11}} + {\bf 2}_{\mathrm{A}\,01\underline{11}} + {\bf 2}_{\mathrm{S}\,01\underline{11}} + {\bf 2}_{\mathrm{S}\,\underline{11}10} + {\bf 2}_{\mathrm{S}\,\underline{11}01} \nonumber \\
&+& {\bf 2}_{2021} + {\bf 2}_{2012} + {\bf 2}_{0221} + {\bf 2}_{0212} + {\bf 2}_{2210} + {\bf 2}_{2201} + {\bf 2}_{1022} + {\bf 2}_{0122} + {\bf 2}_{2120} + {\bf 2}_{2102} + {\bf 2}_{1220} + {\bf 2}_{1202} \nonumber \\
&+& {\bf 2}_{\mathrm{A}\,\underline{11}21} + {\bf 2}_{\mathrm{A}\,\underline{11}12} + {\bf 2}_{\mathrm{A}\,21\underline{11}} + {\bf 2}_{\mathrm{S}\,21\underline{11}} + {\bf 2}_{\mathrm{A}\,12\underline{11}} + {\bf 2}_{\mathrm{S}\,12\underline{11}} + {\bf 2}_{\mathrm{S}\,\underline{11}21} + {\bf 2}_{\mathrm{S}\,\underline{11}12} \nonumber \\
&+& {\bf 2}_{2221} + {\bf 2}_{2212} + {\bf 2}_{2122} + {\bf 2}_{1222} \nonumber \\
&+& {\bf 3}_{1100} + {\bf 3}_{1010} + {\bf 3}_{1001} + {\bf 3}_{0110} + {\bf 3}_{0101} + {\bf 3}_{0011} \nonumber \\
&+& {\bf 3}_{2011} + {\bf 3}_{0211} + {\bf 3}_{1021} + {\bf 3}_{1012} + {\bf 3}_{0121} + {\bf 3}_{0112} + {\bf 3}_{2110} + {\bf 3}_{2101} + {\bf 3}_{1210} + {\bf 3}_{1201} + {\bf 3}_{1120} + {\bf 3}_{1102} \nonumber \\
&+& {\bf 3}_{\mathrm{AS}\,\underline{11}\,\underline{11}} + {\bf 3}_{\mathrm{SA}\,\underline{11}\,\underline{11}} + {\bf 3}_{\mathrm{SS}\,\underline{11}\,\underline{11}} \nonumber \\
&+& {\bf 3}_{2211} + {\bf 3}_{2121} + {\bf 3}_{2112} + {\bf 3}_{1221} + {\bf 3}_{1212} + {\bf 3}_{1122} \nonumber \\
&+& {\bf 4}_{\mathrm{S}\,\underline{11}10} + {\bf 4}_{\mathrm{S}\,\underline{11}01} + {\bf 4}_{\mathrm{S}\,10\underline{11}} + {\bf 4}_{\mathrm{S}\,01\underline{11}}  \nonumber \\
&+& {\bf 4}_{\mathrm{S}\,21\underline{11}} + {\bf 4}_{\mathrm{S}\,12\underline{11}} + {\bf 4}_{\mathrm{S}\,\underline{11}21} + {\bf 4}_{\mathrm{S}\,\underline{11}12} \nonumber \\
&+& {\bf 5}_{\mathrm{S}\,\underline{11}\,\underline{11}},
\end{eqnarray}
where the subscript $n_{k}n_{k+1}n_{k+2}n_{k+3}$ ($n_{k}$, $n_{k+1}$, $n_{k+2}$, $n_{k+3}=0$, $1$, $2$) denotes the number of the Dirac fermions, $n_{k}$, $n_{k+1}$ $n_{k+2}$ and $n_{k+3}$, at the $k$-th, $(k+1)$-th, $(k+2)$-th and $(k+3)$-th vortices, respectively.
One obtains the basis states of the Hilbert space by applying $\hat{\psi}_{\ell}^{a\dag}$ ($\ell=k$, $k+1$, $k+2$, $k+3$ and $a=1$, $2$) successively to the Fock vacuum $| 0 \rangle$ defined by  $\hat{\psi}_{\ell}^{a} | 0 \rangle = 0$ for all $\ell$ and $a=1$, $2$.

For singlet, there are nine Hilbert spaces, $\mathbb{H}^{{\bf 1}_{[0000]}}$, $\mathbb{H}^{{\bf 1}_{[2000]}}$, $\mathbb{H}^{{\bf 1}_{[1100]}}$, $\mathbb{H}^{{\bf 1}_{[2200]}}$, $\mathbb{H}^{{\bf 1}_{[2110]}}$, $\mathbb{H}^{{\bf 1}_{[2220]}}$, $\mathbb{H}^{{\bf 1}_{[1111]}}$, $\mathbb{H}^{{\bf 1}_{[2211]}}$ and $\mathbb{H}^{{\bf 1}_{[2222]}}$.
They are defined as $\mathbb{H}^{{\bf 1}_{[0000]}} \equiv \{ | {\bf 1}_{0000} \rangle \}$ with
\begin{eqnarray}
 |{\bf 1}_{0000} \rangle \equiv | 0 \rangle,
\end{eqnarray}
$\mathbb{H}^{{\bf 1}_{[2000]}} \equiv \{ | {\bf 1}_{0020} \rangle, | {\bf 1}_{0002} \rangle, | {\bf 1}_{2000} \rangle, | {\bf 1}_{0200} \rangle \}$ with
\begin{eqnarray}
 | {\bf 1}_{0020} \rangle &\equiv& \hat{\psi}_{k+2}^{1\dag} \hat{\psi}_{k+2}^{2\dag} | 0 \rangle, \nonumber \\
 | {\bf 1}_{0002} \rangle &\equiv& \hat{\psi}_{k+3}^{1\dag} \hat{\psi}_{k+3}^{2\dag} | 0 \rangle, \nonumber \\
 | {\bf 1}_{2000} \rangle &\equiv& \hat{\psi}_{k}^{1\dag} \hat{\psi}_{k}^{2\dag} | 0 \rangle, \nonumber \\
 | {\bf 1}_{0200} \rangle &\equiv& \hat{\psi}_{k+1}^{1\dag} \hat{\psi}_{k+1}^{2\dag} | 0 \rangle, 
\end{eqnarray}
$\mathbb{H}^{{\bf 1}_{[1100]}} \equiv \{ | {\bf 1}_{0011} \rangle, | {\bf 1}_{1100} \rangle, | {\bf 1}_{1010} \rangle, | {\bf 1}_{1001} \rangle, | {\bf 1}_{0110} \rangle, | {\bf 1}_{0101} \rangle \}$ with
\begin{eqnarray}
 | {\bf 1}_{0011} \rangle &\equiv& \frac{1}{\sqrt{2}} ( \hat{\psi}_{k+2}^{1\dag} \hat{\psi}_{k+3}^{2\dag} - \hat{\psi}_{k+2}^{2\dag} \hat{\psi}_{k+3}^{1\dag} ) | 0 \rangle, \nonumber \\
 | {\bf 1}_{1100} \rangle &\equiv& \frac{1}{\sqrt{2}} ( \hat{\psi}_{k}^{1\dag} \hat{\psi}_{k+1}^{2\dag} - \hat{\psi}_{k}^{2\dag} \hat{\psi}_{k+1}^{1\dag} ) | 0 \rangle, \nonumber \\
 | {\bf 1}_{1010} \rangle &\equiv& \frac{1}{\sqrt{2}} ( \hat{\psi}_{k}^{1\dag} \hat{\psi}_{k+2}^{2\dag} - \hat{\psi}_{k}^{2\dag} \hat{\psi}_{k+2}^{1\dag} ) | 0 \rangle, \nonumber \\
 | {\bf 1}_{1001} \rangle &\equiv& \frac{1}{\sqrt{2}} ( \hat{\psi}_{k}^{1\dag} \hat{\psi}_{k+3}^{2\dag} - \hat{\psi}_{k}^{2\dag} \hat{\psi}_{k+3}^{1\dag} ) | 0 \rangle, \nonumber \\
 | {\bf 1}_{0110} \rangle &\equiv& \frac{1}{\sqrt{2}} ( \hat{\psi}_{k+1}^{1\dag} \hat{\psi}_{k+2}^{2\dag} - \hat{\psi}_{k+1}^{2\dag} \hat{\psi}_{k+2}^{1\dag} ) | 0 \rangle, \nonumber \\
 | {\bf 1}_{0101} \rangle &\equiv& \frac{1}{\sqrt{2}} ( \hat{\psi}_{k+1}^{1\dag} \hat{\psi}_{k+3}^{2\dag} - \hat{\psi}_{k+1}^{2\dag} \hat{\psi}_{k+3}^{1\dag} ) | 0 \rangle,
\end{eqnarray}
$\mathbb{H}^{{\bf 1}_{[2200]}} \equiv \{ | {\bf 1}_{0022} \rangle, | {\bf 1}_{2020} \rangle, | {\bf 1}_{2002} \rangle, | {\bf 1}_{0220} \rangle, | {\bf 1}_{0202} \rangle, | {\bf 1}_{2200} \rangle \}$ with
\begin{eqnarray}
 | {\bf 1}_{0022} \rangle &\equiv& \hat{\psi}_{k+2}^{1\dag} \hat{\psi}_{k+2}^{2\dag} \hat{\psi}_{k+3}^{1\dag} \hat{\psi}_{k+3}^{2\dag} | 0 \rangle, \nonumber \\
 | {\bf 1}_{2020} \rangle &\equiv& \hat{\psi}_{k}^{1\dag} \hat{\psi}_{k}^{2\dag} \hat{\psi}_{k+2}^{1\dag} \hat{\psi}_{k+2}^{2\dag} | 0 \rangle, \nonumber \\
 | {\bf 1}_{2002} \rangle &\equiv& \hat{\psi}_{k}^{1\dag} \hat{\psi}_{k}^{2\dag} \hat{\psi}_{k+3}^{1\dag} \hat{\psi}_{k+3}^{2\dag} | 0 \rangle, \nonumber \\
 | {\bf 1}_{0220} \rangle &\equiv& \hat{\psi}_{k+1}^{1\dag} \hat{\psi}_{k+1}^{2\dag} \hat{\psi}_{k+2}^{1\dag} \hat{\psi}_{k+2}^{2\dag} | 0 \rangle, \nonumber \\
 | {\bf 1}_{0202} \rangle &\equiv& \hat{\psi}_{k+1}^{1\dag} \hat{\psi}_{k+1}^{2\dag} \hat{\psi}_{k+3}^{1\dag} \hat{\psi}_{k+3}^{2\dag} | 0 \rangle, \nonumber \\
 | {\bf 1}_{2200} \rangle &\equiv& \hat{\psi}_{k}^{1\dag} \hat{\psi}_{k}^{2\dag} \hat{\psi}_{k+1}^{1\dag} \hat{\psi}_{k+1}^{2\dag} | 0 \rangle,
\end{eqnarray}
$\mathbb{H}^{{\bf 1}_{[2110]}} \equiv \{ | {\bf 1}_{1120} \rangle, | {\bf 1}_{1102} \rangle, | {\bf 1}_{2011} \rangle, | {\bf 1}_{0211} \rangle, | {\bf 1}_{1021} \rangle, | {\bf 1}_{1012} \rangle, | {\bf 1}_{0121}\rangle, | {\bf 1}_{0112} \rangle, | {\bf 1}_{2110} \rangle, | {\bf 1}_{2101} \rangle, | {\bf 1}_{1210} \rangle, | {\bf 1}_{1201} \rangle \}$ with
\begin{eqnarray}
 | {\bf 1}_{1120} \rangle &\equiv& \frac{1}{\sqrt{2}} ( \hat{\psi}_{k}^{1\dag} \hat{\psi}_{k+1}^{2\dag} - \hat{\psi}_{k}^{2\dag} \hat{\psi}_{k+1}^{1\dag} ) \hat{\psi}_{k+2}^{1\dag} \hat{\psi}_{k+2}^{2\dag} | 0 \rangle, \nonumber \\
 | {\bf 1}_{1102} \rangle &\equiv& \frac{1}{\sqrt{2}} ( \hat{\psi}_{k}^{1\dag} \hat{\psi}_{k+1}^{2\dag} - \hat{\psi}_{k}^{2\dag} \hat{\psi}_{k+1}^{1\dag} ) \hat{\psi}_{k+3}^{1\dag} \hat{\psi}_{k+3}^{2\dag} | 0 \rangle, \nonumber \\
 | {\bf 1}_{2011} \rangle &\equiv& \frac{1}{\sqrt{2}} \hat{\psi}_{k}^{1\dag} \hat{\psi}_{k}^{2\dag} ( \hat{\psi}_{k+2}^{1\dag} \hat{\psi}_{k+3}^{2\dag} - \hat{\psi}_{k+2}^{2\dag} \hat{\psi}_{k+3}^{1\dag} )  | 0 \rangle, \nonumber \\
 | {\bf 1}_{0211} \rangle &\equiv& \frac{1}{\sqrt{2}} \hat{\psi}_{k+1}^{1\dag} \hat{\psi}_{k+1}^{2\dag} ( \hat{\psi}_{k+2}^{1\dag} \hat{\psi}_{k+3}^{2\dag} - \hat{\psi}_{k+2}^{2\dag} \hat{\psi}_{k+3}^{1\dag} )  | 0 \rangle, \nonumber \\
 | {\bf 1}_{1021} \rangle &\equiv& \frac{1}{\sqrt{2}} ( \hat{\psi}_{k}^{1\dag} \hat{\psi}_{k+3}^{2\dag} - \hat{\psi}_{k}^{2\dag} \hat{\psi}_{k+3}^{1\dag} ) \hat{\psi}_{k+2}^{1\dag} \hat{\psi}_{k+2}^{2\dag} | 0 \rangle, \nonumber \\
 | {\bf 1}_{1012} \rangle &\equiv& \frac{1}{\sqrt{2}} ( \hat{\psi}_{k}^{1\dag} \hat{\psi}_{k+2}^{2\dag} - \hat{\psi}_{k}^{2\dag} \hat{\psi}_{k+2}^{1\dag} ) \hat{\psi}_{k+3}^{1\dag} \hat{\psi}_{k+3}^{2\dag} | 0 \rangle, \nonumber \\
 | {\bf 1}_{0121} \rangle &\equiv& \frac{1}{\sqrt{2}} ( \hat{\psi}_{k+1}^{1\dag} \hat{\psi}_{k+3}^{2\dag} - \hat{\psi}_{k+1}^{2\dag} \hat{\psi}_{k+3}^{1\dag} ) \hat{\psi}_{k+2}^{1\dag} \hat{\psi}_{k+2}^{2\dag} | 0 \rangle, \nonumber \\
 | {\bf 1}_{0112} \rangle &\equiv& \frac{1}{\sqrt{2}} ( \hat{\psi}_{k+1}^{1\dag} \hat{\psi}_{k+2}^{2\dag} - \hat{\psi}_{k+1}^{2\dag} \hat{\psi}_{k+2}^{1\dag} ) \hat{\psi}_{k+3}^{1\dag} \hat{\psi}_{k+3}^{2\dag} | 0 \rangle, \nonumber \\
 | {\bf 1}_{2110} \rangle &\equiv& \frac{1}{\sqrt{2}} \hat{\psi}_{k}^{1\dag} \hat{\psi}_{k}^{2\dag} ( \hat{\psi}_{k+1}^{1\dag} \hat{\psi}_{k+2}^{2\dag} - \hat{\psi}_{k+1}^{2\dag} \hat{\psi}_{k+2}^{1\dag} )  | 0 \rangle, \nonumber \\
 | {\bf 1}_{2101} \rangle &\equiv& \frac{1}{\sqrt{2}} \hat{\psi}_{k}^{1\dag} \hat{\psi}_{k}^{2\dag} ( \hat{\psi}_{k+1}^{1\dag} \hat{\psi}_{k+3}^{2\dag} - \hat{\psi}_{k+1}^{2\dag} \hat{\psi}_{k+3}^{1\dag} )  | 0 \rangle, \nonumber \\
 | {\bf 1}_{1210} \rangle &\equiv& \frac{1}{\sqrt{2}} ( \hat{\psi}_{k}^{1\dag} \hat{\psi}_{k+2}^{2\dag} - \hat{\psi}_{k}^{2\dag} \hat{\psi}_{k+2}^{1\dag} ) \hat{\psi}_{k+1}^{1\dag} \hat{\psi}_{k+1}^{2\dag}  | 0 \rangle, \nonumber \\
 | {\bf 1}_{1201} \rangle &\equiv& \frac{1}{\sqrt{2}} ( \hat{\psi}_{k}^{1\dag} \hat{\psi}_{k+3}^{2\dag} - \hat{\psi}_{k}^{2\dag} \hat{\psi}_{k+3}^{1\dag} ) \hat{\psi}_{k+1}^{1\dag} \hat{\psi}_{k+1}^{2\dag}  | 0 \rangle,
\end{eqnarray}
$\mathbb{H}^{{\bf 1}_{[2220]}} \equiv \{ | {\bf 1}_{2022} \rangle, | {\bf 1}_{0222} \rangle, | {\bf 1}_{2220} \rangle, | {\bf 1}_{2202} \rangle \}$ with
\begin{eqnarray}
 | {\bf 1}_{2022} \rangle &\equiv& \hat{\psi}_{k}^{1\dag} \hat{\psi}_{k}^{2\dag} \hat{\psi}_{k+2}^{1\dag} \hat{\psi}_{k+2}^{2\dag} \hat{\psi}_{k+3}^{1\dag} \hat{\psi}_{k+3}^{2\dag} | 0 \rangle, \nonumber \\
 | {\bf 1}_{0222} \rangle &\equiv& \hat{\psi}_{k+1}^{1\dag} \hat{\psi}_{k+1}^{2\dag} \hat{\psi}_{k+2}^{1\dag} \hat{\psi}_{k+2}^{2\dag} \hat{\psi}_{k+3}^{1\dag} \hat{\psi}_{k+3}^{2\dag} | 0 \rangle, \nonumber \\
 | {\bf 1}_{2220} \rangle &\equiv& \hat{\psi}_{k}^{1\dag} \hat{\psi}_{k}^{2\dag} \hat{\psi}_{k+1}^{1\dag} \hat{\psi}_{k+1}^{2\dag} \hat{\psi}_{k+2}^{1\dag} \hat{\psi}_{k+2}^{2\dag} | 0 \rangle, \nonumber \\
 | {\bf 1}_{2202} \rangle &\equiv& \hat{\psi}_{k}^{1\dag} \hat{\psi}_{k}^{2\dag} \hat{\psi}_{k+1}^{1\dag} \hat{\psi}_{k+1}^{2\dag} \hat{\psi}_{k+3}^{1\dag} \hat{\psi}_{k+3}^{2\dag} | 0 \rangle,
\end{eqnarray}
$\mathbb{H}^{{\bf 1}_{[1111]}} \equiv \{ | {\bf
1}_{\mathrm{AA}\,\underline{11}\,\underline{11}} \rangle, | {\bf
1}_{\mathrm{SS}\,\underline{11}\,\underline{11}} \rangle \}$ with
\begin{eqnarray}
| {\bf 1}_{\mathrm{AA}\,\underline{11}\,\underline{11}} \rangle &\equiv& \frac{1}{2} (
 \hat{\psi}_{k}^{1\dag} \hat{\psi}_{k+1}^{2\dag} -
 \hat{\psi}_{k}^{2\dag} \hat{\psi}_{k+1}^{1\dag} ) (
 \hat{\psi}_{k+2}^{1\dag} \hat{\psi}_{k+3}^{2\dag} -
 \hat{\psi}_{k+2}^{2\dag} \hat{\psi}_{k+3}^{1\dag} ) | 0 \rangle, \label{eq:wf_1_1111} \\
 | {\bf 1}_{\mathrm{SS}\,\underline{11}\,\underline{11}} \rangle &\equiv&
  \frac{1}{\sqrt{3}} \left\{ \hat{\psi}_{k}^{1\dag}
		      \hat{\psi}_{k+1}^{1\dag} \hat{\psi}_{k+2}^{2\dag}
		      \hat{\psi}_{k+3}^{2\dag}
 - \frac{1}{2} ( \hat{\psi}_{k}^{1\dag} \hat{\psi}_{k+1}^{2\dag} +
 \hat{\psi}_{k}^{2\dag} \hat{\psi}_{k+1}^{1\dag} ) (
 \hat{\psi}_{k+2}^{1\dag} \hat{\psi}_{k+3}^{2\dag} +
 \hat{\psi}_{k+2}^{2\dag} \hat{\psi}_{k+3}^{1\dag} ) 
 + \hat{\psi}_{k}^{2\dag} \hat{\psi}_{k+1}^{2\dag}
 \hat{\psi}_{k+2}^{1\dag} \hat{\psi}_{k+3}^{1\dag} \right\} | 0 \rangle,
  \nonumber
\end{eqnarray}
with a notation A (S) for an antisymmetric (symmetric) combination
in first and second pair of indices with the underline in the $k$-th and
$(k+1)$-th vortices and in the $(k+2)$-th and $(k+3)$-th vortices.
$\mathbb{H}^{{\bf 1}_{[2211]}} \equiv \{ | {\bf 1}_{1122} \rangle, | {\bf
1}_{2211} \rangle,| {\bf 1}_{2121} \rangle, | {\bf 1}_{2112} \rangle,|
{\bf 1}_{1221} \rangle, | {\bf 1}_{1212} \rangle \}$ with
\begin{eqnarray}
| {\bf 1}_{1122} \rangle &\equiv& \frac{1}{\sqrt{2}} ( \hat{\psi}_{k}^{1\dag}
 \hat{\psi}_{k+1}^{2\dag} - \hat{\psi}_{k}^{2\dag}
 \hat{\psi}_{k+1}^{1\dag} ) \hat{\psi}_{k+2}^{1\dag}
 \hat{\psi}_{k+2}^{2\dag} \hat{\psi}_{k+3}^{1\dag}
 \hat{\psi}_{k+3}^{2\dag} | 0 \rangle, \nonumber \\
| {\bf 1}_{2211} \rangle &\equiv& \frac{1}{\sqrt{2}} \hat{\psi}_{k}^{1\dag}
 \hat{\psi}_{k}^{2\dag} \hat{\psi}_{k+1}^{1\dag}
 \hat{\psi}_{k+1}^{2\dag} ( \hat{\psi}_{k+2}^{1\dag}
 \hat{\psi}_{k+3}^{2\dag} - \hat{\psi}_{k+2}^{2\dag}
 \hat{\psi}_{k+3}^{1\dag} ) | 0 \rangle, \nonumber \\
| {\bf 1}_{2121} \rangle &\equiv& \frac{1}{\sqrt{2}} \hat{\psi}_{k}^{1\dag}
 \hat{\psi}_{k}^{2\dag} ( \hat{\psi}_{k+1}^{1\dag}
 \hat{\psi}_{k+3}^{2\dag} - \hat{\psi}_{k+1}^{2\dag}
 \hat{\psi}_{k+3}^{1\dag} ) \hat{\psi}_{k+2}^{1\dag}
 \hat{\psi}_{k+2}^{2\dag} | 0 \rangle, \nonumber \\
| {\bf 1}_{2112} \rangle &\equiv& \frac{1}{\sqrt{2}} \hat{\psi}_{k}^{1\dag}
 \hat{\psi}_{k}^{2\dag} ( \hat{\psi}_{k+1}^{1\dag}
 \hat{\psi}_{k+2}^{2\dag} - \hat{\psi}_{k+1}^{2\dag}
 \hat{\psi}_{k+2}^{1\dag} ) \hat{\psi}_{k+3}^{1\dag}
 \hat{\psi}_{k+3}^{2\dag} | 0 \rangle, \nonumber \\
| {\bf 1}_{1221} \rangle &\equiv& \frac{1}{\sqrt{2}} ( \hat{\psi}_{k}^{1\dag}
 \hat{\psi}_{k+3}^{2\dag} - \hat{\psi}_{k}^{2\dag}
 \hat{\psi}_{k+3}^{1\dag} ) \hat{\psi}_{k+1}^{1\dag}
 \hat{\psi}_{k+1}^{2\dag} \hat{\psi}_{k+2}^{1\dag}
 \hat{\psi}_{k+2}^{2\dag} | 0 \rangle, \nonumber \\
| {\bf 1}_{1212} \rangle &\equiv& \frac{1}{\sqrt{2}} ( \hat{\psi}_{k}^{1\dag}
 \hat{\psi}_{k+2}^{2\dag} - \hat{\psi}_{k}^{2\dag}
 \hat{\psi}_{k+2}^{1\dag} ) \hat{\psi}_{k+1}^{1\dag}
 \hat{\psi}_{k+1}^{2\dag} \hat{\psi}_{k+3}^{1\dag}
 \hat{\psi}_{k+3}^{2\dag} | 0 \rangle,
\end{eqnarray}
and $\mathbb{H}^{{\bf 1}_{[2222]}} \equiv \{ | {\bf 1}_{2222} \rangle \}$ with
\begin{eqnarray}
| {\bf 1}_{2222} \rangle \equiv \hat{\psi}_{k}^{1\dag} \hat{\psi}_{k}^{2\dag} \hat{\psi}_{k+1}^{1\dag} \hat{\psi}_{k+1}^{2\dag} \hat{\psi}_{k+2}^{1\dag} \hat{\psi}_{k+2}^{2\dag} \hat{\psi}_{k+3}^{1\dag} \hat{\psi}_{k+3}^{2\dag} | 0 \rangle.
\end{eqnarray}
For doublet, there are six Hilbert spaces, $\mathbb{H}^{{\bf 2}_{[1000]}}$, $\mathbb{H}^{{\bf 2}_{[2100]}}$, $\mathbb{H}^{{\bf 2}_{[1110]}}$, $\mathbb{H}^{{\bf 2}_{[2210]}}$, $\mathbb{H}^{{\bf 2}_{[2111]}}$ and $\mathbb{H}^{{\bf 2}_{[2221]}}$.
They are defined as $\mathbb{H}^{{\bf 2}_{[1000]}} \equiv \{ |{\bf 2}_{1000}\rangle, |{\bf 2}_{0100}\rangle, |{\bf 2}_{0010}\rangle, |{\bf 2}_{0001}\rangle \}$ with
\begin{eqnarray}
|{\bf 2}_{1000}\rangle &\equiv&
\left(
\begin{array}{c}
 \hat{\psi}_{k}^{1\dag} \\
 \hat{\psi}_{k}^{2\dag}
\end{array}
\right) | 0 \rangle, \nonumber \\
|{\bf 2}_{0100}\rangle &\equiv&
\left(
\begin{array}{c}
 \hat{\psi}_{k+1}^{1\dag} \\
 \hat{\psi}_{k+1}^{2\dag}
\end{array}
\right) | 0 \rangle, \nonumber \\
|{\bf 2}_{0010}\rangle &\equiv&
\left(
\begin{array}{c}
 \hat{\psi}_{k+2}^{1\dag} \\
 \hat{\psi}_{k+2}^{2\dag}
\end{array}
\right) | 0 \rangle, \nonumber \\
|{\bf 2}_{0001}\rangle &\equiv&
\left(
\begin{array}{c}
 \hat{\psi}_{k+3}^{1\dag} \\
 \hat{\psi}_{k+3}^{2\dag}
\end{array}
\right) | 0 \rangle,
\end{eqnarray}
$\mathbb{H}^{{\bf 2}_{[2100]}} \equiv \{ |{\bf 2}_{0021}\rangle, |{\bf 2}_{0012}\rangle, |{\bf 2}_{2010}\rangle, |{\bf 2}_{2001}\rangle, |{\bf 2}_{0210}\rangle, |{\bf 2}_{0201}\rangle, |{\bf 2}_{1020}\rangle, |{\bf 2}_{1002}\rangle, |{\bf 2}_{0120}\rangle, |{\bf 2}_{0102}\rangle, |{\bf 2}_{2100}\rangle, |{\bf 2}_{1200}\rangle \}$ with
\begin{eqnarray}
|{\bf 2}_{0021}\rangle &\equiv& \hat{\psi}_{k+2}^{1\dag} \hat{\psi}_{k+2}^{2\dag}
\left(
\begin{array}{c}
 \hat{\psi}_{k+3}^{1\dag} \\
 \hat{\psi}_{k+3}^{2\dag}
\end{array}
\right) |0\rangle, \nonumber \\
|{\bf 2}_{0012}\rangle &\equiv&
\left(
\begin{array}{c}
 \hat{\psi}_{k+2}^{1\dag} \\
 \hat{\psi}_{k+2}^{2\dag}
\end{array}
\right) \hat{\psi}_{k+3}^{1\dag} \hat{\psi}_{k+3}^{2\dag} |0\rangle, \nonumber \\
|{\bf 2}_{2010}\rangle &\equiv& \hat{\psi}_{k}^{1\dag} \hat{\psi}_{k}^{2\dag}
\left(
\begin{array}{c}
 \hat{\psi}_{k+2}^{1\dag} \\
 \hat{\psi}_{k+2}^{2\dag}
\end{array}
\right) |0\rangle, \nonumber \\
|{\bf 2}_{2001}\rangle &\equiv& \hat{\psi}_{k}^{1\dag} \hat{\psi}_{k}^{2\dag}
\left(
\begin{array}{c}
 \hat{\psi}_{k+3}^{1\dag} \\
 \hat{\psi}_{k+3}^{2\dag}
\end{array}
\right) |0\rangle, \nonumber \\
|{\bf 2}_{0210}\rangle &\equiv& \hat{\psi}_{k+1}^{1\dag} \hat{\psi}_{k+1}^{2\dag}
\left(
\begin{array}{c}
 \hat{\psi}_{k+2}^{1\dag} \\
 \hat{\psi}_{k+2}^{2\dag}
\end{array}
\right) |0\rangle, \nonumber \\
|{\bf 2}_{0201}\rangle &\equiv& \hat{\psi}_{k+1}^{1\dag} \hat{\psi}_{k+1}^{2\dag}
\left(
\begin{array}{c}
 \hat{\psi}_{k+3}^{1\dag} \\
 \hat{\psi}_{k+3}^{2\dag}
\end{array}
\right) |0\rangle, \nonumber \\
|{\bf 2}_{1020}\rangle &\equiv&
\left(
\begin{array}{c}
 \hat{\psi}_{k}^{1\dag} \\
 \hat{\psi}_{k}^{2\dag}
\end{array}
\right) \hat{\psi}_{k+2}^{1\dag} \hat{\psi}_{k+2}^{2\dag} |0\rangle, \nonumber \\
|{\bf 2}_{1002}\rangle &\equiv&
\left(
\begin{array}{c}
 \hat{\psi}_{k}^{1\dag} \\
 \hat{\psi}_{k}^{2\dag}
\end{array}
\right) \hat{\psi}_{k+3}^{1\dag} \hat{\psi}_{k+3}^{2\dag} |0\rangle, \nonumber \\
|{\bf 2}_{0120}\rangle &\equiv&
\left(
\begin{array}{c}
 \hat{\psi}_{k+1}^{1\dag} \\
 \hat{\psi}_{k+1}^{2\dag}
\end{array}
\right) \hat{\psi}_{k+2}^{1\dag} \hat{\psi}_{k+2}^{2\dag} |0\rangle, \nonumber \\
|{\bf 2}_{0102}\rangle &\equiv&
\left(
\begin{array}{c}
 \hat{\psi}_{k+1}^{1\dag} \\
 \hat{\psi}_{k+1}^{2\dag}
\end{array}
\right) \hat{\psi}_{k+3}^{1\dag} \hat{\psi}_{k+3}^{2\dag} |0\rangle, \nonumber \\
|{\bf 2}_{2100}\rangle &\equiv& \hat{\psi}_{k}^{1\dag} \hat{\psi}_{k}^{2\dag}
\left(
\begin{array}{c}
 \hat{\psi}_{k+1}^{1\dag} \\
 \hat{\psi}_{k+1}^{2\dag}
\end{array}
\right) |0\rangle, \nonumber \\
|{\bf 2}_{1200}\rangle &\equiv& 
\left(
\begin{array}{c}
 \hat{\psi}_{k}^{1\dag} \\
 \hat{\psi}_{k}^{2\dag}
\end{array}
\right) \hat{\psi}_{k+1}^{1\dag} \hat{\psi}_{k+1}^{2\dag} |0\rangle,
\end{eqnarray}
$\mathbb{H}^{{\bf 2}_{[1110]}} \equiv \{ |{\bf
2}_{\mathrm{A}\,\underline{11}10}\rangle, |{\bf 2}_{\mathrm{A}\,\underline{11}01}\rangle,
|{\bf 2}_{\mathrm{A}\,10\underline{11}}\rangle, |{\bf
2}_{\mathrm{S}\,10\underline{11}}\rangle, |{\bf 2}_{\mathrm{A}\,01\underline{11}}\rangle,
|{\bf 2}_{\mathrm{S}\,01\underline{11}}\rangle, |{\bf
2}_{\mathrm{S}\,\underline{11}10}\rangle, |{\bf 2}_{\mathrm{S}\,\underline{11}01}\rangle
\}$ with
\begin{eqnarray}
 |{\bf 2}_{\mathrm{A}\,\underline{11}10}\rangle &\equiv& \frac{1}{\sqrt{2}} (
  \hat{\psi}_{k}^{1\dag} \hat{\psi}_{k+1}^{2\dag} -
  \hat{\psi}_{k}^{2\dag} \hat{\psi}_{k+1}^{1\dag} )
\left(
\begin{array}{c}
 \hat{\psi}_{k+2}^{1\dag} \\
 \hat{\psi}_{k+2}^{2\dag}
\end{array}
\right) |0\rangle, \nonumber \\
 |{\bf 2}_{\mathrm{A}\,\underline{11}01}\rangle &\equiv& \frac{1}{\sqrt{2}} (
  \hat{\psi}_{k}^{1\dag} \hat{\psi}_{k+1}^{2\dag} -
  \hat{\psi}_{k}^{2\dag} \hat{\psi}_{k+1}^{1\dag} )
\left(
\begin{array}{c}
 \hat{\psi}_{k+3}^{1\dag} \\
 \hat{\psi}_{k+3}^{2\dag}
\end{array}
\right) |0\rangle, \nonumber \\
 |{\bf 2}_{\mathrm{A}\,10\underline{11}}\rangle &\equiv& \frac{1}{\sqrt{2}}
\left(
\begin{array}{c}
 \hat{\psi}_{k}^{1\dag} \\
 \hat{\psi}_{k}^{2\dag}
\end{array}
\right) ( \hat{\psi}_{k+2}^{1\dag} \hat{\psi}_{k+3}^{2\dag} -
\hat{\psi}_{k+2}^{2\dag} \hat{\psi}_{k+3}^{1\dag} ) |0\rangle,
\nonumber \\
 |{\bf 2}_{\mathrm{S}\,10\underline{11}}\rangle &\equiv&
 \left(
 \begin{array}{c}
 \frac{1}{\sqrt{3}} \hat{\psi}_{k}^{2\dag} \hat{\psi}_{k+2}^{1\dag}
  \hat{\psi}_{k+3}^{1\dag} - \frac{1}{\sqrt{3}} \hat{\psi}_{k}^{1\dag} (
  \hat{\psi}_{k+2}^{1\dag} \hat{\psi}_{k+3}^{2\dag} +
  \hat{\psi}_{k+2}^{2\dag} \hat{\psi}_{k+3}^{1\dag} ) \\
 \frac{1}{\sqrt{6}} \hat{\psi}_{k}^{2\dag} ( \hat{\psi}_{k+2}^{1\dag}
  \hat{\psi}_{k+3}^{2\dag} + \hat{\psi}_{k+2}^{2\dag}
  \hat{\psi}_{k+3}^{1\dag} ) - \sqrt{\frac{2}{3}}
  \hat{\psi}_{k}^{1\dag} \hat{\psi}_{k+2}^{2\dag}
  \hat{\psi}_{k+3}^{2\dag}
 \end{array}
\right) |0\rangle, \nonumber \\
 |{\bf 2}_{\mathrm{A}\,01\underline{11}}\rangle &\equiv& \frac{1}{\sqrt{2}}
 \left(
\begin{array}{c}
 \hat{\psi}_{k+1}^{1\dag} \\
 \hat{\psi}_{k+1}^{2\dag}
\end{array}
\right) ( \hat{\psi}_{k+2}^{1\dag} \hat{\psi}_{k+3}^{2\dag} -
 \hat{\psi}_{k+2}^{2\dag} \hat{\psi}_{k+3}^{1\dag} ) |0\rangle,
 \nonumber \\
 |{\bf 2}_{\mathrm{S}\,01\underline{11}}\rangle &\equiv&
\left(
 \begin{array}{c}
 \frac{1}{\sqrt{3}} \hat{\psi}_{k+1}^{2\dag} \hat{\psi}_{k+2}^{1\dag} \hat{\psi}_{k+3}^{1\dag} - \frac{1}{\sqrt{3}} \hat{\psi}_{k+1}^{1\dag} ( \hat{\psi}_{k+2}^{1\dag} \hat{\psi}_{k+3}^{2\dag} + \hat{\psi}_{k+2}^{2\dag} \hat{\psi}_{k+3}^{1\dag} ) \\
 \frac{1}{\sqrt{6}} \hat{\psi}_{k+1}^{2\dag} ( \hat{\psi}_{k+2}^{1\dag} \hat{\psi}_{k+3}^{2\dag} + \hat{\psi}_{k+2}^{2\dag} \hat{\psi}_{k+3}^{1\dag} ) - \sqrt{\frac{2}{3}}  \hat{\psi}_{k+1}^{1\dag} \hat{\psi}_{k+2}^{2\dag} \hat{\psi}_{k+3}^{2\dag}
 \end{array}
\right)  |0\rangle, \nonumber \\
 |{\bf 2}_{\mathrm{S}\,\underline{11}10}\rangle &\equiv&
\left(
\begin{array}{c}
\frac{1}{\sqrt{3}} \hat{\psi}_{k}^{1\dag} \hat{\psi}_{k+1}^{1\dag} \hat{\psi}_{k+2}^{2\dag} - \frac{1}{\sqrt{3}} ( \hat{\psi}_{k}^{1\dag} \hat{\psi}_{k+1}^{2\dag} + \hat{\psi}_{k}^{2\dag} \hat{\psi}_{k+1}^{1\dag} ) \hat{\psi}_{k+2}^{1\dag} \\
\frac{1}{\sqrt{6}} ( \hat{\psi}_{k}^{1\dag} \hat{\psi}_{k+1}^{2\dag} + \hat{\psi}_{k}^{2\dag} \hat{\psi}_{k+1}^{1\dag} ) \hat{\psi}_{k+2}^{2\dag} - \sqrt{\frac{2}{3}} \hat{\psi}_{k}^{2\dag} \hat{\psi}_{k+1}^{2\dag} \hat{\psi}_{k+2}^{1\dag}
\end{array}
\right) |0\rangle, \nonumber \\
 |{\bf 2}_{\mathrm{S}\,\underline{11}01}\rangle &\equiv&
\left(
\begin{array}{c}
\frac{1}{\sqrt{3}} \hat{\psi}_{k}^{1\dag} \hat{\psi}_{k+1}^{1\dag} \hat{\psi}_{k+3}^{2\dag} - \frac{1}{\sqrt{3}} ( \hat{\psi}_{k}^{1\dag} \hat{\psi}_{k+1}^{2\dag} + \hat{\psi}_{k}^{2\dag} \hat{\psi}_{k+1}^{1\dag} ) \hat{\psi}_{k+3}^{1\dag} \\
\frac{1}{\sqrt{6}} ( \hat{\psi}_{k}^{1\dag} \hat{\psi}_{k+1}^{2\dag} + \hat{\psi}_{k}^{2\dag} \hat{\psi}_{k+1}^{1\dag} ) \hat{\psi}_{k+3}^{2\dag} - \sqrt{\frac{2}{3}} \hat{\psi}_{k}^{2\dag} \hat{\psi}_{k+1}^{2\dag} \hat{\psi}_{k+3}^{1\dag}
\end{array}
\right) |0\rangle,   \label{eq:H2(1110)}
\end{eqnarray}
with a notation A (S) for an antisymmetric (symmetric) combination in a pair of indices with the underline in the $k$-th and $(k+1)$-th vortices, or in the $(k+2)$-th and $(k+3)$-th vortices.
$\mathbb{H}^{{\bf 2}_{[2210]}} \equiv \{ |{\bf 2}_{2021}\rangle, |{\bf 2}_{2012}\rangle, |{\bf 2}_{0221}\rangle, |{\bf 2}_{0212}\rangle, |{\bf 2}_{2210}\rangle, |{\bf 2}_{2201}\rangle, |{\bf 2}_{1022}\rangle, |{\bf 2}_{0122}\rangle, |{\bf 2}_{2120}\rangle, |{\bf 2}_{2102}\rangle, |{\bf 2}_{1220}\rangle, |{\bf 2}_{1202}\rangle \}$ with
\begin{eqnarray}
|{\bf 2}_{2021}\rangle &\equiv& \hat{\psi}_{k}^{1\dag} \hat{\psi}_{k}^{2\dag} \hat{\psi}_{k+2}^{1\dag} \hat{\psi}_{k+2}^{2\dag}
\left(
\begin{array}{c}
 \hat{\psi}_{k+3}^{1\dag} \\
 \hat{\psi}_{k+3}^{2\dag}
\end{array}
\right) |0\rangle, \nonumber \\
|{\bf 2}_{2012}\rangle &\equiv& \hat{\psi}_{k}^{1\dag} \hat{\psi}_{k}^{2\dag}
\left(
\begin{array}{c}
 \hat{\psi}_{k+2}^{1\dag} \\
 \hat{\psi}_{k+2}^{2\dag}
\end{array}
\right) \hat{\psi}_{k+3}^{1\dag} \hat{\psi}_{k+3}^{2\dag} |0\rangle, \nonumber \\
|{\bf 2}_{0221}\rangle &\equiv& \hat{\psi}_{k+1}^{1\dag} \hat{\psi}_{k+1}^{2\dag} \hat{\psi}_{k+2}^{1\dag} \hat{\psi}_{k+2}^{2\dag} 
\left(
\begin{array}{c}
 \hat{\psi}_{k+3}^{1\dag} \\
 \hat{\psi}_{k+3}^{2\dag}
\end{array}
\right) |0\rangle, \nonumber \\
|{\bf 2}_{0212}\rangle &\equiv& \hat{\psi}_{k+1}^{1\dag} \hat{\psi}_{k+1}^{2\dag} 
\left(
\begin{array}{c}
 \hat{\psi}_{k+2}^{1\dag} \\
 \hat{\psi}_{k+2}^{2\dag}
\end{array}
\right) \hat{\psi}_{k+3}^{1\dag} \hat{\psi}_{k+3}^{2\dag} |0\rangle, \nonumber \\
|{\bf 2}_{2210}\rangle &\equiv& \hat{\psi}_{k}^{1\dag} \hat{\psi}_{k}^{2\dag} \hat{\psi}_{k+1}^{1\dag} \hat{\psi}_{k+1}^{2\dag}
\left(
\begin{array}{c}
 \hat{\psi}_{k+2}^{1\dag} \\
 \hat{\psi}_{k+2}^{2\dag}
\end{array}
\right) |0\rangle, \nonumber \\
|{\bf 2}_{2201}\rangle &\equiv& \hat{\psi}_{k}^{1\dag} \hat{\psi}_{k}^{2\dag} \hat{\psi}_{k+1}^{1\dag} \hat{\psi}_{k+1}^{2\dag}
\left(
\begin{array}{c}
 \hat{\psi}_{k+3}^{1\dag} \\
 \hat{\psi}_{k+3}^{2\dag}
\end{array}
\right) |0\rangle, \nonumber \\
|{\bf 2}_{1022}\rangle &\equiv&
\left(
\begin{array}{c}
 \hat{\psi}_{k}^{1\dag} \\
 \hat{\psi}_{k}^{2\dag}
\end{array}
\right) \hat{\psi}_{k+2}^{1\dag} \hat{\psi}_{k+2}^{2\dag} \hat{\psi}_{k+3}^{1\dag} \hat{\psi}_{k+3}^{2\dag} |0\rangle, \nonumber \\
|{\bf 2}_{0122}\rangle &\equiv&
\left(
\begin{array}{c}
 \hat{\psi}_{k+1}^{1\dag} \\
 \hat{\psi}_{k+1}^{2\dag}
\end{array}
\right) \hat{\psi}_{k+2}^{1\dag} \hat{\psi}_{k+2}^{2\dag} \hat{\psi}_{k+3}^{1\dag} \hat{\psi}_{k+3}^{2\dag} |0\rangle, \nonumber \\
|{\bf 2}_{2120}\rangle &\equiv& \hat{\psi}_{k}^{1\dag} \hat{\psi}_{k}^{2\dag}
\left(
\begin{array}{c}
 \hat{\psi}_{k+1}^{1\dag} \\
 \hat{\psi}_{k+1}^{2\dag}
\end{array}
\right) \hat{\psi}_{k+2}^{1\dag} \hat{\psi}_{k+2}^{2\dag} |0\rangle, \nonumber \\
|{\bf 2}_{2102}\rangle &\equiv& \hat{\psi}_{k}^{1\dag} \hat{\psi}_{k}^{2\dag}
\left(
\begin{array}{c}
 \hat{\psi}_{k+1}^{1\dag} \\
 \hat{\psi}_{k+1}^{2\dag}
\end{array}
\right) \hat{\psi}_{k+3}^{1\dag} \hat{\psi}_{k+3}^{2\dag} |0\rangle, \nonumber \\
|{\bf 2}_{1220}\rangle &\equiv& 
\left(
\begin{array}{c}
 \hat{\psi}_{k}^{1\dag} \\
 \hat{\psi}_{k}^{2\dag}
\end{array}
\right) \hat{\psi}_{k+1}^{1\dag} \hat{\psi}_{k+1}^{2\dag} \hat{\psi}_{k+2}^{1\dag} \hat{\psi}_{k+2}^{2\dag} |0\rangle, \nonumber \\
|{\bf 2}_{1202}\rangle &\equiv& 
\left(
\begin{array}{c}
 \hat{\psi}_{k}^{1\dag} \\
 \hat{\psi}_{k}^{2\dag}
\end{array}
\right) \hat{\psi}_{k+1}^{1\dag} \hat{\psi}_{k+1}^{2\dag} \hat{\psi}_{k+3}^{1\dag} \hat{\psi}_{k+3}^{2\dag} |0\rangle,
\end{eqnarray}
$\mathbb{H}^{{\bf 2}_{[2111]}} \equiv \{ |{\bf
2}_{\mathrm{A}\,\underline{11}21}\rangle, |{\bf 2}_{\mathrm{A}\,\underline{11}12}\rangle,
|{\bf 2}_{\mathrm{A}\,21\underline{11}}\rangle, |{\bf
2}_{\mathrm{S}\,21\underline{11}}\rangle, |{\bf 2}_{\mathrm{A}\,12\underline{11}}\rangle,
|{\bf 2}_{\mathrm{S}\,12\underline{11}}\rangle, |{\bf
2}_{\mathrm{S}\,\underline{11}21}\rangle, |{\bf 2}_{\mathrm{S}\,\underline{11}12}\rangle
\}$ with
\begin{eqnarray}
|{\bf 2}_{\mathrm{A}\,\underline{11}21}\rangle &\equiv&
 \frac{1}{\sqrt{2}} ( \hat{\psi}_{k}^{1\dag} \hat{\psi}_{k+1}^{2\dag} - \hat{\psi}_{k}^{2\dag} \hat{\psi}_{k+1}^{1\dag} ) \hat{\psi}_{k+2}^{1\dag} \hat{\psi}_{k+2}^{2\dag}
\left(
\begin{array}{c}
 \hat{\psi}_{k+3}^{1\dag} \\
 \hat{\psi}_{k+3}^{2\dag}
\end{array}
\right) |0\rangle, \nonumber \\
 |{\bf 2}_{\mathrm{A}\,\underline{11}12}\rangle &\equiv& \frac{1}{\sqrt{2}} ( \hat{\psi}_{k}^{1\dag} \hat{\psi}_{k+1}^{2\dag} - \hat{\psi}_{k}^{2\dag} \hat{\psi}_{k+1}^{1\dag} )
\left(
\begin{array}{c}
 \hat{\psi}_{k+2}^{1\dag} \\
 \hat{\psi}_{k+2}^{2\dag}
\end{array}
\right) \hat{\psi}_{k+3}^{1\dag} \hat{\psi}_{k+3}^{2\dag} |0\rangle, \nonumber \\
 |{\bf 2}_{\mathrm{A}\,21\underline{11}}\rangle &\equiv& \frac{1}{\sqrt{2}} \hat{\psi}_{k}^{1\dag} \hat{\psi}_{k}^{2\dag}
\left(
\begin{array}{c}
 \hat{\psi}_{k+1}^{1\dag} \\
 \hat{\psi}_{k+1}^{2\dag}
\end{array}
\right) ( \hat{\psi}_{k+2}^{1\dag} \hat{\psi}_{k+3}^{2\dag} - \hat{\psi}_{k+2}^{2\dag} \hat{\psi}_{k+3}^{1\dag} )  |0\rangle, \nonumber \\
|{\bf 2}_{\mathrm{S}\,21\underline{11}}\rangle &\equiv& \hat{\psi}_{k}^{1\dag} \hat{\psi}_{k}^{2\dag} \left(
 \begin{array}{c}
 \frac{1}{\sqrt{3}} \hat{\psi}_{k+1}^{2\dag} \hat{\psi}_{k+2}^{1\dag} \hat{\psi}_{k+3}^{1\dag} - \frac{1}{\sqrt{3}} \hat{\psi}_{k+1}^{1\dag} ( \hat{\psi}_{k+2}^{1\dag} \hat{\psi}_{k+3}^{2\dag} + \hat{\psi}_{k+2}^{2\dag} \hat{\psi}_{k+3}^{1\dag} ) \\
 \frac{1}{\sqrt{6}} \hat{\psi}_{k+1}^{2\dag} ( \hat{\psi}_{k+2}^{1\dag} \hat{\psi}_{k+3}^{2\dag} + \hat{\psi}_{k+2}^{2\dag} \hat{\psi}_{k+3}^{1\dag} ) - \sqrt{\frac{2}{3}}  \hat{\psi}_{k+1}^{1\dag} \hat{\psi}_{k+2}^{2\dag} \hat{\psi}_{k+3}^{2\dag}
 \end{array}
\right) |0\rangle, \nonumber \\
|{\bf 2}_{\mathrm{A}\,12\underline{11}}\rangle &\equiv& \frac{1}{\sqrt{2}}
\left(
\begin{array}{c}
 \hat{\psi}_{k}^{1\dag} \\
 \hat{\psi}_{k}^{2\dag}
\end{array}
\right) \hat{\psi}_{k+1}^{1\dag} \hat{\psi}_{k+1}^{2\dag} ( \hat{\psi}_{k+2}^{1\dag} \hat{\psi}_{k+3}^{2\dag} - \hat{\psi}_{k+2}^{2\dag} \hat{\psi}_{k+3}^{1\dag} )  |0\rangle, \nonumber \\
|{\bf 2}_{\mathrm{S}\,12\underline{11}}\rangle &\equiv& \hat{\psi}_{k+1}^{1\dag} \hat{\psi}_{k+1}^{2\dag}
\left(
 \begin{array}{c}
 \frac{1}{\sqrt{3}} \hat{\psi}_{k}^{2\dag} \hat{\psi}_{k+2}^{1\dag} \hat{\psi}_{k+3}^{1\dag} - \frac{1}{\sqrt{3}} \hat{\psi}_{k}^{1\dag} ( \hat{\psi}_{k+2}^{1\dag} \hat{\psi}_{k+3}^{2\dag} + \hat{\psi}_{k+2}^{2\dag} \hat{\psi}_{k+3}^{1\dag} ) \\
 \frac{1}{\sqrt{6}} \hat{\psi}_{k}^{2\dag} ( \hat{\psi}_{k+2}^{1\dag} \hat{\psi}_{k+3}^{2\dag} + \hat{\psi}_{k+2}^{2\dag} \hat{\psi}_{k+3}^{1\dag} ) - \sqrt{\frac{2}{3}}  \hat{\psi}_{k}^{1\dag} \hat{\psi}_{k+2}^{2\dag} \hat{\psi}_{k+3}^{2\dag}
 \end{array}
\right) |0\rangle, \nonumber \\
|{\bf 2}_{\mathrm{S}\,\underline{11}21}\rangle &\equiv&
 \left(
\begin{array}{c}
\frac{1}{\sqrt{3}} \hat{\psi}_{k}^{1\dag} \hat{\psi}_{k+1}^{1\dag} \hat{\psi}_{k+3}^{2\dag} - \frac{1}{\sqrt{3}} ( \hat{\psi}_{k}^{1\dag} \hat{\psi}_{k+1}^{2\dag} + \hat{\psi}_{k}^{2\dag} \hat{\psi}_{k+1}^{1\dag} ) \hat{\psi}_{k+3}^{1\dag} \\
\frac{1}{\sqrt{6}} ( \hat{\psi}_{k}^{1\dag} \hat{\psi}_{k+1}^{2\dag} + \hat{\psi}_{k}^{2\dag} \hat{\psi}_{k+1}^{1\dag} ) \hat{\psi}_{k+3}^{2\dag} - \sqrt{\frac{2}{3}} \hat{\psi}_{k}^{2\dag} \hat{\psi}_{k+1}^{2\dag} \hat{\psi}_{k+3}^{1\dag}
\end{array}
\right) \hat{\psi}_{k+2}^{1\dag} \hat{\psi}_{k+2}^{2\dag} |0\rangle, \nonumber \\
|{\bf 2}_{\mathrm{S}\,\underline{11}12}\rangle &\equiv&
\left(
\begin{array}{c}
\frac{1}{\sqrt{3}} \hat{\psi}_{k}^{1\dag} \hat{\psi}_{k+1}^{1\dag} \hat{\psi}_{k+2}^{2\dag} - \frac{1}{\sqrt{3}} ( \hat{\psi}_{k}^{1\dag} \hat{\psi}_{k+1}^{2\dag} + \hat{\psi}_{k}^{2\dag} \hat{\psi}_{k+1}^{1\dag} ) \hat{\psi}_{k+2}^{1\dag} \\
\frac{1}{\sqrt{6}} ( \hat{\psi}_{k}^{1\dag} \hat{\psi}_{k+1}^{2\dag} + \hat{\psi}_{k}^{2\dag} \hat{\psi}_{k+1}^{1\dag} ) \hat{\psi}_{k+2}^{2\dag} - \sqrt{\frac{2}{3}} \hat{\psi}_{k}^{2\dag} \hat{\psi}_{k+1}^{2\dag} \hat{\psi}_{k+2}^{1\dag}
\end{array}
\right) \hat{\psi}_{k+3}^{1\dag} \hat{\psi}_{k+3}^{2\dag} |0\rangle,
\end{eqnarray}
with a notation A (S) for an antisymmetric (symmetric) combination in a pair of indices with the underline in the $k$-th and $(k+1)$-th vortices, or in the $(k+2)$-th and $(k+3)$-th vortices.
$\mathbb{H}^{{\bf 2}_{[2221]}} \equiv \{ |{\bf 2}_{2221}\rangle, |{\bf 2}_{2212}\rangle, |{\bf 2}_{2122}\rangle, |{\bf 2}_{1222}\rangle \}$ with
\begin{eqnarray}
|{\bf 2}_{2221}\rangle &\equiv& \hat{\psi}_{k}^{1\dag} \hat{\psi}_{k}^{2\dag} \hat{\psi}_{k+1}^{1\dag} \hat{\psi}_{k+1}^{2\dag} \hat{\psi}_{k+2}^{1\dag} \hat{\psi}_{k+2}^{2\dag}
\left(
\begin{array}{c}
 \hat{\psi}_{k+3}^{1\dag} \\
 \hat{\psi}_{k+3}^{2\dag}
\end{array}
\right) |0\rangle, \nonumber \\
|{\bf 2}_{2212}\rangle &\equiv& \hat{\psi}_{k}^{1\dag} \hat{\psi}_{k}^{2\dag} \hat{\psi}_{k+1}^{1\dag} \hat{\psi}_{k+1}^{2\dag} 
\left(
\begin{array}{c}
 \hat{\psi}_{k+2}^{1\dag} \\
 \hat{\psi}_{k+2}^{2\dag}
\end{array}
\right) \hat{\psi}_{k+3}^{1\dag} \hat{\psi}_{k+3}^{2\dag} |0\rangle, \nonumber \\
|{\bf 2}_{2122}\rangle &\equiv& \hat{\psi}_{k}^{1\dag} \hat{\psi}_{k}^{2\dag}
\left(
\begin{array}{c}
 \hat{\psi}_{k+1}^{1\dag} \\
 \hat{\psi}_{k+1}^{2\dag}
\end{array}
\right) \hat{\psi}_{k+2}^{1\dag} \hat{\psi}_{k+2}^{2\dag} \hat{\psi}_{k+3}^{1\dag} \hat{\psi}_{k+3}^{2\dag} |0\rangle, \nonumber \\
|{\bf 2}_{1222}\rangle &\equiv& 
\left(
\begin{array}{c}
 \hat{\psi}_{k}^{1\dag} \\
 \hat{\psi}_{k}^{2\dag}
\end{array}
\right) \hat{\psi}_{k+1}^{1\dag} \hat{\psi}_{k+1}^{2\dag} \hat{\psi}_{k+2}^{1\dag} \hat{\psi}_{k+2}^{2\dag} \hat{\psi}_{k+3}^{1\dag} \hat{\psi}_{k+3}^{2\dag} |0\rangle.
\end{eqnarray}
For triplet, there are four Hilbert spaces, $\mathbb{H}^{{\bf 3}_{[1100]}}$, $\mathbb{H}^{{\bf 3}_{[2110]}}$, $\mathbb{H}^{{\bf 3}_{[1111]}}$ and $\mathbb{H}^{{\bf 3}_{[2211]}}$.
They are defined as $\mathbb{H}^{{\bf 3}_{[1100]}} \equiv \{ |{\bf 3}_{1100}\rangle, |{\bf 3}_{1010}\rangle, |{\bf 3}_{1001}\rangle, |{\bf 3}_{0110}\rangle, |{\bf 3}_{0101}\rangle, |{\bf 3}_{0011}\rangle \}$ with
\begin{eqnarray}
|{\bf 3}_{1100}\rangle &\equiv&
\left(
\begin{array}{c}
 \hat{\psi}_{k}^{1\dag} \hat{\psi}_{k+1}^{1\dag} \\
 \frac{1}{\sqrt{2}} ( \hat{\psi}_{k}^{1\dag} \hat{\psi}_{k+1}^{2\dag} + \hat{\psi}_{k}^{2\dag} \hat{\psi}_{k+1}^{1\dag} ) \\
 \hat{\psi}_{k}^{2\dag} \hat{\psi}_{k+1}^{2\dag}
\end{array}
\right) |0\rangle, \nonumber \\
|{\bf 3}_{1010}\rangle &\equiv&
\left(
\begin{array}{c}
 \hat{\psi}_{k}^{1\dag} \hat{\psi}_{k+2}^{1\dag} \\
 \frac{1}{\sqrt{2}} ( \hat{\psi}_{k}^{1\dag} \hat{\psi}_{k+2}^{2\dag} + \hat{\psi}_{k}^{2\dag} \hat{\psi}_{k+2}^{1\dag} ) \\
 \hat{\psi}_{k}^{2\dag} \hat{\psi}_{k+2}^{2\dag}
\end{array}
\right) |0\rangle, \nonumber \\
|{\bf 3}_{1001}\rangle &\equiv&
\left(
\begin{array}{c}
 \hat{\psi}_{k}^{1\dag} \hat{\psi}_{k+3}^{1\dag} \\
 \frac{1}{\sqrt{2}} ( \hat{\psi}_{k}^{1\dag} \hat{\psi}_{k+3}^{2\dag} + \hat{\psi}_{k}^{2\dag} \hat{\psi}_{k+3}^{1\dag} ) \\
 \hat{\psi}_{k}^{2\dag} \hat{\psi}_{k+3}^{2\dag}
\end{array}
\right) |0\rangle, \nonumber \\
|{\bf 3}_{0110}\rangle &\equiv&
\left(
\begin{array}{c}
 \hat{\psi}_{k+1}^{1\dag} \hat{\psi}_{k+2}^{1\dag} \\
 \frac{1}{\sqrt{2}} ( \hat{\psi}_{k+1}^{1\dag} \hat{\psi}_{k+2}^{2\dag} + \hat{\psi}_{k+1}^{2\dag} \hat{\psi}_{k+2}^{1\dag} ) \\
 \hat{\psi}_{k+1}^{2\dag} \hat{\psi}_{k+2}^{2\dag}
\end{array}
\right) |0\rangle, \nonumber \\
|{\bf 3}_{0101}\rangle &\equiv&
\left(
\begin{array}{c}
 \hat{\psi}_{k+1}^{1\dag} \hat{\psi}_{k+3}^{1\dag} \\
 \frac{1}{\sqrt{2}} ( \hat{\psi}_{k+1}^{1\dag} \hat{\psi}_{k+3}^{2\dag} + \hat{\psi}_{k+1}^{2\dag} \hat{\psi}_{k+3}^{1\dag} ) \\
 \hat{\psi}_{k+1}^{2\dag} \hat{\psi}_{k+3}^{2\dag}
\end{array}
\right) |0\rangle, \nonumber \\
|{\bf 3}_{0011}\rangle &\equiv&
\left(
\begin{array}{c}
 \hat{\psi}_{k+2}^{1\dag} \hat{\psi}_{k+3}^{1\dag} \\
 \frac{1}{\sqrt{2}} ( \hat{\psi}_{k+2}^{1\dag} \hat{\psi}_{k+3}^{2\dag} + \hat{\psi}_{k+2}^{2\dag} \hat{\psi}_{k+3}^{1\dag} ) \\
 \hat{\psi}_{k+2}^{2\dag} \hat{\psi}_{k+3}^{2\dag}
\end{array}
\right) |0\rangle,
\end{eqnarray}
$\mathbb{H}^{{\bf 3}_{[2110]}} \equiv \{ |{\bf 3}_{2011}\rangle, |{\bf 3}_{0211}\rangle, |{\bf 3}_{1021}\rangle, |{\bf 3}_{1012}\rangle, |{\bf 3}_{0121}\rangle, |{\bf 3}_{0112}\rangle, |{\bf 3}_{2110}\rangle, |{\bf 3}_{2101}\rangle, |{\bf 3}_{1210}\rangle, |{\bf 3}_{1201}\rangle, |{\bf 3}_{1120}\rangle, |{\bf 3}_{1102}\rangle \}$ with
\begin{eqnarray}
|{\bf 3}_{2011}\rangle &\equiv& \hat{\psi}_{k}^{1\dag} \hat{\psi}_{k}^{2\dag}
\left(
\begin{array}{c}
 \hat{\psi}_{k+2}^{1\dag} \hat{\psi}_{k+3}^{1\dag} \\
 \frac{1}{\sqrt{2}} ( \hat{\psi}_{k+2}^{1\dag} \hat{\psi}_{k+3}^{2\dag} + \hat{\psi}_{k+2}^{2\dag} \hat{\psi}_{k+3}^{1\dag} ) \\
 \hat{\psi}_{k+2}^{2\dag} \hat{\psi}_{k+3}^{2\dag}
\end{array}
\right) |0\rangle, \nonumber \\
|{\bf 3}_{0211}\rangle &\equiv& \hat{\psi}_{k+1}^{1\dag} \hat{\psi}_{k+1}^{2\dag}
\left(
\begin{array}{c}
 \hat{\psi}_{k+2}^{1\dag} \hat{\psi}_{k+3}^{1\dag} \\
 \frac{1}{\sqrt{2}} ( \hat{\psi}_{k+2}^{1\dag} \hat{\psi}_{k+3}^{2\dag} + \hat{\psi}_{k+2}^{2\dag} \hat{\psi}_{k+3}^{1\dag} ) \\
 \hat{\psi}_{k+2}^{2\dag} \hat{\psi}_{k+3}^{2\dag}
\end{array}
\right) |0\rangle, \nonumber \\
|{\bf 3}_{1021}\rangle &\equiv& 
\left(
\begin{array}{c}
 \hat{\psi}_{k}^{1\dag} \hat{\psi}_{k+3}^{1\dag} \\
 \frac{1}{\sqrt{2}} ( \hat{\psi}_{k}^{1\dag} \hat{\psi}_{k+3}^{2\dag} + \hat{\psi}_{k}^{2\dag} \hat{\psi}_{k+3}^{1\dag} ) \\
 \hat{\psi}_{k}^{2\dag} \hat{\psi}_{k+3}^{2\dag}
\end{array}
\right) \hat{\psi}_{k+2}^{1\dag} \hat{\psi}_{k+2}^{2\dag} |0\rangle, \nonumber \\
|{\bf 3}_{1012}\rangle &\equiv& 
\left(
\begin{array}{c}
 \hat{\psi}_{k}^{1\dag} \hat{\psi}_{k+2}^{1\dag} \\
 \frac{1}{\sqrt{2}} ( \hat{\psi}_{k}^{1\dag} \hat{\psi}_{k+2}^{2\dag} + \hat{\psi}_{k}^{2\dag} \hat{\psi}_{k+2}^{1\dag} ) \\
 \hat{\psi}_{k}^{2\dag} \hat{\psi}_{k+2}^{2\dag}
\end{array}
\right) \hat{\psi}_{k+3}^{1\dag} \hat{\psi}_{k+3}^{2\dag} |0\rangle, \nonumber \\
|{\bf 3}_{0121}\rangle &\equiv&
\left(
\begin{array}{c}
 \hat{\psi}_{k+1}^{1\dag} \hat{\psi}_{k+3}^{1\dag} \\
 \frac{1}{\sqrt{2}} ( \hat{\psi}_{k+1}^{1\dag} \hat{\psi}_{k+3}^{2\dag} + \hat{\psi}_{k+1}^{2\dag} \hat{\psi}_{k+3}^{1\dag} ) \\
 \hat{\psi}_{k+1}^{2\dag} \hat{\psi}_{k+3}^{2\dag}
\end{array}
\right) \hat{\psi}_{k+2}^{1\dag} \hat{\psi}_{k+2}^{2\dag} |0\rangle, \nonumber \\
|{\bf 3}_{0112}\rangle &\equiv& 
\left(
\begin{array}{c}
 \hat{\psi}_{k+1}^{1\dag} \hat{\psi}_{k+2}^{1\dag} \\
 \frac{1}{\sqrt{2}} ( \hat{\psi}_{k+1}^{1\dag} \hat{\psi}_{k+2}^{2\dag} + \hat{\psi}_{k+1}^{2\dag} \hat{\psi}_{k+2}^{1\dag} ) \\
 \hat{\psi}_{k+1}^{2\dag} \hat{\psi}_{k+2}^{2\dag}
\end{array}
\right) \hat{\psi}_{k+3}^{1\dag} \hat{\psi}_{k+3}^{2\dag} |0\rangle, \nonumber \\
|{\bf 3}_{2110}\rangle &\equiv& \hat{\psi}_{k}^{1\dag} \hat{\psi}_{k}^{2\dag}
\left(
\begin{array}{c}
 \hat{\psi}_{k+1}^{1\dag} \hat{\psi}_{k+2}^{1\dag} \\
 \frac{1}{\sqrt{2}} ( \hat{\psi}_{k+1}^{1\dag} \hat{\psi}_{k+2}^{2\dag} + \hat{\psi}_{k+1}^{2\dag} \hat{\psi}_{k+2}^{1\dag} ) \\
 \hat{\psi}_{k+1}^{2\dag} \hat{\psi}_{k+2}^{2\dag}
\end{array}
\right) |0\rangle, \nonumber \\
|{\bf 3}_{2101}\rangle &\equiv& \hat{\psi}_{k}^{1\dag} \hat{\psi}_{k}^{2\dag}
\left(
\begin{array}{c}
 \hat{\psi}_{k+1}^{1\dag} \hat{\psi}_{k+3}^{1\dag} \\
 \frac{1}{\sqrt{2}} ( \hat{\psi}_{k+1}^{1\dag} \hat{\psi}_{k+3}^{2\dag} + \hat{\psi}_{k+1}^{2\dag} \hat{\psi}_{k+3}^{1\dag} ) \\
 \hat{\psi}_{k+1}^{2\dag} \hat{\psi}_{k+3}^{2\dag}
\end{array}
\right) |0\rangle, \nonumber \\
|{\bf 3}_{1210}\rangle &\equiv&
\left(
\begin{array}{c}
 \hat{\psi}_{k}^{1\dag} \hat{\psi}_{k+2}^{1\dag} \\
 \frac{1}{\sqrt{2}} ( \hat{\psi}_{k}^{1\dag} \hat{\psi}_{k+2}^{2\dag} + \hat{\psi}_{k}^{2\dag} \hat{\psi}_{k+2}^{1\dag} ) \\
 \hat{\psi}_{k}^{2\dag} \hat{\psi}_{k+2}^{2\dag}
\end{array}
\right) \hat{\psi}_{k+1}^{1\dag} \hat{\psi}_{k+1}^{2\dag} |0\rangle, \nonumber \\
|{\bf 3}_{1201}\rangle &\equiv&
\left(
\begin{array}{c}
 \hat{\psi}_{k}^{1\dag} \hat{\psi}_{k+3}^{1\dag} \\
 \frac{1}{\sqrt{2}} ( \hat{\psi}_{k}^{1\dag} \hat{\psi}_{k+3}^{2\dag} + \hat{\psi}_{k}^{2\dag} \hat{\psi}_{k+3}^{1\dag} ) \\
 \hat{\psi}_{k}^{2\dag} \hat{\psi}_{k+3}^{2\dag}
\end{array}
\right) \hat{\psi}_{k+1}^{1\dag} \hat{\psi}_{k+1}^{2\dag} |0\rangle, \nonumber \\
|{\bf 3}_{1120}\rangle &\equiv&
\left(
\begin{array}{c}
 \hat{\psi}_{k}^{1\dag} \hat{\psi}_{k+1}^{1\dag} \\
 \frac{1}{\sqrt{2}} ( \hat{\psi}_{k}^{1\dag} \hat{\psi}_{k+1}^{2\dag} + \hat{\psi}_{k}^{2\dag} \hat{\psi}_{k+1}^{1\dag} ) \\
 \hat{\psi}_{k}^{2\dag} \hat{\psi}_{k+1}^{2\dag}
\end{array}
\right) \hat{\psi}_{k+2}^{1\dag} \hat{\psi}_{k+2}^{2\dag} |0\rangle, \nonumber \\
|{\bf 3}_{1102}\rangle &\equiv& 
\left(
\begin{array}{c}
 \hat{\psi}_{k}^{1\dag} \hat{\psi}_{k+1}^{1\dag} \\
 \frac{1}{\sqrt{2}} ( \hat{\psi}_{k}^{1\dag} \hat{\psi}_{k+1}^{2\dag} + \hat{\psi}_{k}^{2\dag} \hat{\psi}_{k+1}^{1\dag} ) \\
 \hat{\psi}_{k}^{2\dag} \hat{\psi}_{k+1}^{2\dag}
\end{array}
\right) \hat{\psi}_{k+3}^{1\dag} \hat{\psi}_{k+3}^{2\dag} |0\rangle,
\end{eqnarray}
$\mathbb{H}^{{\bf 3}_{[1111]}} \equiv \{ |{\bf 3}_{\mathrm{AS}\,\underline{11}\,\underline{11}}\rangle, |{\bf 3}_{\mathrm{SA}\,\underline{11}\,\underline{11}}\rangle, |{\bf 3}_{\mathrm{SS}\,\underline{11}\,\underline{11}}\rangle \}$ with
\begin{eqnarray}
|{\bf 3}_{\mathrm{AS}\,\underline{11}\,\underline{11}}\rangle &\equiv& \frac{1}{\sqrt{2}} ( \hat{\psi}_{k}^{1\dag} \hat{\psi}_{k+1}^{2\dag} - \hat{\psi}_{k}^{2\dag} \hat{\psi}_{k+1}^{1\dag} )
\left(
\begin{array}{c}
 \hat{\psi}_{k+2}^{1\dag} \hat{\psi}_{k+3}^{1\dag} \\
 \frac{1}{\sqrt{2}} ( \hat{\psi}_{k+2}^{1\dag} \hat{\psi}_{k+3}^{2\dag} + \hat{\psi}_{k+2}^{2\dag} \hat{\psi}_{k+3}^{1\dag} ) \\
 \hat{\psi}_{k+2}^{2\dag} \hat{\psi}_{k+3}^{2\dag} 
\end{array}
\right) |0\rangle, \nonumber \\
|{\bf 3}_{\mathrm{SA}\,\underline{11}\,\underline{11}}\rangle &\equiv& 
\left(
\begin{array}{c}
 \hat{\psi}_{k}^{1\dag} \hat{\psi}_{k+1}^{1\dag} \\
 \frac{1}{\sqrt{2}} ( \hat{\psi}_{k}^{1\dag} \hat{\psi}_{k+1}^{2\dag} + \hat{\psi}_{k}^{2\dag} \hat{\psi}_{k+1}^{1\dag} ) \\
 \hat{\psi}_{k}^{2\dag} \hat{\psi}_{k+1}^{2\dag} 
\end{array}
\right) \frac{1}{\sqrt{2}} ( \hat{\psi}_{k+2}^{1\dag} \hat{\psi}_{k+3}^{2\dag} - \hat{\psi}_{k+2}^{2\dag} \hat{\psi}_{k+3}^{1\dag} ) |0\rangle, \nonumber \\
|{\bf 3}_{\mathrm{SS}\,\underline{11}\,\underline{11}}\rangle &\equiv&
\left(
\begin{array}{c}
 \frac{1}{2} \hat{\psi}_{k}^{1\dag} \hat{\psi}_{k+1}^{1\dag} ( \hat{\psi}_{k+2}^{1\dag} \hat{\psi}_{k+3}^{2\dag} + \hat{\psi}_{k+2}^{2\dag} \hat{\psi}_{k+3}^{1\dag} ) - \frac{1}{2} ( \hat{\psi}_{k}^{1\dag} \hat{\psi}_{k+1}^{2\dag} + \hat{\psi}_{k}^{2\dag} \hat{\psi}_{k+1}^{1\dag} ) \hat{\psi}_{k+2}^{1\dag} \hat{\psi}_{k+3}^{1\dag} \\
 \frac{1}{\sqrt{2}} \hat{\psi}_{k}^{1\dag} \hat{\psi}_{k+1}^{1\dag} \hat{\psi}_{k+2}^{2\dag} \hat{\psi}_{k+3}^{2\dag} - \frac{1}{\sqrt{2}} \hat{\psi}_{k}^{2\dag} \hat{\psi}_{k+1}^{2\dag} \hat{\psi}_{k+2}^{1\dag} \hat{\psi}_{k+3}^{1\dag} \\
 \frac{1}{2} ( \hat{\psi}_{k}^{1\dag} \hat{\psi}_{k+1}^{2\dag} + \hat{\psi}_{k}^{2\dag} \hat{\psi}_{k+1}^{1\dag} ) \hat{\psi}_{k+2}^{2\dag} \hat{\psi}_{k+3}^{2\dag} - \frac{1}{2} \hat{\psi}_{k}^{2\dag} \hat{\psi}_{k+1}^{2\dag} ( \hat{\psi}_{k+2}^{1\dag} \hat{\psi}_{k+3}^{2\dag} + \hat{\psi}_{k+2}^{2\dag} \hat{\psi}_{k+3}^{1\dag} )
 \end{array}
\right) |0\rangle, \label{eq:wf_3_1111}
\end{eqnarray}
with a notation A (S) for an antisymmetric (symmetric) combination in first or second pair of indices with the underline in the $k$-th and $(k+1)$-th vortices, and in the $(k+2)$-th and $(k+3)$-th vortices.
$\mathbb{H}^{{\bf 3}_{[2211]}} \equiv \{ |{\bf 3}_{2211}\rangle, |{\bf 3}_{2121}\rangle, |{\bf 3}_{2112}\rangle, |{\bf 3}_{1221}\rangle, |{\bf 3}_{1212}\rangle, |{\bf 3}_{1122}\rangle \}$ with
\begin{eqnarray}
|{\bf 3}_{2211}\rangle &\equiv& \hat{\psi}_{k}^{1\dag} \hat{\psi}_{k}^{2\dag} \hat{\psi}_{k+1}^{1\dag} \hat{\psi}_{k+1}^{2\dag}
\left(
\begin{array}{c}
 \hat{\psi}_{k+2}^{1\dag} \hat{\psi}_{k+3}^{1\dag} \\
 \frac{1}{\sqrt{2}} ( \hat{\psi}_{k+2}^{1\dag} \hat{\psi}_{k+3}^{2\dag} + \hat{\psi}_{k+2}^{2\dag} \hat{\psi}_{k+3}^{1\dag} ) \\
 \hat{\psi}_{k+2}^{2\dag} \hat{\psi}_{k+3}^{2\dag}
\end{array}
\right) |0\rangle, \nonumber \\
|{\bf 3}_{2121}\rangle &\equiv& \hat{\psi}_{k}^{1\dag} \hat{\psi}_{k}^{2\dag} \hat{\psi}_{k+2}^{1\dag} \hat{\psi}_{k+2}^{2\dag}
\left(
\begin{array}{c}
 \hat{\psi}_{k+1}^{1\dag} \hat{\psi}_{k+3}^{1\dag} \\
 \frac{1}{\sqrt{2}} ( \hat{\psi}_{k+1}^{1\dag} \hat{\psi}_{k+3}^{2\dag} + \hat{\psi}_{k+1}^{2\dag} \hat{\psi}_{k+3}^{1\dag} ) \\
 \hat{\psi}_{k+1}^{2\dag} \hat{\psi}_{k+3}^{2\dag}
\end{array}
\right) |0\rangle, \nonumber \\
|{\bf 3}_{2112}\rangle &\equiv& \hat{\psi}_{k}^{1\dag} \hat{\psi}_{k}^{2\dag} \hat{\psi}_{k+3}^{1\dag} \hat{\psi}_{k+3}^{2\dag}
\left(
\begin{array}{c}
 \hat{\psi}_{k+1}^{1\dag} \hat{\psi}_{k+2}^{1\dag} \\
 \frac{1}{\sqrt{2}} ( \hat{\psi}_{k+1}^{1\dag} \hat{\psi}_{k+2}^{2\dag} + \hat{\psi}_{k+1}^{2\dag} \hat{\psi}_{k+2}^{1\dag} ) \\
 \hat{\psi}_{k+1}^{2\dag} \hat{\psi}_{k+2}^{2\dag}
\end{array}
\right) |0\rangle, \nonumber \\
|{\bf 3}_{1221}\rangle &\equiv&
\left(
\begin{array}{c}
 \hat{\psi}_{k}^{1\dag} \hat{\psi}_{k+3}^{1\dag} \\
 \frac{1}{\sqrt{2}} ( \hat{\psi}_{k}^{1\dag} \hat{\psi}_{k+3}^{2\dag} + \hat{\psi}_{k}^{2\dag} \hat{\psi}_{k+3}^{1\dag} ) \\
 \hat{\psi}_{k}^{2\dag} \hat{\psi}_{k+3}^{2\dag}
\end{array}
\right) \hat{\psi}_{k+1}^{1\dag} \hat{\psi}_{k+1}^{2\dag} \hat{\psi}_{k+2}^{1\dag} \hat{\psi}_{k+2}^{2\dag} |0\rangle, \nonumber \\
|{\bf 3}_{1212}\rangle &\equiv&
\left(
\begin{array}{c}
 \hat{\psi}_{k}^{1\dag} \hat{\psi}_{k+2}^{1\dag} \\
 \frac{1}{\sqrt{2}} ( \hat{\psi}_{k}^{1\dag} \hat{\psi}_{k+2}^{2\dag} + \hat{\psi}_{k}^{2\dag} \hat{\psi}_{k+2}^{1\dag} ) \\
 \hat{\psi}_{k}^{2\dag} \hat{\psi}_{k+2}^{2\dag}
\end{array}
\right) \hat{\psi}_{k+1}^{1\dag} \hat{\psi}_{k+1}^{2\dag} \hat{\psi}_{k+3}^{1\dag} \hat{\psi}_{k+3}^{2\dag} |0\rangle, \nonumber \\
|{\bf 3}_{1122}\rangle &\equiv&
\left(
\begin{array}{c}
 \hat{\psi}_{k}^{1\dag} \hat{\psi}_{k+1}^{1\dag} \\
 \frac{1}{\sqrt{2}} ( \hat{\psi}_{k}^{1\dag} \hat{\psi}_{k+1}^{2\dag} + \hat{\psi}_{k}^{2\dag} \hat{\psi}_{k+1}^{1\dag} ) \\
 \hat{\psi}_{k}^{2\dag} \hat{\psi}_{k+1}^{2\dag}
\end{array}
\right) \hat{\psi}_{k+2}^{1\dag} \hat{\psi}_{k+2}^{2\dag} \hat{\psi}_{k+3}^{1\dag} \hat{\psi}_{k+3}^{2\dag} |0\rangle.
\end{eqnarray}
For quartet, there are two Hilbert spaces, $\mathbb{H}^{{\bf 4}_{[1110]}}$ and  $\mathbb{H}^{{\bf 4}_{[2111]}}$.
They are defined as $\mathbb{H}^{{\bf 4}_{[1110]}} \equiv \{ |{\bf 4}_{\mathrm{S}\,\underline{11}10}\rangle, |{\bf 4}_{\mathrm{S}\,\underline{11}01}\rangle, |{\bf 4}_{\mathrm{S}\,10\underline{11}}\rangle, |{\bf 4}_{\mathrm{S}\,01\underline{11}}\rangle \}$ with
\begin{eqnarray}
|{\bf 4}_{\mathrm{S}\,\underline{11}10}\rangle &\equiv&
\left(
\begin{array}{c}
 \hat{\psi}_{k}^{1\dag} \hat{\psi}_{k+1}^{1\dag} \hat{\psi}_{k+2}^{1\dag} \\
 \frac{1}{\sqrt{3}} \hat{\psi}_{k}^{1\dag} \hat{\psi}_{k+1}^{1\dag} \hat{\psi}_{k+2}^{2\dag}  + \frac{1}{\sqrt{3}} ( \hat{\psi}_{k}^{1\dag} \hat{\psi}_{k+1}^{2\dag} + \hat{\psi}_{k}^{2\dag} \hat{\psi}_{k+1}^{1\dag} ) \hat{\psi}_{k+2}^{1\dag}  \\
 \frac{1}{\sqrt{3}} ( \hat{\psi}_{k}^{1\dag} \hat{\psi}_{k+1}^{2\dag} + \hat{\psi}_{k}^{2\dag} \hat{\psi}_{k+1}^{1\dag} ) \hat{\psi}_{k+2}^{2\dag} + \frac{1}{\sqrt{3}}  \hat{\psi}_{k}^{2\dag} \hat{\psi}_{k+1}^{2\dag} \hat{\psi}_{k+2}^{1\dag} \\
 \hat{\psi}_{k}^{2\dag} \hat{\psi}_{k+1}^{2\dag} \hat{\psi}_{k+2}^{2\dag}
\end{array}
\right) |0\rangle, \nonumber \\
|{\bf 4}_{\mathrm{S}\,\underline{11}01}\rangle &\equiv&
\left(
\begin{array}{c}
 \hat{\psi}_{k}^{1\dag} \hat{\psi}_{k+1}^{1\dag} \hat{\psi}_{k+3}^{1\dag} \\
 \frac{1}{\sqrt{3}} \hat{\psi}_{k}^{1\dag} \hat{\psi}_{k+1}^{1\dag} \hat{\psi}_{k+3}^{2\dag}  + \frac{1}{\sqrt{3}} ( \hat{\psi}_{k}^{1\dag} \hat{\psi}_{k+1}^{2\dag} + \hat{\psi}_{k}^{2\dag} \hat{\psi}_{k+1}^{1\dag} ) \hat{\psi}_{k+3}^{1\dag}  \\
 \frac{1}{\sqrt{3}} ( \hat{\psi}_{k}^{1\dag} \hat{\psi}_{k+1}^{2\dag} + \hat{\psi}_{k}^{2\dag} \hat{\psi}_{k+1}^{1\dag} ) \hat{\psi}_{k+3}^{2\dag} + \frac{1}{\sqrt{3}}  \hat{\psi}_{k}^{2\dag} \hat{\psi}_{k+1}^{2\dag} \hat{\psi}_{k+3}^{1\dag} \\
 \hat{\psi}_{k}^{2\dag} \hat{\psi}_{k+1}^{2\dag} \hat{\psi}_{k+3}^{2\dag}
\end{array}
\right) |0\rangle, \nonumber \\
|{\bf 4}_{\mathrm{S}\,10\underline{11}}\rangle &\equiv&
\left(
\begin{array}{c}
 \hat{\psi}_{k}^{1\dag} \hat{\psi}_{k+2}^{1\dag} \hat{\psi}_{k+3}^{1\dag} \\
 \frac{1}{\sqrt{3}} \hat{\psi}_{k}^{2\dag} \hat{\psi}_{k+2}^{1\dag} \hat{\psi}_{k+3}^{1\dag} + \frac{1}{\sqrt{3}} \hat{\psi}_{k}^{1\dag} ( \hat{\psi}_{k+2}^{1\dag} \hat{\psi}_{k+3}^{2\dag} + \hat{\psi}_{k+2}^{2\dag} \hat{\psi}_{k+3}^{1\dag} ) \\
 \frac{1}{\sqrt{3}} \hat{\psi}_{k}^{2\dag} ( \hat{\psi}_{k+2}^{1\dag} \hat{\psi}_{k+3}^{2\dag} + \hat{\psi}_{k+2}^{2\dag} \hat{\psi}_{k+3}^{1\dag} ) + \frac{1}{\sqrt{3}} \hat{\psi}_{k}^{1\dag} \hat{\psi}_{k+2}^{2\dag} \hat{\psi}_{k+3}^{2\dag} \\
 \hat{\psi}_{k}^{2\dag} \hat{\psi}_{k+2}^{2\dag} \hat{\psi}_{k+3}^{2\dag}
\end{array}
\right) |0\rangle, \nonumber \\
|{\bf 4}_{\mathrm{S}\,01\underline{11}}\rangle &\equiv&
\left(
\begin{array}{c}
 \hat{\psi}_{k+1}^{1\dag} \hat{\psi}_{k+2}^{1\dag} \hat{\psi}_{k+3}^{1\dag} \\
 \frac{1}{\sqrt{3}} \hat{\psi}_{k+1}^{2\dag} \hat{\psi}_{k+2}^{1\dag} \hat{\psi}_{k+3}^{1\dag} + \frac{1}{\sqrt{3}} \hat{\psi}_{k+1}^{1\dag} ( \hat{\psi}_{k+2}^{1\dag} \hat{\psi}_{k+3}^{2\dag} + \hat{\psi}_{k+2}^{2\dag} \hat{\psi}_{k+3}^{1\dag} ) \\
 \frac{1}{\sqrt{3}} \hat{\psi}_{k+1}^{2\dag} ( \hat{\psi}_{k+2}^{1\dag} \hat{\psi}_{k+3}^{2\dag} + \hat{\psi}_{k+2}^{2\dag} \hat{\psi}_{k+3}^{1\dag} ) + \frac{1}{\sqrt{3}} \hat{\psi}_{k+1}^{1\dag} \hat{\psi}_{k+2}^{2\dag} \hat{\psi}_{k+3}^{2\dag} \\
 \hat{\psi}_{k+1}^{2\dag} \hat{\psi}_{k+2}^{2\dag} \hat{\psi}_{k+3}^{2\dag}
\end{array}
\right) |0\rangle,
\end{eqnarray}
and $\mathbb{H}^{{\bf 4}_{[2111]}} \equiv \{ |{\bf 4}_{\mathrm{S}\,21\underline{11}}\rangle, |{\bf 4}_{\mathrm{S}\,12\underline{11}}\rangle, |{\bf 4}_{\mathrm{S}\,\underline{11}21}\rangle, |{\bf 4}_{\mathrm{S}\,\underline{11}12}\rangle \}$ with
\begin{eqnarray}
|{\bf 4}_{\mathrm{S}\,21\underline{11}}\rangle &\equiv& \hat{\psi}_{k}^{1\dag} \hat{\psi}_{k}^{2\dag}
\left(
\begin{array}{c}
 \hat{\psi}_{k+1}^{1\dag} \hat{\psi}_{k+2}^{1\dag} \hat{\psi}_{k+3}^{1\dag} \\
 \frac{1}{\sqrt{3}} \hat{\psi}_{k+1}^{2\dag} \hat{\psi}_{k+2}^{1\dag} \hat{\psi}_{k+3}^{1\dag} + \frac{1}{\sqrt{3}} \hat{\psi}_{k+1}^{1\dag} ( \hat{\psi}_{k+2}^{1\dag} \hat{\psi}_{k+3}^{2\dag} + \hat{\psi}_{k+2}^{2\dag} \hat{\psi}_{k+3}^{1\dag} ) \\
 \frac{1}{\sqrt{3}} \hat{\psi}_{k+1}^{2\dag} ( \hat{\psi}_{k+2}^{1\dag} \hat{\psi}_{k+3}^{2\dag} + \hat{\psi}_{k+2}^{2\dag} \hat{\psi}_{k+3}^{1\dag} ) + \frac{1}{\sqrt{3}} \hat{\psi}_{k+1}^{1\dag} \hat{\psi}_{k+2}^{2\dag} \hat{\psi}_{k+3}^{2\dag} \\
 \hat{\psi}_{k+1}^{2\dag} \hat{\psi}_{k+2}^{2\dag} \hat{\psi}_{k+3}^{2\dag}
\end{array}
\right) |0\rangle, \nonumber \\
|{\bf 4}_{\mathrm{S}\,12\underline{11}}\rangle &\equiv&
\left(
\begin{array}{c}
 \hat{\psi}_{k}^{1\dag} \hat{\psi}_{k+2}^{1\dag} \hat{\psi}_{k+3}^{1\dag} \\
 \frac{1}{\sqrt{3}} \hat{\psi}_{k}^{2\dag} \hat{\psi}_{k+2}^{1\dag} \hat{\psi}_{k+3}^{1\dag} + \frac{1}{\sqrt{3}} \hat{\psi}_{k}^{1\dag} ( \hat{\psi}_{k+2}^{1\dag} \hat{\psi}_{k+3}^{2\dag} + \hat{\psi}_{k+2}^{2\dag} \hat{\psi}_{k+3}^{1\dag} ) \\
 \frac{1}{\sqrt{3}} \hat{\psi}_{k}^{2\dag} ( \hat{\psi}_{k+2}^{1\dag} \hat{\psi}_{k+3}^{2\dag} + \hat{\psi}_{k+2}^{2\dag} \hat{\psi}_{k+3}^{1\dag} ) + \frac{1}{\sqrt{3}} \hat{\psi}_{k}^{1\dag} \hat{\psi}_{k+2}^{2\dag} \hat{\psi}_{k+3}^{2\dag} \\
 \hat{\psi}_{k}^{2\dag} \hat{\psi}_{k+2}^{2\dag} \hat{\psi}_{k+3}^{2\dag}
\end{array}
\right) \hat{\psi}_{k+1}^{1\dag} \hat{\psi}_{k+1}^{2\dag} |0\rangle, \nonumber \\
|{\bf 4}_{\mathrm{S}\,\underline{11}21}\rangle &\equiv&
\left(
\begin{array}{c}
 \hat{\psi}_{k}^{1\dag} \hat{\psi}_{k+1}^{1\dag} \hat{\psi}_{k+3}^{1\dag} \\
 \frac{1}{\sqrt{3}} \hat{\psi}_{k}^{1\dag} \hat{\psi}_{k+1}^{1\dag} \hat{\psi}_{k+3}^{2\dag}  + \frac{1}{\sqrt{3}} ( \hat{\psi}_{k}^{1\dag} \hat{\psi}_{k+1}^{2\dag} + \hat{\psi}_{k}^{2\dag} \hat{\psi}_{k+1}^{1\dag} ) \hat{\psi}_{k+3}^{1\dag}  \\
 \frac{1}{\sqrt{3}} ( \hat{\psi}_{k}^{1\dag} \hat{\psi}_{k+1}^{2\dag} + \hat{\psi}_{k}^{2\dag} \hat{\psi}_{k+1}^{1\dag} ) \hat{\psi}_{k+3}^{2\dag} + \frac{1}{\sqrt{3}}  \hat{\psi}_{k}^{2\dag} \hat{\psi}_{k+1}^{2\dag} \hat{\psi}_{k+3}^{1\dag} \\
 \hat{\psi}_{k}^{2\dag} \hat{\psi}_{k+1}^{2\dag} \hat{\psi}_{k+3}^{2\dag}
\end{array}
\right) \hat{\psi}_{k+2}^{1\dag} \hat{\psi}_{k+2}^{2\dag} |0\rangle, \nonumber \\
|{\bf 4}_{\mathrm{S}\,\underline{11}12}\rangle &\equiv&
\left(
\begin{array}{c}
 \hat{\psi}_{k}^{1\dag} \hat{\psi}_{k+1}^{1\dag} \hat{\psi}_{k+2}^{1\dag} \\
 \frac{1}{\sqrt{3}} \hat{\psi}_{k}^{1\dag} \hat{\psi}_{k+1}^{1\dag} \hat{\psi}_{k+2}^{2\dag}  + \frac{1}{\sqrt{3}} ( \hat{\psi}_{k}^{1\dag} \hat{\psi}_{k+1}^{2\dag} + \hat{\psi}_{k}^{2\dag} \hat{\psi}_{k+1}^{1\dag} ) \hat{\psi}_{k+2}^{1\dag}  \\
 \frac{1}{\sqrt{3}} ( \hat{\psi}_{k}^{1\dag} \hat{\psi}_{k+1}^{2\dag} + \hat{\psi}_{k}^{2\dag} \hat{\psi}_{k+1}^{1\dag} ) \hat{\psi}_{k+2}^{2\dag} + \frac{1}{\sqrt{3}}  \hat{\psi}_{k}^{2\dag} \hat{\psi}_{k+1}^{2\dag} \hat{\psi}_{k+2}^{1\dag} \\
 \hat{\psi}_{k}^{2\dag} \hat{\psi}_{k+1}^{2\dag} \hat{\psi}_{k+2}^{2\dag}
\end{array}
\right) \hat{\psi}_{k+3}^{1\dag} \hat{\psi}_{k+3}^{2\dag} |0\rangle,
\end{eqnarray}
with a notation A (S) for an antisymmetric (symmetric) combination in a pair of indices with the underline in the $k$-th and $(k+1)$-th vortices, or in the $(k+2)$-th and $(k+3)$-th vortices.

For quintet, there is one Hilbert space $\mathbb{H}^{{\bf 5}_{[1111]}}$ which is defined as $\mathbb{H}^{{\bf 5}_{[1111]}} \equiv \{ |{\bf 5}_{\mathrm{SS}\,\underline{11}\,\underline{11}}\rangle \}$ with
\begin{equation}
|{\bf 5}_{\mathrm{SS}\,\underline{11}\,\underline{11}}\rangle \equiv
\left(
\begin{array}{c}
 \hat{\psi}_{k}^{1\dag} \hat{\psi}_{k+1}^{1\dag} \hat{\psi}_{k+2}^{1\dag} \hat{\psi}_{k+3}^{1\dag} \\
 \frac{1}{2} \hat{\psi}_{k}^{1\dag} \hat{\psi}_{k+1}^{1\dag} ( \hat{\psi}_{k+2}^{1\dag} \hat{\psi}_{k+3}^{2\dag} + \hat{\psi}_{k+2}^{2\dag} \hat{\psi}_{k+3}^{1\dag} ) + \frac{1}{2} ( \hat{\psi}_{k}^{1\dag} \hat{\psi}_{k+1}^{2\dag} + \hat{\psi}_{k}^{2\dag} \hat{\psi}_{k+1}^{1\dag} )
\hat{\psi}_{k+2}^{1\dag} \hat{\psi}_{k+3}^{1\dag} \\
 \frac{1}{\sqrt{6}} \hat{\psi}_{k}^{1\dag} \hat{\psi}_{k+1}^{1\dag} \hat{\psi}_{k+2}^{2\dag} \hat{\psi}_{k+3}^{2\dag} + \frac{1}{\sqrt{6}} ( \hat{\psi}_{k}^{1\dag} \hat{\psi}_{k+1}^{2\dag} + \hat{\psi}_{k}^{2\dag} \hat{\psi}_{k+1}^{1\dag} ) ( \hat{\psi}_{k+2}^{1\dag} \hat{\psi}_{k+3}^{2\dag} + \hat{\psi}_{k+2}^{2\dag} \hat{\psi}_{k+3}^{1\dag} ) + \frac{1}{\sqrt{6}} \hat{\psi}_{k}^{2\dag} \hat{\psi}_{k+1}^{2\dag} \hat{\psi}_{k+2}^{1\dag} \hat{\psi}_{k+3}^{1\dag} \\
 \frac{1}{2} ( \hat{\psi}_{k}^{1\dag} \hat{\psi}_{k+1}^{2\dag} + \hat{\psi}_{k}^{2\dag} \hat{\psi}_{k+1}^{1\dag} ) \hat{\psi}_{k+2}^{2\dag} \hat{\psi}_{k+3}^{2\dag} + \frac{1}{2} \hat{\psi}_{k}^{2\dag} \hat{\psi}_{k+1}^{2\dag} ( \hat{\psi}_{k+2}^{1\dag} \hat{\psi}_{k+3}^{2\dag} + \hat{\psi}_{k+2}^{2\dag} \hat{\psi}_{k+3}^{1\dag} ) \\
 \hat{\psi}_{k}^{2\dag} \hat{\psi}_{k+1}^{2\dag} \hat{\psi}_{k+2}^{2\dag} \hat{\psi}_{k+3}^{2\dag}
\end{array}
\right) |0\rangle,
\end{equation}
with a notation S for a symmetric combination in first and second pair of indices with the underline in the $k$-th and $(k+1)$-th vortices, and in the $(k+2)$-th and $(k+3)$-th vortices.
Therefore, the Hilbert space is totally given as a direct sum,
\begin{eqnarray}
\mathbb{H}^{\{n=4\}} 
&=& \mathbb{H}^{{\bf 1}_{[0000]}} \oplus \mathbb{H}^{{\bf 1}_{[2000]}} \oplus \mathbb{H}^{{\bf 1}_{[1100]}} \oplus \mathbb{H}^{{\bf 1}_{[2200]}} \oplus \mathbb{H}^{{\bf 1}_{[2110]}} \oplus \mathbb{H}^{{\bf 1}_{[2220]}} \oplus \mathbb{H}^{{\bf 1}_{[1111]}} \oplus \mathbb{H}^{{\bf 1}_{[2211]}} \oplus \mathbb{H}^{{\bf 1}_{[2222]}} \nonumber \\
& \oplus & \mathbb{H}^{{\bf 2}_{[1000]}} \oplus \mathbb{H}^{{\bf 2}_{[2100]}} \oplus \mathbb{H}^{{\bf 2}_{[1110]}} \oplus \mathbb{H}^{{\bf 2}_{[2210]}}  \oplus \mathbb{H}^{{\bf 2}_{[2111]}} \oplus \mathbb{H}^{{\bf 2}_{[2221]}} \nonumber \\
& \oplus & \mathbb{H}^{{\bf 3}_{[1100]}} \oplus \mathbb{H}^{{\bf 3}_{[2110]}} \oplus \mathbb{H}^{{\bf 3}_{[1111]}} \oplus \mathbb{H}^{{\bf 3}_{[2211]}} \nonumber \\
& \oplus & \mathbb{H}^{{\bf 4}_{[1110]}} \oplus \mathbb{H}^{{\bf 4}_{[2111]}} \nonumber \\
& \oplus & \mathbb{H}^{{\bf 5}_{[1111]}}.
\end{eqnarray}

With these basis states in the Hilbert spaces, the operators $\hat{\tau}_{k}$, $\hat{\tau}_{k+1}$ and $\hat{\tau}_{k+2}$ are expressed as matrices.
For singlet, the matrices are
\begin{eqnarray}
\tau_{k}^{{\bf 1}_{[0000]}} = \tau_{k+1}^{{\bf 1}_{[0000]}} = \tau_{k+2}^{{\bf 1}_{[0000]}} = 1,
\end{eqnarray}
for $\mathbb{H}^{{\bf 1}_{[0000]}}$,
\begin{eqnarray}
\tau_{k}^{{\bf 1}_{[2000]}} =
\left(
\begin{array}{cccc}
 1 & 0 & 0 & 0 \\
 0 & 1 & 0 & 0 \\
 0 & 0 & 0 & 1 \\
 0 & 0 & 1 & 0 
\end{array}
\right), \,
\tau_{k+1}^{{\bf 1}_{[2000]}} =
\left(
\begin{array}{cccc}
 0 & 0 & 0 & 1 \\
 0 & 1 & 0 & 0 \\
 0 & 0 & 1 & 0 \\
 1 & 0 & 0 & 0 
\end{array}
\right), \,
\tau_{k+2}^{{\bf 1}_{[2000]}} =
\left(
\begin{array}{cccc}
 0 & 1 & 0 & 0 \\
 1 & 0 & 0 & 0 \\
 0 & 0 & 1 & 0 \\
 0 & 0 & 0 & 1 
\end{array}
\right),
\end{eqnarray}
for $\mathbb{H}^{{\bf 1}_{[2000]}}$,
\begin{eqnarray}
\tau_{k}^{{\bf 1}_{[1100]}} =
\left(
\begin{array}{cccccc}
 1 & 0 & 0 & 0 & 0 & 0 \\
 0 & -1 & 0 & 0 & 0 & 0 \\
 0 & 0 & 0 & 0 & -1 & 0 \\
 0 & 0 & 0 & 0 & 0 & -1 \\
 0 & 0 & 1 & 0 & 0 & 0 \\
 0 & 0 & 0 & 1 & 0 & 0
\end{array}
\right), \,
\tau_{k+1}^{{\bf 1}_{[1100]}} =
\left(
\begin{array}{cccccc}
 0 & 0 & 0 & 0 & 0 & 1 \\
 0 & 0 & -1 & 0 & 0 & 0 \\
 0 & 1 & 0 & 0 & 0 & 0 \\
 0 & 0 & 0 & 1 & 0 & 0 \\
 0 & 0 & 0 & 0 & -1 & 0 \\
-1 & 0 & 0 & 0 & 0 & 0
\end{array}
\right), \,
\tau_{k+2}^{{\bf 1}_{[1100]}} =
\left(
\begin{array}{cccccc}
 -1 & 0 & 0 & 0 & 0 & 0 \\
 0 & 1 & 0 & 0 & 0 & 0 \\
 0 & 0 & 0 & -1 & 0 & 0 \\
 0 & 0 & 1 & 0 & 0 & 0 \\
 0 & 0 & 0 & 0 & 0 & -1 \\
 0 & 0 & 0 & 0 & 1 & 0
\end{array}
\right),
\end{eqnarray}
for $\mathbb{H}^{{\bf 1}_{[1100]}}$,
\begin{eqnarray}
&& \tau_{k}^{{\bf 1}_{[2200]}} =
\left(
\begin{array}{cccccc}
 1 & 0 & 0 & 0 & 0 & 0 \\
 0 & 0 & 0 & 1 & 0 & 0 \\
 0 & 0 & 0 & 0 & 1 & 0 \\
 0 & 1 & 0 & 0 & 0 & 0 \\
 0 & 0 & 1 & 0 & 0 & 0 \\
 0 & 0 & 0 & 0 & 0 & 1
\end{array}
\right), \,
 \tau_{k+1}^{{\bf 1}_{[2200]}} =
\left(
\begin{array}{cccccc}
 0 & 0 & 0 & 0 & 1 & 0 \\
 0 & 0 & 0 & 0 & 0 & 1 \\
 0 & 0 & 1 & 0 & 0 & 0 \\
 0 & 0 & 0 & 1 & 0 & 0 \\
 1 & 0 & 0 & 0 & 0 & 0 \\
 0 & 1 & 0 & 0 & 0 & 0
\end{array}
\right), \,
 \tau_{k+2}^{{\bf 1}_{[2200]}} =
\left(
\begin{array}{cccccc}
 1 & 0 & 0 & 0 & 0 & 0 \\
 0 & 0 & 1 & 0 & 0 & 0 \\
 0 & 1 & 0 & 0 & 0 & 0 \\
 0 & 0 & 0 & 0 & 1 & 0 \\
 0 & 0 & 0 & 1 & 0 & 0 \\
 0 & 0 & 0 & 0 & 0 & 1
\end{array}
\right),
\end{eqnarray}
for $\mathbb{H}^{{\bf 1}_{[2200]}}$,
\begin{eqnarray}
&& \tau_{k}^{{\bf 1}_{[2110]}} =
\left(
\begin{array}{cccccccccccc}
 -1 & 0 & 0 & 0 & 0 & 0 & 0 & 0 & 0 & 0 & 0 & 0	\\
 0 & -1 & 0 & 0 & 0 & 0 & 0 & 0 & 0 & 0 & 0 & 0 \\
 0 & 0 & 0 & 1 & 0 & 0 & 0 & 0 & 0 & 0 & 0 & 0 \\
 0 & 0 & 1 & 0 & 0 & 0 & 0 & 0 & 0 & 0 & 0 & 0 \\
 0 & 0 & 0 & 0 & 0 & 0 & -1 & 0 & 0 & 0 & 0 & 0 \\
 0 & 0 & 0 & 0 & 0 & 0 & 0 & -1 & 0 & 0 & 0 & 0 \\
 0 & 0 & 0 & 0 & 1 & 0 & 0 & 0 & 0 & 0 & 0 & 0 \\
 0 & 0 & 0 & 0 & 0 & 1 & 0 & 0 & 0 & 0 & 0 & 0 \\
 0 & 0 & 0 & 0 & 0 & 0 & 0 & 0 & 0 & 0 & 1 & 0 \\
 0 & 0 & 0 & 0 & 0 & 0 & 0 & 0 & 0 & 0 & 0 & 1 \\
 0 & 0 & 0 & 0 & 0 & 0 & 0 & 0 & -1 & 0 & 0 & 0 \\
 0 & 0 & 0 & 0 & 0 & 0 & 0 & 0 & 0 & -1 & 0 & 0
\end{array}
\right),\,
\tau_{k+1}^{{\bf 1}_{[2110]}} =
\left(
\begin{array}{cccccccccccc}
 0 & 0 & 0 & 0 & 0 & 0 & 0 & 0 & 0 & 0 & -1 & 0 \\
 0 & 0 & 0 & 0 & 0 & -1 & 0 & 0 & 0 & 0 & 0 & 0 \\
 0 & 0 & 0 & 0 & 0 & 0 & 0 & 0 & 0 & 1 & 0 & 0 \\
 0 & 0 & 0 & 0 & 0 & 0 & 1 & 0 & 0 & 0 & 0 & 0 \\
 0 & 0 & 0 & 0 & 0 & 0 & 0 & 0 & 0 & 0 & 0 & 1 \\
 0 & 1 & 0 & 0 & 0 & 0 & 0 & 0 & 0 & 0 & 0 & 0 \\
 0 & 0 & 0 & -1 & 0 & 0 & 0 & 0 & 0 & 0 & 0 & 0 \\
 0 & 0 & 0 & 0 & 0 & 0 & 0 & -1 & 0 & 0 & 0 & 0 \\
 0 & 0 & 0 & 0 & 0 & 0 & 0 & 0 & -1 & 0 & 0 & 0 \\
 0 & 0 & -1 & 0 & 0 & 0 & 0 & 0 & 0 & 0 & 0 & 0 \\
 1 & 0 & 0 & 0 & 0 & 0 & 0 & 0 & 0 & 0 & 0 & 0 \\
 0 & 0 & 0 & 0 & 1 & 0 & 0 & 0 & 0 & 0 & 0 & 0
\end{array}
\right), \nonumber \\
&& \tau_{k+2}^{{\bf 1}_{[2110]}} =
\left(
\begin{array}{cccccccccccc}
 0 & 1 & 0 & 0 & 0 & 0 & 0 & 0 & 0 & 0 & 0 & 0 \\
 1 & 0 & 0 & 0 & 0 & 0 & 0 & 0 & 0 & 0 & 0 & 0 \\
 0 & 0 & -1 & 0 & 0 & 0 & 0 & 0 & 0 & 0 & 0 & 0 \\
 0 & 0 & 0 & -1 & 0 & 0 & 0 & 0 & 0 & 0 & 0 & 0 \\
 0 & 0 & 0 & 0 & 0 & 1 & 0 & 0 & 0 & 0 & 0 & 0 \\
 0 & 0 & 0 & 0 & -1 & 0 & 0 & 0 & 0 & 0 & 0 & 0 \\
 0 & 0 & 0 & 0 & 0 & 0 & 0 & 1 & 0 & 0 & 0 & 0 \\
 0 & 0 & 0 & 0 & 0 & 0 & -1 & 0 & 0 & 0 & 0 & 0 \\
 0 & 0 & 0 & 0 & 0 & 0 & 0 & 0 & 0 & -1 & 0 & 0 \\
 0 & 0 & 0 & 0 & 0 & 0 & 0 & 0 & 1 & 0 & 0 & 0 \\
 0 & 0 & 0 & 0 & 0 & 0 & 0 & 0 & 0 & 0 & 0 & -1 \\
 0 & 0 & 0 & 0 & 0 & 0 & 0 & 0 & 0 & 0 & 1 & 0
\end{array}
\right),
\end{eqnarray}
for $\mathbb{H}^{{\bf 1}_{[2110]}}$,
\begin{eqnarray}
\tau_{k}^{{\bf 1}_{[2220]}} =
\left(
\begin{array}{cccc}
 0 & 1 & 0 & 0 \\
 1 & 0 & 0 & 0 \\
 0 & 0 & 1 & 0 \\
 0 & 0 & 0 & 1 
\end{array}
\right), \,
\tau_{k+1}^{{\bf 1}_{[2220]}} =
\left(
\begin{array}{cccc}
 0 & 0 & 0 & 1 \\
 0 & 1 & 0 & 0 \\
 0 & 0 & 1 & 0 \\
 1 & 0 & 0 & 0 
\end{array}
\right), \,
\tau_{k+2}^{{\bf 1}_{[2220]}} =
\left(
\begin{array}{cccc}
 1 & 0 & 0 & 0 \\
 0 & 1 & 0 & 0 \\
 0 & 0 & 0 & 1 \\
 0 & 0 & 1 & 0 
\end{array}
\right),
\end{eqnarray}
for $\mathbb{H}^{{\bf 1}_{[2220]}}$,
\begin{eqnarray}
\tau_{k}^{{\bf 1}_{[1111]}} =
\left(
\begin{array}{cc}
 -1 & 0 \\
 0 & 1
\end{array}
\right), \,
\tau_{k+1}^{{\bf 1}_{[1111]}} =
\left(
\begin{array}{cc}
 \frac{1}{2} & \frac{\sqrt{3}}{2} \\
 \frac{\sqrt{3}}{2} & -\frac{1}{2}
\end{array}
\right), \,
\tau_{k+2}^{{\bf 1}_{[1111]}} =
\left(
\begin{array}{cc}
 -1 & 0 \\
 0 & 1
\end{array}
\right), \label{eq:matrix_1_1111}
\end{eqnarray}
for $\mathbb{H}^{{\bf 1}_{[1111]}}$,
\begin{eqnarray}
\tau_{k}^{{\bf 1}_{[2211]}} =
\left(
\begin{array}{cccccc}
 -1 & 0 & 0 & 0 & 0 & 0 \\
 0 & 1 & 0 & 0 & 0 & 0 \\
 0 & 0 & 0 & 0 & 1 & 0 \\
 0 & 0 & 0 & 0 & 0 & 1 \\
 0 & 0 & -1 & 0 & 0 & 0 \\
 0 & 0 & 0 & -1 & 0 & 0
\end{array}
\right), \,
\tau_{k+1}^{{\bf 1}_{[2211]}} =
\left(
\begin{array}{cccccc}
 0 & 0 & 0 & 0 & 0 & -1 \\
 0 & 0 & 1 & 0 & 0 & 0 \\
 0 & 1 & 0 & 0 & 0 & 0 \\
 0 & 0 & 0 & -1 & 0 & 0 \\
 0 & 0 & 0 & 0 & 1 & 0 \\
 1 & 0 & 0 & 0 & 0 & 0
\end{array}
\right), \,
\tau_{k+2}^{{\bf 1}_{[2211]}} =
\left(
\begin{array}{cccccc}
 1 & 0 & 0 & 0 & 0 & 0 \\
 0 & -1 & 0 & 0 & 0 & 0 \\
 0 & 0 & 0 & 1 & 0 & 0 \\
 0 & 0 & -1 & 0 & 0 & 0 \\
 0 & 0 & 0 & 0 & 0 & 1 \\
 0 & 0 & 0 & 0 & -1 & 0
\end{array}
\right),
\end{eqnarray}
for $\mathbb{H}^{{\bf 1}_{[2211]}}$,
\begin{eqnarray}
\tau_{k}^{{\bf 1}_{[2222]}} = \tau_{k+1}^{{\bf 1}_{[2222]}} = \tau_{k+2}^{{\bf 1}_{[2222]}} = 1,
\end{eqnarray}
for $\mathbb{H}^{{\bf 1}_{[2222]}}$.
For doublet, the matrices are
\begin{eqnarray}
\tau_{k}^{{\bf 2}_{[1000]}} = 
\left(
\begin{array}{cccc}
 0 & -1 & 0 & 0 \\
 1 & 0 & 0 & 0 \\
 0 & 0 & 1 & 0 \\
 0 & 0 & 0 & 1
\end{array}
\right), \,
\tau_{k+1}^{{\bf 2}_{[1000]}} = 
\left(
\begin{array}{cccc}
 1 & 0 & 0 & 0 \\
 0 & 0 & -1 & 0 \\
 0 & 1 & 0 & 0 \\
 0 & 0 & 0 & 1
\end{array}
\right), \,
\tau_{k+2}^{{\bf 2}_{[1000]}} = 
\left(
\begin{array}{cccc}
 1 & 0 & 0 & 0 \\
 0 & 1 & 0 & 0 \\
 0 & 0 & 0 & -1 \\
 0 & 0 & 1 & 0
\end{array}
\right),
\end{eqnarray}
for $\mathbb{H}^{{\bf 2}_{[1000]}}$,
\begin{eqnarray}
&& \tau_{k}^{{\bf 2}_{[2100]}} =
\left(
\begin{array}{cccccccccccc}
 1 & 0 & 0 & 0 & 0 & 0 & 0 & 0 & 0 & 0 & 0 & 0 \\
 0 & 1 & 0 & 0 & 0 & 0 & 0 & 0 & 0 & 0 & 0 & 0 \\
 0 & 0 & 0 & 0 & 1 & 0 & 0 & 0 & 0 & 0 & 0 & 0 \\
 0 & 0 & 0 & 0 & 0 & 1 & 0 & 0 & 0 & 0 & 0 & 0 \\
 0 & 0 & 1 & 0 & 0 & 0 & 0 & 0 & 0 & 0 & 0 & 0 \\
 0 & 0 & 0 & 1 & 0 & 0 & 0 & 0 & 0 & 0 & 0 & 0 \\
 0 & 0 & 0 & 0 & 0 & 0 & 0 & 0 & -1 & 0 & 0 & 0 \\
 0 & 0 & 0 & 0 & 0 & 0 & 0 & 0 & 0 & -1 & 0 & 0 \\
 0 & 0 & 0 & 0 & 0 & 0 & 1 & 0 & 0 & 0 & 0 & 0 \\
 0 & 0 & 0 & 0 & 0 & 0 & 0 & 1 & 0 & 0 & 0 & 0 \\
 0 & 0 & 0 & 0 & 0 & 0 & 0 & 0 & 0 & 0 & 0 & 1 \\
 0 & 0 & 0 & 0 & 0 & 0 & 0 & 0 & 0 & 0 & -1 & 0
\end{array}
\right), \,
 \tau_{k+1}^{{\bf 2}_{[2100]}} =
\left(
\begin{array}{cccccccccccc}
 0 & 0 & 0 & 0 & 0 & 1 & 0 & 0 & 0 & 0 & 0 & 0 \\
 0 & 0 & 0 & 0 & 0 & 0 & 0 & 0 & 0 & 1 & 0 & 0 \\
 0 & 0 & 0 & 0 & 0 & 0 & 0 & 0 & 0 & 0 & 1 & 0 \\
 0 & 0 & 0 & 1 & 0 & 0 & 0 & 0 & 0 & 0 & 0 & 0 \\
 0 & 0 & 0 & 0 & 0 & 0 & 0 & 0 & 1 & 0 & 0 & 0 \\
 1 & 0 & 0 & 0 & 0 & 0 & 0 & 0 & 0 & 0 & 0 & 0 \\
 0 & 0 & 0 & 0 & 0 & 0 & 0 & 0 & 0 & 0 & 0 & 1 \\
 0 & 0 & 0 & 0 & 0 & 0 & 0 & 1 & 0 & 0 & 0 & 0 \\
 0 & 0 & 0 & 0 & -1 & 0 & 0 & 0 & 0 & 0 & 0 & 0 \\
 0 & -1 & 0 & 0 & 0 & 0 & 0 & 0 & 0 & 0 & 0 & 0 \\
 0 & 0 & -1 & 0 & 0 & 0 & 0 & 0 & 0 & 0 & 0 & 0 \\
 0 & 0 & 0 & 0 & 0 & 0 & 1 & 0 & 0 & 0 & 0 & 0
\end{array}
\right), \nonumber \\
&& \tau_{k+2}^{{\bf 2}_{[2100]}} =
\left(
\begin{array}{cccccccccccc}
 0 & 1 & 0 & 0 & 0 & 0 & 0 & 0 & 0 & 0 & 0 & 0 \\
 -1 & 0 & 0 & 0 & 0 & 0 & 0 & 0 & 0 & 0 & 0 & 0 \\
 0 & 0 & 0 & -1 & 0 & 0 & 0 & 0 & 0 & 0 & 0 & 0 \\
 0 & 0 & 1 & 0 & 0 & 0 & 0 & 0 & 0 & 0 & 0 & 0 \\
 0 & 0 & 0 & 0 & 0 & -1 & 0 & 0 & 0 & 0 & 0 & 0 \\
 0 & 0 & 0 & 0 & 1 & 0 & 0 & 0 & 0 & 0 & 0 & 0 \\
 0 & 0 & 0 & 0 & 0 & 0 & 0 & 1 & 0 & 0 & 0 & 0 \\
 0 & 0 & 0 & 0 & 0 & 0 & 1 & 0 & 0 & 0 & 0 & 0 \\
 0 & 0 & 0 & 0 & 0 & 0 & 0 & 0 & 0 & 1 & 0 & 0 \\
 0 & 0 & 0 & 0 & 0 & 0 & 0 & 0 & 1 & 0 & 0 & 0 \\
 0 & 0 & 0 & 0 & 0 & 0 & 0 & 0 & 0 & 0 & 1 & 0 \\
 0 & 0 & 0 & 0 & 0 & 0 & 0 & 0 & 0 & 0 & 0 & 1
\end{array}
\right),
\end{eqnarray}
for $\mathbb{H}^{{\bf 2}_{[2100]}}$,
\begin{eqnarray}
&& \tau_{k}^{{\bf 2}_{[1110]}} =
\left(
\begin{array}{cccccccc}
 -1 & 0 & 0 & 0 & 0 & 0 & 0 & 0 \\
 0 & -1 & 0 & 0 & 0 & 0 & 0 & 0 \\
 0 & 0 & 0 & 0 & -1 & 0 & 0 & 0 \\
 0 & 0 & 0 & 0 & 0 & -1 & 0 & 0 \\
 0 & 0 & 1 & 0 & 0 & 0 & 0 & 0 \\
 0 & 0 & 0 & 1 & 0 & 0 & 0 & 0 \\
 0 & 0 & 0 & 0 & 0 & 0 & 1 & 0 \\
 0 & 0 & 0 & 0 & 0 & 0 & 0 & 1
\end{array}
\right),\,
\tau_{k+1}^{{\bf 2}_{[1110]}} =
\left(
\begin{array}{cccccccc}
 \frac{1}{2} & 0 & 0 & 0 & 0 & 0 & \frac{\sqrt{3}}{2} & 0 \\
 0 & 0 & \frac{1}{2} & \frac{\sqrt{3}}{2} & 0 & 0 & 0 & 0 \\
 0 & -\frac{1}{2} & 0 & 0 & 0 & 0 & 0 & \frac{\sqrt{3}}{2} \\
 0 & -\frac{\sqrt{3}}{2} & 0 & 0 & 0 & 0 & 0 & -\frac{1}{2} \\
 0 & 0 & 0 & 0 & \frac{1}{2} & \frac{\sqrt{3}}{2} & 0 & 0 \\
 0 & 0 & 0 & 0 & \frac{\sqrt{3}}{2} & -\frac{1}{2} & 0 & 0 \\
 \frac{\sqrt{3}}{2} & 0 & 0 & 0 & 0 & 0 & -\frac{1}{2} & 0 \\
 0 & 0 & -\frac{\sqrt{3}}{2} & \frac{1}{2} & 0 & 0 & 0 & 0
\end{array}
\right),\nonumber \\
&& \tau_{k+2}^{{\bf 2}_{[1110]}} =
\left(
\begin{array}{cccccccc}
 0 & -1 & 0 & 0 & 0 & 0 & 0 & 0 \\
 1 & 0 & 0 & 0 & 0 & 0 & 0 & 0 \\
 0 & 0 & -1 & 0 & 0 & 0 & 0 & 0 \\
 0 & 0 & 0 & 1 & 0 & 0 & 0 & 0 \\
 0 & 0 & 0 & 0 & -1 & 0 & 0 & 0 \\
 0 & 0 & 0 & 0 & 0 & 1 & 0 & 0 \\
 0 & 0 & 0 & 0 & 0 & 0 & 0 & -1 \\
 0 & 0 & 0 & 0 & 0 & 0 & 1 & 0
\end{array}
\right),  \label{eq:tau(1110)}
\end{eqnarray}
for $\mathbb{H}^{{\bf 2}_{[1110]}}$,
\begin{eqnarray}
&& \tau_{k}^{{\bf 2}_{[2210]}} =
\left(
\begin{array}{cccccccccccc}
 0 & 0 & 1 & 0 & 0 & 0 & 0 & 0 & 0 & 0 & 0 & 0 \\
 0 & 0 & 0 & 1 & 0 & 0 & 0 & 0 & 0 & 0 & 0 & 0 \\
 1 & 0 & 0 & 0 & 0 & 0 & 0 & 0 & 0 & 0 & 0 & 0 \\
 0 & 1 & 0 & 0 & 0 & 0 & 0 & 0 & 0 & 0 & 0 & 0 \\
 0 & 0 & 0 & 0 & 1 & 0 & 0 & 0 & 0 & 0 & 0 & 0 \\
 0 & 0 & 0 & 0 & 0 & 1 & 0 & 0 & 0 & 0 & 0 & 0 \\
 0 & 0 & 0 & 0 & 0 & 0 & 0 & -1 & 0 & 0 & 0 & 0 \\
 0 & 0 & 0 & 0 & 0 & 0 & 1 & 0 & 0 & 0 & 0 & 0 \\
 0 & 0 & 0 & 0 & 0 & 0 & 0 & 0 & 0 & 0 & 1 & 0 \\
 0 & 0 & 0 & 0 & 0 & 0 & 0 & 0 & 0 & 0 & 0 & 1 \\
 0 & 0 & 0 & 0 & 0 & 0 & 0 & 0 & -1 & 0 & 0 & 0 \\
 0 & 0 & 0 & 0 & 0 & 0 & 0 & 0 & 0 & -1 & 0 & 0
\end{array}
\right), \,
\tau_{k+1}^{{\bf 2}_{[2210]}} =
\left(
\begin{array}{cccccccccccc}
 0 & 0 & 0 & 0 & 0 & 1 & 0 & 0 & 0 & 0 & 0 & 0 \\
 0 & 0 & 0 & 0 & 0 & 0 & 0 & 0 & 0 & 1 & 0 & 0 \\
 0 & 0 & 1 & 0 & 0 & 0 & 0 & 0 & 0 & 0 & 0 & 0 \\
 0 & 0 & 0 & 0 & 0 & 0 & 0 & 1 & 0 & 0 & 0 & 0 \\
 0 & 0 & 0 & 0 & 0 & 0 & 0 & 0 & 1 & 0 & 0 & 0 \\
 1 & 0 & 0 & 0 & 0 & 0 & 0 & 0 & 0 & 0 & 0 & 0 \\
 0 & 0 & 0 & 0 & 0 & 0 & 0 & 0 & 0 & 0 & 0 & 1 \\
 0 & 0 & 0 & -1 & 0 & 0 & 0 & 0 & 0 & 0 & 0 & 0 \\
 0 & 0 & 0 & 0 & -1 & 0 & 0 & 0 & 0 & 0 & 0 & 0 \\
 0 & -1 & 0 & 0 & 0 & 0 & 0 & 0 & 0 & 0 & 0 & 0 \\
 0 & 0 & 0 & 0 & 0 & 0 & 0 & 0 & 0 & 0 & 1 & 0 \\
 0 & 0 & 0 & 0 & 0 & 0 & 1 & 0 & 0 & 0 & 0 & 0
\end{array}
\right), \nonumber \\
&& \tau_{k+2}^{{\bf 2}_{[2210]}} =
\left(
\begin{array}{cccccccccccc}
 0 & 1 & 0 & 0 & 0 & 0 & 0 & 0 & 0 & 0 & 0 & 0 \\
 -1 & 0 & 0 & 0 & 0 & 0 & 0 & 0 & 0 & 0 & 0 & 0 \\
 0 & 0 & 0 & 1 & 0 & 0 & 0 & 0 & 0 & 0 & 0 & 0 \\
 0 & 0 & -1 & 0 & 0 & 0 & 0 & 0 & 0 & 0 & 0 & 0 \\
 0 & 0 & 0 & 0 & 0 & -1 & 0 & 0 & 0 & 0 & 0 & 0 \\
 0 & 0 & 0 & 0 & 1 & 0 & 0 & 0 & 0 & 0 & 0 & 0 \\
 0 & 0 & 0 & 0 & 0 & 0 & 1 & 0 & 0 & 0 & 0 & 0 \\
 0 & 0 & 0 & 0 & 0 & 0 & 0 & 1 & 0 & 0 & 0 & 0 \\
 0 & 0 & 0 & 0 & 0 & 0 & 0 & 0 & 0 & 1 & 0 & 0 \\
 0 & 0 & 0 & 0 & 0 & 0 & 0 & 0 & 1 & 0 & 0 & 0 \\
 0 & 0 & 0 & 0 & 0 & 0 & 0 & 0 & 0 & 0 & 0 & 1 \\
 0 & 0 & 0 & 0 & 0 & 0 & 0 & 0 & 0 & 0 & 1 & 0
\end{array}
\right),
\end{eqnarray}
for $\mathbb{H}^{{\bf 2}_{[2210]}}$,
\begin{eqnarray}
&& \tau_{k}^{{\bf 2}_{[2111]}} =
\left(
\begin{array}{cccccccc}
 -1 & 0 & 0 & 0 & 0 & 0 & 0 & 0 \\
 0 & -1 & 0 & 0 & 0 & 0 & 0 & 0 \\
 0 & 0 & 0 & 0 & 1 & 0 & 0 & 0 \\
 0 & 0 & 0 & 0 & 0 & 1 & 0 & 0 \\
 0 & 0 & -1 & 0 & 0 & 0 & 0 & 0 \\
 0 & 0 & 0 & -1 & 0 & 0 & 0 & 0 \\
 0 & 0 & 0 & 0 & 0 & 0 & 1 & 0 \\
 0 & 0 & 0 & 0 & 0 & 0 & 0 & 1
\end{array}
\right),\,
 \tau_{k+1}^{{\bf 2}_{[2111]}} =
\left(
\begin{array}{cccccccc}
 0 & 0 & 0 & 0 & \frac{1}{2} & \frac{\sqrt{3}}{2} & 0 & 0 \\
 0 & \frac{1}{2} & 0 & 0 & 0 & 0 & 0 & \frac{\sqrt{3}}{2} \\
 0 & 0 & \frac{1}{2} & \frac{\sqrt{3}}{2} & 0 & 0 & 0 & 0 \\
 0 & 0 & \frac{\sqrt{3}}{2} & -\frac{1}{2} & 0 & 0 & 0 & 0 \\
 -\frac{1}{2} & 0 & 0 & 0 & 0 & 0 & \frac{\sqrt{3}}{2} & 0 \\
 -\frac{\sqrt{3}}{2} & 0 & 0 & 0 & 0 & 0 & -\frac{1}{2} & 0 \\
 0 & 0 & 0 & 0 & -\frac{\sqrt{3}}{2} & \frac{1}{2} & 0 & 0 \\
 0 & \frac{\sqrt{3}}{2} & 0 & 0 & 0 & 0 & 0 & -\frac{1}{2}
\end{array}
\right), \nonumber \\
&& \tau_{k+2}^{{\bf 2}_{[2111]}} =
\left(
\begin{array}{cccccccc}
 0 & 1 & 0 & 0 & 0 & 0 & 0 & 0 \\
 -1 & 0 & 0 & 0 & 0 & 0 & 0 & 0 \\
 0 & 0 & -1 & 0 & 0 & 0 & 0 & 0 \\
 0 & 0 & 0 & 1 & 0 & 0 & 0 & 0 \\
 0 & 0 & 0 & 0 & -1 & 0 & 0 & 0 \\
 0 & 0 & 0 & 0 & 0 & 1 & 0 & 0 \\
 0 & 0 & 0 & 0 & 0 & 0 & 0 & 1 \\
 0 & 0 & 0 & 0 & 0 & 0 & -1 & 0
\end{array}
\right),
\end{eqnarray}
for $\mathbb{H}^{{\bf 2}_{[2111]}}$,
\begin{eqnarray}
\tau_{k}^{{\bf 2}_{[2221]}} = 
\left(
\begin{array}{cccc}
 1 & 0 & 0 & 0 \\
 0 & 1 & 0 & 0 \\
 0 & 0 & 0 & 1 \\
 0 & 0 & -1 & 0
\end{array}
\right), \,
\tau_{k+1}^{{\bf 2}_{[2221]}} = 
\left(
\begin{array}{cccc}
 1 & 0 & 0 & 0 \\
 0 & 0 & 1 & 0 \\
 0 & -1 & 0 & 0 \\
 0 & 0 & 0 & 1
\end{array}
\right), \,
\tau_{k+2}^{{\bf 2}_{[2221]}} = 
\left(
\begin{array}{cccc}
 0 & 1 & 0 & 0 \\
 -1 & 0 & 0 & 0 \\
 0 & 0 & 1 & 0 \\
 0 & 0 & 0 & 1
\end{array}
\right),
\end{eqnarray}
for $\mathbb{H}^{{\bf 2}_{[2221]}}$.
For triplet, the matrices are
\begin{eqnarray}
\tau_{k}^{{\bf 3}_{[1100]}} =
\left(
\begin{array}{cccccc}
 1 & 0 & 0 & 0 & 0 & 0 \\
 0 & 0 & 0 & -1 & 0 & 0 \\
 0 & 0 & 0 & 0 & -1 & 0 \\
 0 & 1 & 0 & 0 & 0 & 0 \\
 0 & 0 & 1 & 0 & 0 & 0 \\
 0 & 0 & 0 & 0 & 0 & 1
\end{array}
\right), \,
\tau_{k+1}^{{\bf 3}_{[1100]}} =
\left(
\begin{array}{cccccc}
 0 & -1 & 0 & 0 & 0 & 0 \\
 1 & 0 & 0 & 0 & 0 & 0 \\
 0 & 0 & 1 & 0 & 0 & 0 \\
 0 & 0 & 0 & 1 & 0 & 0 \\
 0 & 0 & 0 & 0 & 0 & -1 \\
 0 & 0 & 0 & 0 & 1 & 0
\end{array}
\right), \,
\tau_{k+2}^{{\bf 3}_{[1100]}} =
\left(
\begin{array}{cccccc}
 1 & 0 & 0 & 0 & 0 & 0 \\
 0 & 0 & -1 & 0 & 0 & 0 \\
 0 & 1 & 0 & 0 & 0 & 0 \\
 0 & 0 & 0 & 0 & -1 & 0 \\
 0 & 0 & 0 & 1 & 0 & 0 \\
 0 & 0 & 0 & 0 & 0 & 1
\end{array}
\right),
\end{eqnarray}
for $\mathbb{H}^{{\bf 3}_{[1100]}}$,
\begin{eqnarray}
&& \tau_{k}^{{\bf 3}_{[2110]}} =
\left(
\begin{array}{cccccccccccc}
 0 & 1 & 0 & 0 & 0 & 0 & 0 & 0 & 0 & 0 & 0 & 0 \\
 1 & 0 & 0 & 0 & 0 & 0 & 0 & 0 & 0 & 0 & 0 & 0 \\
 0 & 0 & 0 & 0 & -1 & 0 & 0 & 0 & 0 & 0 & 0 & 0 \\
 0 & 0 & 0 & 0 & 0 & -1 & 0 & 0 & 0 & 0 & 0 & 0 \\
 0 & 0 & 1 & 0 & 0 & 0 & 0 & 0 & 0 & 0 & 0 & 0 \\
 0 & 0 & 0 & 1 & 0 & 0 & 0 & 0 & 0 & 0 & 0 & 0 \\
 0 & 0 & 0 & 0 & 0 & 0 & 0 & 0 & 1 & 0 & 0 & 0 \\
 0 & 0 & 0 & 0 & 0 & 0 & 0 & 0 & 0 & 1 & 0 & 0 \\
 0 & 0 & 0 & 0 & 0 & 0 & -1 & 0 & 0 & 0 & 0 & 0 \\
 0 & 0 & 0 & 0 & 0 & 0 & 0 & -1 & 0 & 0 & 0 & 0 \\
 0 & 0 & 0 & 0 & 0 & 0 & 0 & 0 & 0 & 0 & 1 & 0 \\
 0 & 0 & 0 & 0 & 0 & 0 & 0 & 0 & 0 & 0 & 0 & 1
\end{array}
\right), \,
\tau_{k+1}^{{\bf 3}_{[2110]}} =
\left(
\begin{array}{cccccccccccc}
 0 & 0 & 0 & 0 & 0 & 0 & 0 & 1 & 0 & 0 & 0 & 0 \\
 0 & 0 & 0 & 0 & 1 & 0 & 0 & 0 & 0 & 0 & 0 & 0 \\
 0 & 0 & 0 & 0 & 0 & 0 & 0 & 0 & 0 & 1 & 0 & 0 \\
 0 & 0 & 0 & 0 & 0 & 0 & 0 & 0 & 0 & 0 & 0 & 1 \\
 0 & -1 & 0 & 0 & 0 & 0 & 0 & 0 & 0 & 0 & 0 & 0 \\
 0 & 0 & 0 & 0 & 0 & 1 & 0 & 0 & 0 & 0 & 0 & 0 \\
 0 & 0 & 0 & 0 & 0 & 0 & 1 & 0 & 0 & 0 & 0 & 0 \\
 -1 & 0 & 0 & 0 & 0 & 0 & 0 & 0 & 0 & 0 & 0 & 0 \\
 0 & 0 & 0 & 0 & 0 & 0 & 0 & 0 & 0 & 0 & 1 & 0 \\
 0 & 0 & 1 & 0 & 0 & 0 & 0 & 0 & 0 & 0 & 0 & 0 \\
 0 & 0 & 0 & 0 & 0 & 0 & 0 & 0 & -1 & 0 & 0 & 0 \\
 0 & 0 & 0 & -1 & 0 & 0 & 0 & 0 & 0 & 0 & 0 & 0
\end{array}
\right), \nonumber \\
&& \tau_{k+2}^{{\bf 3}_{[2110]}} =
\left(
\begin{array}{cccccccccccc}
 1 & 0 & 0 & 0 & 0 & 0 & 0 & 0 & 0 & 0 & 0 & 0 \\
 0 & 1 & 0 & 0 & 0 & 0 & 0 & 0 & 0 & 0 & 0 & 0 \\
 0 & 0 & 0 & 1 & 0 & 0 & 0 & 0 & 0 & 0 & 0 & 0 \\
 0 & 0 & -1 & 0 & 0 & 0 & 0 & 0 & 0 & 0 & 0 & 0 \\
 0 & 0 & 0 & 0 & 0 & 1 & 0 & 0 & 0 & 0 & 0 & 0 \\
 0 & 0 & 0 & 0 & -1 & 0 & 0 & 0 & 0 & 0 & 0 & 0 \\
 0 & 0 & 0 & 0 & 0 & 0 & 0 & -1 & 0 & 0 & 0 & 0 \\
 0 & 0 & 0 & 0 & 0 & 0 & 1 & 0 & 0 & 0 & 0 & 0 \\
 0 & 0 & 0 & 0 & 0 & 0 & 0 & 0 & 0 & -1 & 0 & 0 \\
 0 & 0 & 0 & 0 & 0 & 0 & 0 & 0 & 1 & 0 & 0 & 0 \\
 0 & 0 & 0 & 0 & 0 & 0 & 0 & 0 & 0 & 0 & 0 & 1 \\
 0 & 0 & 0 & 0 & 0 & 0 & 0 & 0 & 0 & 0 & 1 & 0
\end{array}
\right),
\end{eqnarray}
for $\mathbb{H}^{{\bf 3}_{[2110]}}$,
\begin{eqnarray}
\tau_{k}^{{\bf 3}_{[1111]}} = 
\left(
\begin{array}{ccc}
 -1 & 0 & 0 \\
 0 & 1 & 0 \\
 0 & 0 & 1
\end{array}
\right), \,
\tau_{k+1}^{{\bf 3}_{[1111]}} = 
\left(
\begin{array}{ccc}
 \frac{1}{2} & -\frac{1}{2} & \frac{1}{\sqrt{2}} \\
 -\frac{1}{2} & \frac{1}{2} & \frac{1}{\sqrt{2}} \\
 \frac{1}{\sqrt{2}} & \frac{1}{\sqrt{2}} & 0
\end{array}
\right), \,
\tau_{k+2}^{{\bf 3}_{[1111]}} = 
\left(
\begin{array}{ccc}
 1 & 0 & 0 \\
 0 & -1 & 0 \\
 0 & 0 & 1
\end{array}
\right), \label{eq:matrix_3_1111}
\end{eqnarray}
for $\mathbb{H}^{{\bf 3}_{[1111]}}$,
\begin{eqnarray}
\tau_{k}^{{\bf 3}_{[2211]}} =
\left(
\begin{array}{cccccc}
 1 & 0 & 0 & 0 & 0 & 0 \\
 0 & 0 & 0 & 1 & 0 & 0 \\
 0 & 0 & 0 & 0 & 1 & 0 \\
 0 & -1 & 0 & 0 & 0 & 0 \\
 0 & 0 & -1 & 0 & 0 & 0 \\
 0 & 0 & 0 & 0 & 0 & 1
\end{array}
\right), \,
\tau_{k+1}^{{\bf 3}_{[2211]}} =
\left(
\begin{array}{cccccc}
 0 & 1 & 0 & 0 & 0 & 0 \\
 -1 & 0 & 0 & 0 & 0 & 0 \\
 0 & 0 & 1 & 0 & 0 & 0 \\
 0 & 0 & 0 & 1 & 0 & 0 \\
 0 & 0 & 0 & 0 & 0 & 1 \\
 0 & 0 & 0 & 0 & -1 & 0
\end{array}
\right), \,
\tau_{k+2}^{{\bf 3}_{[2211]}} =
\left(
\begin{array}{cccccc}
 1 & 0 & 0 & 0 & 0 & 0 \\
 0 & 0 & 1 & 0 & 0 & 0 \\
 0 & -1 & 0 & 0 & 0 & 0 \\
 0 & 0 & 0 & 0 & 1 & 0 \\
 0 & 0 & 0 & -1 & 0 & 0 \\
 0 & 0 & 0 & 0 & 0 & 1
\end{array}
\right),
\end{eqnarray}
for $\mathbb{H}^{{\bf 3}_{[2211]}}$.
For quartet, the matrices are
\begin{eqnarray}
\tau_{k}^{{\bf 4}_{[1110]}} = 
\left(
\begin{array}{cccc}
 1 & 0 & 0 & 0 \\
 0 & 1 & 0 & 0 \\
 0 & 0 & 0 & -1 \\
 0 & 0 & 1 & 0
\end{array}
\right), \,
\tau_{k+1}^{{\bf 4}_{[1110]}} = 
\left(
\begin{array}{cccc}
 1 & 0 & 0 & 0 \\
 0 & 0 & -1 & 0 \\
 0 & 1 & 0 & 0 \\
 0 & 0 & 0 & 1
\end{array}
\right), \,
\tau_{k+2}^{{\bf 4}_{[1110]}} = 
\left(
\begin{array}{cccc}
 0 & -1 & 0 & 0 \\
 1 & 0 & 0 & 0 \\
 0 & 0 & 1 & 0 \\
 0 & 0 & 0 & 1
\end{array}
\right),
\end{eqnarray}
for $\mathbb{H}^{{\bf 4}_{[1110]}}$,
\begin{eqnarray}
\tau_{k}^{{\bf 4}_{[2111]}} = 
\left(
\begin{array}{cccc}
 0 & 1 & 0 & 0 \\
 -1 & 0 & 0 & 0 \\
 0 & 0 & 1 & 0 \\
 0 & 0 & 0 & 1
\end{array}
\right), \,
\tau_{k+1}^{{\bf 4}_{[2111]}} = 
\left(
\begin{array}{cccc}
 1 & 0 & 0 & 0 \\
 0 & 0 & 1 & 0 \\
 0 & -1 & 0 & 0 \\
 0 & 0 & 0 & 1
\end{array}
\right), \,
\tau_{k+2}^{{\bf 4}_{[2111]}} = 
\left(
\begin{array}{cccc}
 1 & 0 & 0 & 0 \\
 0 & 1 & 0 & 0 \\
 0 & 0 & 0 & 1 \\
 0 & 0 & -1 & 0
\end{array}
\right),
\end{eqnarray}
for $\mathbb{H}^{{\bf 4}_{[2111]}}$.
For quintet, the matrices are
\begin{eqnarray}
\tau_{k}^{{\bf 5}_{[1111]}} = \tau_{k+1}^{{\bf 5}_{[1111]}} = \tau_{k+2}^{{\bf 5}_{[1111]}} = 1,
\end{eqnarray}
for $\mathbb{H}^{{\bf 5}_{[1111]}}$.
Interestingly, we find again the non-Abelian matrices in several Hilbert subspaces.
For example, the matrices $\tau_{k}^{{\bf 1}_{[2000]}}$, $\tau_{k+1}^{{\bf 1}_{[2000]}}$ and $\tau_{k+2}^{{\bf 1}_{[2000]}}$ in the Hilbert subspace $\mathbb{H}^{{\bf 1}_{[2000]}}$ are non-commutative; $\tau_{\ell}^{{\bf 1}_{[2000]}} \tau_{\ell+1}^{{\bf 1}_{[2000]}} \neq \tau_{\ell+1}^{{\bf 1}_{[2000]}} \tau_{\ell}^{{\bf 1}_{[2000]}}$ for $\ell=k$, $k+1$.
Therefore, the exchange of the $\ell$-th and $(\ell+1)$-th vortices ($\ell=k$, $k+1$) induces the non-Abelian representation of the braid group in $\mathbb{H}^{{\bf 1}_{[2000]}}$.
Similarly, the non-Abelian representation of the braid group is realized in the following Hilbert subspaces; $\mathbb{H}^{{\bf 1}_{[1100]}}$, $\mathbb{H}^{{\bf 1}_{[2200]}}$, $\mathbb{H}^{{\bf 1}_{[2110]}}$, $\mathbb{H}^{{\bf 1}_{[2220]}}$, $\mathbb{H}^{{\bf 1}_{[1111]}}$ and $\mathbb{H}^{{\bf 1}_{[2211]}}$ for singlet, $\mathbb{H}^{{\bf 2}_{[1000]}}$, $\mathbb{H}^{{\bf 2}_{[2100]}}$, $\mathbb{H}^{{\bf 2}_{[1110]}}$, $\mathbb{H}^{{\bf 2}_{[2210]}}$, $\mathbb{H}^{{\bf 2}_{[2111]}}$ and $\mathbb{H}^{{\bf 2}_{[2221]}}$ for doublet, $\mathbb{H}^{{\bf 3}_{[1100]}}$, $\mathbb{H}^{{\bf 3}_{[2110]}}$, $\mathbb{H}^{{\bf 3}_{[1111]}}$ and $\mathbb{H}^{{\bf 3}_{[2211]}}$ for triplet, $\mathbb{H}^{{\bf 4}_{[1110]}}$ and $\mathbb{H}^{{\bf 4}_{[2111]}}$ for quartet.

As we have discussed in the text, the U(1) Dirac fermions are embedded in the U(2) Dirac fermions.
When $\hat{\psi}_{\ell}^{2}$ ($\ell=k,k+1,k+2,k+3$) are set to zero,
the Hilbert subspaces $\mathbb{H}^{{\bf 1}_{[0000]}}$, $\mathbb{H}^{{\bf 2}_{[1000]}}$, $\mathbb{H}^{{\bf 3}_{[1100]}}$, $\mathbb{H}^{{\bf 4}_{[1110]}}$ and $\mathbb{H}^{{\bf 5}_{[1111]}}$ in U(2) Dirac vortices coincide with the Hilbert subspaces $\mathbb{H}^{(4,0)}$, $\mathbb{H}^{(4,1)}$, $\mathbb{H}^{(4,2)}$, $\mathbb{H}^{(4,3)}$ and $\mathbb{H}^{(4,4)}$ in U(1) Dirac vortices, respectively.
The matrices between the two are equivalent;
\begin{eqnarray}
 \tau_{\ell}^{{\bf 1}_{[0000]}} &=& \tau_{\ell}^{(4,0)}, \\
 \tau_{\ell}^{{\bf 2}_{[1000]}} &=& \tau_{\ell}^{(4,1)}, \\
 \tau_{\ell}^{{\bf 3}_{[1100]}} &=& \tau_{\ell}^{(4,2)}, \\
 \tau_{\ell}^{{\bf 4}_{[1110]}} &=& \tau_{\ell}^{(4,3)}, \\
 \tau_{\ell}^{{\bf 5}_{[1111]}} &=& \tau_{\ell}^{(4,4)},
\end{eqnarray}
with $\ell=k$, $k+1$, $k+2$.

\section{Subspaces with $(\tau_{k})^{2}=1$ in U(2) Dirac vortices}
\label{sec:square}

\renewcommand{\theequation}{C.\arabic{equation}}

\setcounter{equation}{0}

We recall that, for both cases with U(1) and U(2) Dirac fermions, 
{\it four}-time exchange of vortices is equivalent to the identity; 
$(T_{k})^{4}=1$. The same relation holds at the operator level: 
$(\hat{\tau}_{k}^{\mathrm{s}})^{4}=1$ for U(1) Dirac vortices and
$(\hat{\tau}_{k})^{4}=1$ for U(2) Dirac vortices. 
However, a matrix $\tau_{k}$ representing $\hat{\tau}_{k}^{\mathrm{s}}$
or $\hat{\tau}_{k}$ happens to satisfy a stronger relation, 
$(\tau_{k})^{2}=1$, in some Hilbert subspaces. 
There, {\it two}-time exchange of vortices is equivalent to identity.
In this Appendix, we explain how such a relation can be satisfied,
and check if the representation of the braid group is still non-Abelian.

First, we consider the case of U(1) Dirac fermions.
From Eq.~(\ref{eq:second}), we find that
$(\hat{\tau}_{k}^{\mathrm{s}})^{2}$ is expressed in terms of the number
operator of Dirac fermions in the $\ell$-th ($\ell=k$, $k+1$) vortices,
$\hat{\psi}_{\ell}^{\mathrm{s}\dag} \hat{\psi}_{\ell}^{\mathrm{s}}$.
If we define $N_{\ell}^{\mathrm{s}}=0$, 1 as an expectation 
value of the number operator $\hat{\psi}_{\ell}^{\mathrm{s}\dag}
\hat{\psi}_{\ell}^{\mathrm{s}}$, we find
that the matrix $\tau^{\mathrm{s}}_{k}$ representing the operator
$\hat{\tau}^{\mathrm{s}}_{k}$ yields
\begin{eqnarray}
(\tau_{k}^{\mathrm{s}})^{2} = 
(1-2N_{k}^{\mathrm{s}}) (1-2N_{k+1}^{\mathrm{s}})\, .
\end{eqnarray}
The right-hand-side reduces to 1 only when 
$(N_{k}^{\mathrm{s}}, N_{k+1}^{\mathrm{s}})=(0, 0)$ or $(1,1)$. 
Therefore, we conclude that the relation 
$(\tau_{k}^{\mathrm{s}})^{2}=1$ holds only when all the vortices are empty 
or fully occupied: $(N_{1}^{\mathrm{s}}, \cdots, N_{n}^{\mathrm{s}})=(0, \cdots, 0)$ or $(1, \cdots, 1)$. 
In both cases, $\tau_{\ell}^{\mathrm{s}}=1$ for any $\ell=1, \cdots, n-1$, 
and hence it gives just trivial representation of the braid group.
In fact, as shown in Appendix \ref{sec:U(1)tau}, the matrices for the 
empty and fully-occupied states are 
$\tau_{\ell}^{(n,0)}=\tau_{\ell}^{(n,n)}=1$ with 
$\ell = 1, \cdots, n-1$ for $n=2$, 3, 4. 

Second, let us consider the case of U(2) Dirac fermions.
In this case, there are non-Abelian matrices satisfying $(\tau_{\ell})^{2}=1$ for any $\ell=1,\cdots,n-1$.
From Eq.~(\ref{eq:tau^2}), we find again that $(\hat{\tau}_{k})^{2}$ is expressed in terms of the number operator of Dirac fermions in the $\ell$-th ($\ell=k$, $k+1$) vortices, $\hat{\psi}_{\ell}^{a\dag} \hat{\psi}_{\ell}^{a}$ with $a=1$, 2.
Note that the indices $a=1$, 2 of the pseudo-spin are introduced.
Then, defining $N^{a}_{\ell}=0$, 1 as an expectation value of the number operator $\hat{\psi}_{\ell}^{a\dag} \hat{\psi}_{\ell}^{a}$, we find that the matrix $\tau_{k}$ representing the operator $\hat{\tau}_{k}$ yields
\begin{eqnarray}
(\tau_{k})^{2} = (1-2N_{k}^{1}) (1-2N_{k}^{2}) (1-2N_{k+1}^{1}) (1-2N_{k+1}^{2}).
\end{eqnarray}
The relation $(\tau_{k})^{2}=1$ is fulfilled by the following combinations,
\begin{eqnarray}
(N_{k}^{1}, N_{k}^{2}, N_{k+1}^{1}, N_{k+1}^{2}) &=& (0,0,0,0), \quad (1,1,1,1), \quad (1,1,0,0), \quad (0,0,1,1), \nonumber \\
&& (1,0,1,0), \quad (1,0,0,1), \quad (0,1,1,0), \quad (0,1,0,1).
\end{eqnarray}
When we define $N_{\ell}=N_{\ell}^{1}+N_{\ell}^{2}$, the above combinations are further rewritten as
\begin{eqnarray}
(N_{k}, N_{k+1}) &=& (0,0), \quad (2,2), \quad (2,0), \quad (0,2), \quad (1,1).
\end{eqnarray}
Therefore, we conclude that the relation $(\tau_{k})^{2}=1$ is
satisfied when $N_{k}+N_{k+1}$ is an even number.
This conclusion is consistent with the expectation from the
transformation properties of the Dirac fermions under the operation $(T_k)^2$.
Under the two successive exchanges of $k$-th an $(k+1)$-th vortices, 
the Dirac fermion operators 
$\hat{\psi}_k^a$ and $\hat{\psi}_{k+1}^a$ are multiplied by $-1$.
If a state is composed of an even number of $k$-th and $(k+1)$-th
fermions, the minus signs cancel and the state is unchanged under $(T_k)^2$.
Therefore, in order for that the condition $(\tau_{\ell})^{2}=1$ holds for
any $\ell = 1, \cdots, n-1$, a sum of the Dirac fermion number in the
every neighboring vortices, $N_{\ell}+N_{\ell+1}$, has to be an even
number.
We note that $(N_{1}, \cdots, N_{n})=(0, \cdots, 0)$ and $(2, \cdots,
2)$ corresponding to the empty state and fully-occupied state,
respectively, give just trivial representation of the braid group $\tau_{\ell}=1$, like the
case of U(1) Dirac fermions as discussed above.
It is also the case for the state with the highest dimension in
pseudo-spin representation. 
However, the other combinations of $(N_{1}, \cdots, N_{n})$ induce
non-Abelian matrices, namely non-Abelian representation of the braid group.

Let us see examples in U(2) Dirac fermions with $n=3$ and $4$.

a) For $n=3$, the Hilbert subspaces, in which $N_{\ell}+N_{\ell+1}$ is an even number for any $\ell = 1$, $2$, are
\begin{eqnarray}
 \mathbb{H}^{{\bf 1}_{[000]}}, \quad \mathbb{H}^{{\bf 1}_{[200]}}, \quad \mathbb{H}^{{\bf 1}_{[220]}}, \quad \mathbb{H}^{{\bf 1}_{[222]}}, \quad \mathbb{H}^{{\bf 2}_{[111]}} \quad \mbox{and} \quad \mathbb{H}^{{\bf 4}_{[111]}}.
\end{eqnarray}
The matrices in each Hilbert subspace are
\begin{eqnarray}
 \tau_{\ell}^{{\bf 1}_{[000]}}, \quad \tau_{\ell}^{{\bf 1}_{[200]}}, \quad \tau_{\ell}^{{\bf 1}_{[220]}}, \quad \tau_{\ell}^{{\bf 1}_{[222]}}, \quad \tau_{\ell}^{{\bf 2}_{[111]}} \quad \mbox{and} \quad \tau_{\ell}^{{\bf 4}_{[111]}}.
\end{eqnarray}
Among them, the matrices $\tau_{\ell}^{{\bf 1}_{[000]}}$, $\tau_{\ell}^{{\bf 1}_{[222]}}$ and $\tau_{\ell}^{{\bf 4}_{[111]}}$ are trivial, because they correspond to the empty state, full-occupied state and the state with highest dimension in pseudo-spin representation, respectively.
The other matrices $\tau_{\ell}^{{\bf 1}_{[200]}}$, $\tau_{\ell}^{{\bf 1}_{[220]}}$ and $\tau_{\ell}^{{\bf 2}_{[111]}}$ are non-Abelian matrices, and hence the Hilbert subspaces $\mathbb{H}^{{\bf 1}_{[200]}}$, $\mathbb{H}^{{\bf 1}_{[220]}}$ and $\mathbb{H}^{{\bf 2}_{[111]}}$ lead to the non-Abelian representation of the braid group satisfying $(\tau_{\ell})^{2}=1$ for $\ell=1$, $2$.

b) For $n=4$, from Appendix \ref{sec:n=4}, the Hilbert subspaces, in which $N_{\ell}+N_{\ell+1}$ is an even number for any $\ell = 1$, $2$, $3$, are
\begin{eqnarray}
 \mathbb{H}^{{\bf 1}_{[0000]}}, \quad \mathbb{H}^{{\bf 1}_{[2000]}}, \quad \mathbb{H}^{{\bf 1}_{[2200]}}, \quad \mathbb{H}^{{\bf 1}_{[2220]}}, \quad \mathbb{H}^{{\bf 1}_{[1111]}}, \quad \mathbb{H}^{{\bf 1}_{[2222]}}, \quad \mathbb{H}^{{\bf 3}_{[1111]}}\quad \mbox{and} \quad \mathbb{H}^{{\bf 5}_{[1111]}}.
\end{eqnarray}
The matrices in each Hilbert subspace are
\begin{eqnarray}
 \tau_{\ell}^{{\bf 1}_{[0000]}}, \quad \tau_{\ell}^{{\bf 1}_{[2000]}}, \quad \tau_{\ell}^{{\bf 1}_{[2200]}}, \quad \tau_{\ell}^{{\bf 1}_{[2220]}}, \quad \tau_{\ell}^{{\bf 1}_{[1111]}}, \quad \tau_{\ell}^{{\bf 1}_{[2222]}}, \quad \tau_{\ell}^{{\bf 3}_{[1111]}}\quad \mbox{and} \quad \tau_{\ell}^{{\bf 5}_{[1111]}}.
\end{eqnarray}
Among them, the matrices $\tau_{\ell}^{{\bf 1}_{[0000]}}$, $\tau_{\ell}^{{\bf 1}_{[2222]}}$ and $\tau_{\ell}^{{\bf 5}_{[1111]}}$ are trivial, because they correspond to the empty state, full-occupied state and the state with highest dimension in pseudo-spin representation, respectively.
The other matrices $\tau_{\ell}^{{\bf 1}_{[2000]}}$, $\tau_{\ell}^{{\bf 1}_{[2200]}}$, $\tau_{\ell}^{{\bf 1}_{[2220]}}$, $\tau_{\ell}^{{\bf 1}_{[1111]}}$ and $\tau_{\ell}^{{\bf 3}_{[1111]}}$ are non-Abelian matrices, and hence the Hilbert subspaces $\mathbb{H}^{{\bf 1}_{[2000]}}$, $\mathbb{H}^{{\bf 1}_{[2200]}}$, $\mathbb{H}^{{\bf 1}_{[2220]}}$, $\mathbb{H}^{{\bf 1}_{[1111]}}$ and $\mathbb{H}^{{\bf 3}_{[1111]}}$ lead to the non-Abelian representation of the braid group satisfying $(\tau_{\ell})^{2}=1$ for $\ell=1$, $2$, $3$.

We note that the condition $(\tau_{k})^{2}=1$ with the braid relations (i) $\tau_{k}\tau_{\ell}\tau_{k}=\tau_{\ell}\tau_{k}\tau_{\ell}$ for $|k-\ell|=1$ and (ii) $\tau_{k}\tau_{\ell}=\tau_{\ell}\tau_{k}$ for $|k-\ell|>1$ leads to the relations, (i') $(\tau_{k}\tau_{\ell})^{3} =1$ for $|k-\ell|=1$ and (ii') $(\tau_{k}\tau_{\ell})^{2}=1$ for $|k-\ell|>1$ for $k, \ell=1, \cdots, n-1$.
The relations (i') and (ii')
 imply that such matrices $\tau_{\ell}$ are regarded as generators of the symmetric group or the Coxeter group of the type $A_{n-1}$.
We recall that the Coxeter group was obtained in the SO(3) or more generally SO($N$) ($N$ odd numbers) symmetric Majorana vortices \cite{Yasui:2010yh,Hirono:2012ad}.
There, the matrices for exchanging vortices were tensor product of ``the Ivanov matrices'' found by Ivanov \cite{Ivanov:2001} and the generators of the Coxeter group of $A_{2m-1}$ type ($2m$ the number of vortices).
In the present case of U(2) Dirac vortices, however, such tensor structure was not found.

\end{appendix}

%%%%%%%%%%%%%%%%%%%%%%%%%%%%%%%%%%%%%%%%%%%%%%%%%

\end{document}